\documentclass[useAMS,usenatbib]{mn2e}
\usepackage{graphicx,amssymb,times,amsmath}
\usepackage{longtable}
\usepackage{cleveref}
\usepackage{float}

\usepackage{hyperref}
\hypersetup{colorlinks=true, urlcolor=blue, citecolor=cyan, pdfborder={0 0 0},}
\usepackage{soul}

\crefname{section}{§}{§§}
\Crefname{section}{§}{§§}
\newcommand{\kms}{\,km\,s$^{-1}$~}
\newcommand{\kmss}{\,km\,s$^{-1}$}

\title[A 7 mm line survey of Sgr~B2]{An ATCA Survey of Sagittarius~B2 at 7~mm: Chemical Complexity Meets Broadband Interferometry}
\author[J. Corby et al.]{Joanna F. Corby$^{1}$
\thanks{jfc2113@gmail.com},
Paul A. Jones$^{2}$, Maria R. Cunningham$^{2}$, 
\newauthor{Karl M. Menten$^{3}$, Arnaud Belloche$^{3}$, Frederic R. Schwab$^{4}$, Andrew J. Walsh$^{5}$,} 
\newauthor{Egon Balnozan$^{2}$, Leonardo Bronfman$^{6}$, Nadia Lo$^{6}$, Anthony J. Remijan$^{4}$}
\\
$^{1}$ Department of Astronomy, University of Virginia,  Charlottesville, VA 22903 \\
$^{2}$ School of Physics, University of New South Wales, NSW 2052, Australia \\
$^{3}$ Max-Planck-Institut f{\"u}r Radioastronomie, D-53121 Bonn, Germany \\
$^{4}$ National Radio Astronomy Observatory, Charlottesville, VA 22903, USA \\
$^{5}$ Department of Physics and Astronomy, Curtin University, WA 6102, Australia\\
$^{6}$ Departamento de Astronom{\'{\i}}a, Universidad de Chile, Santiago, Chile \\
}

\begin{document}

\date{Accepted 2015 July 3. Received 2015 June 16; in original form 2015 March 24}

\pagerange{\pageref{firstpage}--\pageref{lastpage}} \pubyear{2015}

\maketitle

\label{firstpage}

\begin{abstract}

We present a 30 - 50~GHz survey of Sagittarius B2(N) conducted with the Australia Telescope Compact Array (ATCA)
with 5 - 10 arcsec resolution.  This work releases the survey data and demonstrates the utility of scripts 
that perform automated spectral 
line fitting on broadband line data.  We describe the line-fitting procedure, evaluate the performance of the 
method, and provide access to all data and scripts.  The scripts are used to characterize the spectra at the positions 
of three H{\sc ii} regions, each with recombination line emission and molecular line absorption.  Towards the most 
line-dense of the three regions characterised in this work, we detect $\sim$500 spectral line components of which
$\sim$90\% are confidently assigned to H and He recombination lines and to 53 molecular species and their isotopologues.  

The data reveal extremely subthermally excited molecular gas absorbing against the continuum background at two 
primary velocity components.  Based on the line radiation over the full spectra, the molecular abundances 
and line excitation in the 
absorbing components appear to vary substantially towards the different positions, possibly indicating that the two gas 
clouds are located proximate to the star forming cores instead of within the envelope of Sgr B2.  
Furthermore, the spatial distributions of species including CS, OCS, SiO, and HNCO indicate that the absorbing 
gas components likely have high UV-flux. 

Finally, the data contain line-of-sight absorption by $\sim$15 molecules observed in translucent gas in the 
Galactic Center, bar, and intervening spiral arm clouds, revealing the complex chemistry and clumpy 
structure of this gas.  Formamide (NH$_2$CHO) is detected for the first time in a translucent cloud.

\end{abstract}

\begin{keywords}
astrochemistry -- ISM: H\sc{ii} regions -- ISM: molecules -- radio lines:ISM -- 
stars: formation -- techniques: interferometric -- ISM:individual (Sagittarius B2) 
\end{keywords}

\section[]{Introduction}
\label{sec:intro}

The complex high mass star forming region Sagittarius B2 (Sgr B2) is a uniquely interesting
source for studying star formation and molecular chemistry.  Sgr B2 is an exceptionally massive
cloud complex at $\sim$3$\times$10$^6$ M$_{\odot}$ \citep{SgrMass}, 
an order of magnitude higher than most star-forming Giant Molecular Clouds, and is in 
the Central Molecular Zone of the Galactic Center, where widescale shocks
and supersonic turbulence are observed to be nearly ubiquitous \citep{MartinP01, Martin08}. 
Furthermore, the region has an exceptionally high X-ray flux \citep{SgrXray}, 
cosmic ray ionization rate \citep{CRio}, and magnetic field strength \citep{Bfield}, making it 
an exceptional case of star formation that is perhaps more similar to the central regions of 
starburst galaxies than it is to most star forming regions in the Galaxy.  In fact, the star formation
rate in Sgr B2 qualifies the region as a mini-starburst \citep{Belloche2013}.

Sgr B2 has a complex physical structure with more than 50 known extended, compact, ultra-compact, and hypercompact H{\sc ii} regions embedded in an
extended envelope of molecular, atomic, and ionized gas \citep{Gaume95, Envelope, Jones08}.  Most of the star forming activity
in the region is concentrated in the North, Main, and South cores.  The three cores of
star formation have a v$_{\text{LSR}}$ of $\sim$55 - 70 \kmss, and the star formation is thought to be
triggered by the collision of a cloud of gas at $\sim$80 - 90 \kms with the main body of Sgr B2 at 
$\sim$40~\kms \citep{Hasegawa94, Sato00}.  
Possible ejecta from Sgr B2 have been identified via line-of-sight molecular absorption 
at 0 - 20 \kmss, although recent observations indicate that
this may correspond to a more extended feature across the Galactic Center \citep{WirstromNH3, Jones12, MarcCar}.
Additional foreground components of molecular gas associated with the Galactic Center and spiral arms
are similarly observed in line-of-sight absorption towards Sgr B2 (Greaves \& Nyman 1996; Gerin et. al 2010 and references therein).

Towards the North core of Sgr B2 (Sgr B2(N)), the focus of this study, continuum emission is produced by a complex
of H{\sc ii} regions, as observed and discussed in \cite{Gaume95}.  
Two $\sim$10 arcsec \,(= 0.4 pc) concentric shell-shaped regions -- K5 and K6 -- are thought to be expanding ionization fronts (Figure \ref{regionsFig}).
To the NW of these, the more compact region K4 is thought to be a similar system at an earlier stage of development.  
Three sub-arcsecond cores of continuum emission -- K1, K2, and K3 -- are located to the SW of K5 and K6.  
Finally, a cometary H{\sc ii} region, L, is located $\sim$35 arcsec NE of the North core.

Molecular gas is observed in the foreground of the H{\sc ii} regions and in extended material surrounding the cores
\citep{HuttemeisterNH3,Jones08}.   Additionally, the Large Molecule Heimat (LMH) is a $\sim$5~arcsec
region surrounding K2 that is continuum-free at radio wavelengths, but 
bright in submillimeter continuum \citep{Miao95, Qin11}.  The LMH is a hot core that is home to the most diverse 
molecular chemistry observed in the ISM, and a significant fraction of the molecular lines detected towards
the LMH remain unidentified \citep{Belloche2013}.  In addition to the LMH, Sgr B2(N) has a second hot core named ``h'' \citep{Belloche2013}. 
The ``h'' region is located on the K5 shell and was first identified as a quasi-thermal methanol core \citep{M-M97}.  Like the LMH, 
``h'' produces submillimeter continuum emission \citep{Qin11} and line emission by many complex organic species, including the first branched
alkyl species detected in the ISM \citep{Hollis03, Belloche2013, Belloche2014Sci}.

Sgr B2 has been targeted by spectral line surveys from centimeter to sub-millimeter
wavelengths \citep{Turner89, Nummelin98, PRIMOS, Belloche2013,HEXOSsgr}, however, most surveys have been conducted with
single dish telescopes at frequencies \textgreater~80~GHz.  At these frequencies, there
is significant line confusion (a single feature can be assigned to multiple possible known or yet unknown carriers)
and line blending (a single feature consists of multiple transitions of known molecules), as the data are
line confusion limited rather than noise limited.  On the other hand, data are noise-limited at centimeter wavelengths,
with significantly less line confusion and blending, enabling more confident line identifications
and measurements of spectral line parameters (i.e. shape, velocity, width, and flux) \citep{McGuire2012}.   

In Sgr B2 and other sources with
chemical variation on size scales smaller than the beam of a single dish telescope, interferometric observations
are critically important for understanding the chemical environments.  Observing the same portion of the spectrum
with both an interferometer and a single dish telescope provides sensitivity to weak lines in extended gas (single dish)
and resolves fine spatial variations to distinguish chemically distinct regions (interferometer).
In molecule-rich sources, interferometric observations can also help to mitigate line confusion and blending, as
lines from different molecules can peak at distinct positions.  This further enables interferometric observations to be 
useful for demonstrating that multiple lines originate from a common molecular carrier, particularly for weak features (sub-100~mK)
produced by species with low abundance, low dipole moments, or high partition functions.

To date, only a few molecular lines mapped with the spatial resolution required to distinguish
the distinct chemical environments in Sgr B2(N) have been published \citep{LiuSnyder99,Hollis03,Belloche08}.  
With the arrival of broadband radio interferometers including the Australia
Telescope Compact Array (ATCA) with the CABB backend, the Karl G. Jansky Very Large Array (VLA), and the Atacama Large
Milimeter Array (ALMA), interferometric line surveys are now efficient, providing access to the spatial
distributions of hundreds of molecular lines within a reasonable observing time.  With the new wealth
of information available, however, come new challenges in data analysis that require automated methods.
To date, these are not adequately developed, and large scale efforts to provide improved
visualization and analysis tools for rich spectral line data cubes are underway \citep{Teuben13}.

In this work, we describe a 30 - 50 GHz spectral line survey conducted with the ATCA.  
The survey provides detailed information on chemical differentiation in Sgr B2(N), the 
most extensive catalog of radio recombination lines collected with an interferometer, and 
novel information on the structure and molecular content of clouds observed in the line-of-sight to Sgr B2. 
The survey serves as a pathfinder for a planned science verification project with ALMA Band 1 (35-50 GHz), presently being developed,
and showcases methods that will be useful for this and other future projects.
We describe the observations (\S\ref{sec:obs}),
present an automated line fitting and identification routine (\S\ref{sec:dataHandling}), and apply the algorithms
to selected regions of the Sgr B2(N) complex (\S\ref{sec:results}).  In \S\ref{sec:molComp}, we characterise the molecular gas
and highlight the distributions of a few molecular species showing distinct morphologies.  
We focus specifically on line identifications
and parameter fitting towards K6, K4, and L, and do not provide full line identifications towards the LMH and h.
The primary aim of this work is not to perform a complete spectral line analysis but rather to
test the functionality of the fitting routine on regions that show 
line absorption and emission at multiple velocity components. 
The type of data analysis routines described 
will be essential for characterizing and analyzing the spectra of complex molecular line sources 
at higher frequencies.  
Furthermore, we release all data and routines to the community to facilitate more complete analyses 
on the many projects that the data set enables.

\section[]{Observations}
\label{sec:obs}

\subsection[]{Telescope Configurations}
\label{sec:arr}

The observations were conducted with the Australia Telecope Compact Array (ATCA) with the Compact Array Broadband Backend (CABB) 
\citep{wi+11} in broadband mode (CFB 1M-0.5k), providing two simultaneous spectra with 2048 channels of 1MHz width.
Eleven different tunings with a separation of 1.85 GHz were required to cover the full 30 - 50~GHz range available with the 7~mm receiver (Table \ref{obs_summary}).
The 1~MHz channels provide a velocity resolution of 10 - 6~\kms from 30 - 50~GHz. 
Even given the broad linewidths in Sgr~B2, the spectral resolution is sub-optimal, so that lines may not be Nyquist sampled. 
However, the broadband mode of the CABB was selected as it makes a line survey over the whole 7~mm band feasible within a relatively short observing time.

The observations were made in three $\sim$8~hour sessions in 2011 October and 2013 April, with hybrid arrays H75 and H214, respectively,
as shown in Table \ref{obs_summary}. 
In each session, two pairs of tunings were observed, and the 43.70~GHz tuning was repeated in each configuration.
The H214 hybrid array enables good (u,v)-coverage of Sgr B2(N), providing sufficient resolution to separate the LMH 
from the shell-shaped H{\sc ii} regions without losing significant surface-brightness sensitivity. 
The H75 array is more compact than ideal but this is somewhat countered by using this array for the highest frequency tunings.
Resulting angular resolutions, ranging from 3.4 to 13 arcsec are provided in Table \ref{rms_extracted}.

The telescope was pointed towards Sgr B2(N) at $\alpha$ = $17^h47^m20\fs4$, $\delta$ = $-28\degr22\arcmin12\arcsec$
(J2000). We observed a cycle of one minute on the complex-gain calibrator (1714-336), ten minutes on source, one minute 
on the gain calibrator with one pair of tunings (A1 and A2 in Table \ref{obs_summary}) and then repeated the
calibrator-source-calibrator cycle with the second pair of tunings (B1 and B2).
The telescope pointing was updated approximately every
hour towards the gain calibrator.  Uranus and 3C~279 were observed as the primary flux and bandpass calibrators respectively.

\subsection[]{Data Reduction}
\label{sec:dataRed}

The raw data were reduced into calibrated data cubes using mostly standard
techniques of bandpass, absolute flux, and complex gain calibration in the {\sc miriad} package \citep{satewr95}. 
To produce continuum-free line data, the continuum was subtracted
in the (u,v)-domain and the fitted continuum was output into separate files.  
A single continuum image was generated from each of the three observing sessions 
and deconvolved with the {\sc clean} algorithm. 
The {\sc clean} model was input to determine a phase-only self-calibration solution, 
which was then applied to the line data and the continuum data.  A final continuum 
image was generated using the {\sc clean} algorithm from the 7.5~GHz bandwidth data obtained in each observing session.

As this is among the first  
spectral line surveys completed with a broadband interferometer, we highlight 
our method for baseline subtraction and challenges for bandpass stability. 
We performed baseline subtraction using the {\sc miriad} task {\sc uvlin}, 
generating a linear fit to the baseline of each 2~GHz tuning in the (u,v)-domain.  
The continuum images were generated from the continuum solutions output by {\sc uvlin}. Because Sgr B2 
has a very high line density and because it is difficult to define line-free channels in the (u,v)-domain,
we only excluded channels covering strongly masing lines from the fit. As very strong lines 
covered a small fraction of the bandwidth of each image, this had a very small effect 
on the continuum subtraction, resulting in a $\sim$1~mJy/beam offset that is \textless 0.1\% of the continuum level.  
Such a small effect might be deemed negligible in most data sets, however given the sensitivity of this data, it is non-negligible.

A larger effect was noticed in the form of a baseline wiggle in the images with a maximum amplitude of \textless1.5\% of the 
continuum strength.  This resulted from a limited bandpass stability due to the very large bandwidth in each tuning.
The bandpass shape is quite complex, and while the selected bandpass calibrator was very strong ($\sim$25~Jy), 
the strong continuum of Sgr B2 ($\sim$1~Jy) made it difficult to 
model the continuum to the dynamic range (about 3 orders of magnitude) required.  Furthermore,
the continuum positions of Sgr~B2(N) and (M)
are the positions that contain most of the molecular line features, so that the residual continuum preferentially affected
areas that are important for the line study.

To correct for these effects, we applied a continuum correction to the data in the image domain. 
Lines were masked at the 3$\sigma$ level, and each 2~GHz cube was fitted with a high order
polynomial to make a baselevel offset spectrum which was subtracted from the data cube.  
We note further processing of the baseline applied to the published spectra in 
\S\ref{sec:sou}.

We note that the CABB flagged a 
consistent set of channels in each observation.  In the publically released images, 
in which the first and last 50 channels of each 2~GHz wide module are cut, the zero-indexed channels 
affected include channels 0, 1, 78, 106, 206, 462, 590, 718, 974, 1102, 1126, 1230, 
1358, 1486, 1742, 1870, and 1947. 
All data cubes and continuum images are available at the CDS database\citep{CDS} 
via anonymous ftp to cdsarc.u-strasbg.fr (130.79.128.5)
or via http://cdsarc.u-strasbg.fr/viz-bin/qcat?J/MNRAS/.
The continuum-subtracted data cubes provided do not have 
a primary beam correction applied.  Models for the ATCA primary beam can be computed from 
information provided on the ATCA documentation, or by correspondence with the authors.  

\section[]{Automated Spectral Line Fitting and Line Identification}
\label{sec:dataHandling}
\subsection[]{Sources and Spectrum Extraction}
\label{sec:sou}

The ATCA line survey contains of order a thousand spectral lines with significant spatial and kinematic structure.
In order to handle the wealth of information in the dataset, we extracted mean spectra from 
elliptical regions placed over physically distinct regions
of Sgr B2(N), transforming a 3-d problem into a set of 1-d problems.    We extracted spectra over the entire
29.8 - 50.2 GHz bandwidth from a radio continuum peak along the K6 shell-shaped H{\sc ii} region, the cometary H{\sc ii} 
region L located 35 arcsec \,NE of Sgr B2(N), and the LMH hot core.   In the higher spatial resolution section of the spectrum (from 29.8 -
44.6 GHz due to the observed configurations), we additionally extracted mean spectra towards K4 and ``h''.
Figure \ref{regionsFig} shows the elliptical regions from which the spectra were extracted and Table \ref{regionsTable}
reports their positions. Note that we did not vary the size of the elliptical regions in different images depending on the beam size.
The elliptical regions are typically larger than the beam sizes in the high spatial resolution data (from 29.8 - 44.6~GHz), 
but smaller than the beam sizes in the low spatial resolution data observed with the compact array configuration (from 44.6 - 50.2~GHz). 
The spectra are sensitive to the spatial resolution; for example, low spatial resolution data towards the K6 elliptical
region includes flux originating on the K5 shell and in the LMH as the regions are poorly resolved.  
In the case of the spatially isolated source L, the elliptical region from which the spectrum was extracted is similar to 
the source size of L and well matched to the $\sim$5 arcsec \,beam size of the high spatial resolution data \citep{Gaume95}.  In the low spatial 
resolution data, the flux is spread over the larger beam, which has the 
effect of decreasing the flux in the low spatial resolution section of the spectrum.

After extracting spectra, we interpolated over bad spectrometer channels listed in \S\ref{sec:dataRed} using linear interpolation, and performed additional 
baseline smoothing by applying a Hodrick-Prescott filter to remove a low frequency baseline wiggle \citep{HPfilter0}. 
Figure \ref{HPfilter} shows the baseline with and without the Hodrick-Prescott filter applied.
A Hodrick-Prescott filter is a commonly used tool for removing long-term growth trends from economics data and is 
similar in function to a high-pass filter.  It is simple to implement and proved a useful means of removing residual baseline without degraded results at the band edges.
We used a modified version of the HP filter, as described in \cite{HPfilter1}, that can use datasets with segments of missing data.  
This was useful in order to extrapolate over spectral lines, 
particularly over broad spectral lines, such as the absorption feature of CH$_2$CN at 40230-40270~MHz.
Applying the filter achieved a 10 - 25\% reduction in the root mean squared (rms) noise in the spectra extracted from different images.  

We then converted from units of Jy beam$^{-1}$ to mean flux density per square arcsec (or equivalently, specific flux density) 
in order to inspect images with distinct beam sizes on the same scale. Finally, we performed a primary beam correction to the extracted spectra.  
We applied the beam correction to the extracted data instead of simply extracting the spectra from beam-corrected images for the following reason.
For each image, the standard technique in {\sc miriad} applies a single primary beam (PB) correction based on the central frequency tuning of the image, and does not include frequency 
dependence within the image.  However, over the 2~GHz bandwidth of each image, frequency-dependence is significant, especially towards L.
To compute appropriate frequency-dependent beam corrections, we compared the spectra extracted from the 
regions in the non-PB corrected cubes to the spectra extracted from the same regions in PB corrected cubes.  We fit a fifth order polynomial to the 
ratio of PB corrected to non-PB corrected data across the 20~GHz of bandwidth.  This fit provides the PB corrections.  They are substantial (up to a factor of 2.6) 
for L but are very small (\textless 1.04) for K4, K6, ``h'', and the LMH.

The half power beam widths at all frequency tunings and the rms noise levels in the final extracted spectra are reported in Table \ref{rms_extracted}.
The supplemental material to this work provides the 
extracted, HP-filtered, and PB-corrected spectra towards all 5 regions, and the line fitting and identification {\sc python} code described in \S\ref{sec:codeMethod}.
This material is additionally available at https://github.com/jfc2113/MicrowaveLineFitter.

\subsection[]{Line Fitting Methodology}
\label{sec:codeMethod}

The line fitting routines operate on the extracted 1-d spectra in order to fit Gaussians and perform line identification by the following procedure:
 \begin{enumerate}
  \item Identify ``detected'' features as those with a single-channel flux $I_{chan} > 3.5\sigma_{\text {region}}$,
  where $\sigma_{\text {region}}$ is the rms noise of line-free sections of a region's extracted spectrum.
  The value of $\sigma_{\text {region}}$ is determined independently for each frequency tuning.
  With a threshold of 3.5$\sigma_{\text {region}}$, one channel in 2149 should be a false detection assuming perfect baseline
  subtraction and constant noise within an image. The fitter selects channels that meet this criteria
  and 5 channels on either side.
  \item Compensate for the poor spectral resolution by using Fourier domain zero padding to interpolate data 
  in the channels selected in step (i).  This generates interpolated data with 0.33~MHz channels within segments of the spectrum that contain lines.
  While the number of independent data points remains the same, this triples the number of points used to constrain the fits,
  improving the fits significantly. This is essential for the ATCA data because in many cases, only 2-3 original channels sample the lines, 
  and a Gaussian fit to so few channels can return a wildly unreasonable fit.  For example, if a line profile is sampled by 2 channels and surrounded 
  by a zero-value baseline, one can imagine many different Gaussian profiles capable of fitting the data perfectly with highly 
  varying heights in particular. 

  \item Fit one or more Gaussians to each feature iteratively.  The spectral line fitter first determines an unconstrained best-fit
      Gaussian to a spectral line using a least squares routine.  The best-fit parameters are then used as input guesses for the
      least squares fitter to re-fit a single Gaussian to the channels within 0.8 $\times$ FWHM of the line center.  A cutoff of 
      0.8 $\times$ FWHM is selected in order to minimize the effects of the baseline uncertainty, which preferentially affects line wings.
      If a 1-component Gaussian shape is appropriate, this includes 94\% of the power of the line.
  \item  Evaluate the 1-component fit against a set of criteria designed to determine whether a multi-component fit is required.
      The criteria were determined empirically for this dataset and are described in Appendix A.
      If the segment of data meets the criteria, a 2-component fit is determined and evaluated.  In the event that neither a 1-component
      nor a 2-component
      fit is deemed sufficient, the fitter grabs a slightly different segment to fit and retries both a 1- and 2-component fit.  If the result 
      remains insufficient, the reported fit is marked as poor, pointing out where further attention is required.
  \item  The best 1- or 2-component Gaussian fit is then subtracted from the raw (as opposed to the interpolated) data.  If the residual
      spectrum contains channels with $I_{\text {chan}} > 3.5\sigma_{\text {region}}$, the residual spectrum re-cycles through steps (ii)-(v).  
      Non-Gaussian line shapes
      (i.e. lines with wing broadening) are fit in this process, so that the iterative solutions do not provide a full characterisation of the line
      shape, but do adequately account for the total flux of the line.  The treatment of line wings is discussed further in Appendix A.  
  \item  Identify lines by comparing the Gaussian fit parameters to spectral line catalogue data.\footnote{All 
  line data were accessed through the online ALMA Spectral Line Catalogue\,-\,Splatalogue
      (available at www.splatalogue.net; \cite{Remijan07}); original line data were compiled in the Cologne Database 
      for Molecular Spectroscopy \citep[CDMS;][]{CDMS}; the NASA Jet Propulsion Laboratory 
      catalogue \citep[JPL;][]{JPL}; the National Institute for Standards in Technology (NIST) 
      Recommended Rest Frequencies for Observed Interstellar Molecular Microwave Transitions - 2002 Revision 
      \citep[the Lovas/NIST list;][]{LOVAS}; and the Spectral Line Atlas of Interstellar Molecules 
      (SLAIM; F. J. Lovas, private communication) and references therein accessible on Splatalogue.}
      Because the code is optimized to this dataset, the line-ID component is not very sophisticated, with the primarily kinematics-based
      rather than chemistry-based consistency checks.
 \end{enumerate}
Because a large number of steps were required to obtain the resulting fits, we do not adopt the errors to individual fits.  Instead, we utilize 
the power of broadband line surveys to derive upper limits to the errors empirically from the variation in line parameters amongst a large number of fits 
({\S\ref{sec:error})

Note that the code does not include a model for the molecular composition and radiative transfer.
This enables us to measure properties of the gas purely empirically, which is particularly appropriate at centimeter wavelengths,
where a large number of lines from various molecules have been shown to be non-thermally excited \citep{Menten04, McGuire2012, F+14}.
The code is written modularly, enabling adaptation for different datasets.  The {\sc python} scripts are further described in Appendix A and 
available in the online journal and at https://github.com/jfc2113/MicrowaveLineFitter.

\section[]{Spectral Line Results}
\label{sec:results}

 The data contain recombination lines and molecular transitions.  Most detected molecular transitions are 
 either observed in absorption preferentially along the K5 and K6 
  shells, L, and K4, or in emission towards the LMH and ``h'' hot cores (Figure \ref{fig:distrs}).   Figure \ref{segmentSpectra} shows 
 a representative segment of spectra from all five regions from which spectra were extracted.   Detected absorption lines
are primarily from low energy states (E$_L \lesssim $\,20K),
whereas most emission lines from the LMH and ``h'' have higher energies (E$_U \gtrsim $\,80K). The LMH has the most line-dense
spectrum as anticipated, and ``h'' has the second richest spectrum.   The excitation and 
 gas phase molecular abundances of ``h'' appear significantly more similar to the LMH than other regions in Sgr B2(N),
 confirming that it is a second hot core in Sgr B2(N) \citep{Belloche2013}.

The spectra towards K6, L, and K4 are similar to one another, predominantly showing emission by
recombination lines and absorption by molecules. 
Of all detected features originating at the positions of K6, L, and K4, $\sim$90\% were confidently 
assigned to recombination line or molecular carriers.  The remaining 10\% including unidentified transitions
are discussed in Appendix B.  
Figure \ref{fig:FullSpectrum_K6} illustrates the raw data
spectrum extracted towards K6 with the output of the line fitter overlaid and line identifications labeled. 
Figures in this format containing the full spectra towards K6, L, and K4 are provided in Appendix B.  
Additionally, tables providing line identifications and
Gaussian fit parameters, including velocity center, height, width, and integrated flux towards K6, L,
and K4 are available in Appendix C.  Table \ref{bigTable_K6} is a sample page of the line 
parameters towards K6.  We do not report errors on individual Gaussian fits in the tables, but discuss errors in \S\ref{sec:error}.
As mentioned in \S\ref{sec:codeMethod}, lines with wing broadening are approximated by a primary Gaussian component
and additional components with lower amplitude.  For clarity, we only list the
primary Gaussian component parameters in the tables, however, the reported integrated flux values
include wing components.  It follows that the reported flux is not always equivalent
to what you would obtain from the primary Gaussian component alone.

%Figure \ref{fig:FullSpectrum_K6} and Table \ref{bigTable_K6} 
%\hl{include firmly identified features, contamination 
%from flux originating from other locations on the image (``contam''), tentatively
%identified features (marked with a ``?''), unidentified lines (U-lines), and tentative U-lines (marked with ``U?'').
%The tentative U-lines (U?-lines) meet the detection threshold, but upon inspecting the spectra and images, 
%the authors remain unconvinced that a real line is present.  As noted in }
%\textsection \ref{sec:codeMethod}, 
%\hl{with the established 
%threshold of 3.5$\sigma$, one channel in 2149 line-free channels should meet the detection threshold assuming 
%perfect baseline subtraction and frequency-independent noise within each tuning.  With minor baseline residuals, it 
%is reasonable to expect an average of $\sim$1 falsely detected feature per tuning.  In the spectra of K6 and L, composed of 11 tunings, 
%we report 11 and 10 U?-lines respectively consistent with this expectation; in the spectrum of K4, composed of 7 tunings, we report 11 U?-lines.
%The number in emission and absorption are similar, as would be expected for random noise and baseline variations.}

\subsection[]{Recombination Lines}
\label{sec:recombs}

In order to assess the performance of the line fitting and identification routine,
the output of the code was evaluated using hydrogen and helium recombination lines
towards K6, L, and K4.  Recombination lines, particularly at higher principal 
quantum numbers (n \textgreater 90), are not well described by simple Gaussian line shapes but exhibit
Voigt profiles \citep{vP10}.  The line fitting routine applies only Gaussian shapes, which 
adequately characterise the bulk motion of ionized gas via the velocity center, particularly
because the $\alpha$ and $\beta$ transitions have lower quantum numbers with less collisional broadening.  However, towards K6 in particular,
most H\,$\alpha$ lines require primary components and weaker 
components to recover the flux in the wings.  

The hydrogen and helium 51\,-\,59\,$\alpha$, 64\,-\,75\,$\beta$, 72\,-\,85\,$\gamma$, 79\,-\,93\,$\delta$, 85\,-\,100\,$\epsilon$, 
and 90\,-\,106\,$\zeta$ recombination transitions fall in the observed band, enabling measurements of recombination line
strengths, widths, and H{\sc ii} region kinematic centers.  Figures  \ref{alphaK6} and \ref{betaK6} present H and 
He\,$\alpha$ and $\beta$ transitions towards K6, and Figures \ref{alphaL} and \ref{alphaK4}
present H and He\,$\alpha$ transitions towards L and K4 respectively.  Each panel shows the H and He data with a single quantum number, with the Gaussian fits
reported in the online journal overlaid.
The final panel of each figure shows a composite averaged spectrum overlaid with a fit to the averaged spectrum.  The best fits to the
averaged spectra are provided in Table \ref{alphaFits}.  Additionally, recombination line kinematic measurements generated from
H~$\alpha$~-~$\gamma$ transitions are provided in Table \ref{meanRecombFits} and a histogram of the line centers is shown in Figure \ref{fig:histRecombs}.

Towards K6, the automated line fitting routine detected 64 hydrogen recombination transitions 
including H\,$\alpha$ - H\,$\zeta$ and 13 helium transitions including
He\,$\alpha$ - He\,$\beta$ lines.  The ionized gas towards K6 has two primary velocity 
components, at $\sim$59 and $\sim$84~\kms (Figure \ref{alphaK6}). 
 The automated line fitter determined that a 2-component fit is required for all H\,$\alpha$ transitions
 and for nearly all H\,$\beta$ and He\,$\alpha$ transitions.  Of the 25 unblended H\,$\alpha$ - $\gamma$ transitions for
which the automated fitting routine used a 2-component fit, it obtained a mean velocity of 58.9\,$\pm$\,0.8~\kms and width 28.7\,$\pm$\,1.1~\kms
for the low velocity component and a mean velocity of 84.7\,$\pm$\,0.8~\kms and width 27.2\,$\pm$\,1.2~\kms for the high velocity component.
The best fits to the composite averaged spectra of H\,$\alpha$ and H\,$\beta$ transitions are both consistent with these parameters (Table \ref{alphaFits}). 
 As is evident in Figures \ref{alphaK6} and  \ref{betaK6}, hydrogen recombination transitions at higher spatial resolution,
 (namely the H(56) - (53)$\alpha$ and H(70) - (66)$\beta$ transitions) have more pronounced double-peaked line shapes than 
 the moderate resolution transitions (H(59) - (57)$\alpha$ and H(75) - (71)$\beta$), which in turn 
 are more double-peaked than the lowest resolution transitions (H(52) - (51)$\alpha$ and H(65) - (64)$\beta$).
 The line profiles are thus highly sensitive to the spatial resolution.

 The double-peaked structure towards K6 was first reported by \cite{dePree95} in an image of the H(66)$\alpha$ transition 
 observed by the VLA with 1\farcs4 resolution. This work reported a 2-component recombination line profile located internal 
 to the K6 shell at a position roughly 2.5 arcsec north of
 our selected K6 region.  The components were reported at 50 and 93~\kmss, and  \cite{dePree95} argued that the H(66)$\alpha$ line
 profile provides a
 direct detection of the two sides of a shell with an expansion velocity of $\sim$21~\kmss.  In the ATCA data, a double-peaked profile with
 components at $\sim$50 and 90~\kms is observed towards the K6-internal position noted in that work.
However, the spatial positions of the peak 50 and 90~\kms recombination line emission are offset, with the 50~kms component centered on the
base of the K6 shell within our K6 elliptical region, and the 90~\kms gas centered $\sim$4 arcsec \,to the 
NW (Figure \ref{fig:H53A}).  If the two components are
part of the same structure,
the data are consistent with the ionization front expanding into a clumpy molecular cloud and preferentially illuminating at the
position of a clumps on the front and back sides of the shell.

Towards L, the automated line fitting routine detected 52 hydrogen transitions from H\,$\alpha$ to H\,$\zeta$ and 8 He\,$\alpha$ lines.   As is
evident in Figure \ref{alphaL}, higher specific flux densities are observed in the high resolution data, which covers
the H(59) - H(53)$\alpha$ transitions, due to the effect of the varying beam size discussed in \S \ref{sec:sou}.
The automated line fitting routine typically fit each recombination line with a single Gaussian.  Of 26 unblended 1-component H\,$\alpha$ - $\gamma$
lines detected by the automated fitting routine,
the mean center velocity is 76.3\,$\pm$\,0.2~\kms and width is 26.3\,$\pm$\,0.7~\kmss, consistent with the Gaussian fit to the composite average
spectrum of H\,$\alpha$ lines (Table
\ref{alphaFits}).  Towards L, \cite{dePree95} reported a recombination line velocity of 75.8\,$\pm$\,0.3~\kms and width 31.7\,$\pm$\,0.7~\kmss, so that the line widths in
particular somewhat disagree with the ATCA data. This is a result of the differences in spatial resolution.
In the ATCA observations, L appears as a single unresolved source; on the other hand, the higher spatial resolution
observations in \cite{dePree95} treat L and
L13.30 located $\sim$3 arcsec ~east as separate sources.  Towards L13.30, de Pree et al. obtain a velocity of 76.5\,$\pm$\,1.2~\kms and width
20.6\,$\pm$\,2.8~\kmss. Our line parameters are intermediate between those of the two unresolved sources.

Recombination lines are weaker towards K4 compared to K6 and L.  As a result, primarily H\,$\alpha$ - $\gamma$ transitions were detected (24
total) and only 2 He\,$\alpha$ transitions were detected.
As is apparent in Figure \ref{alphaK4}, the line profiles suggest a wing on the high velocity side of the line center, near 100~\kmss.
Of the seven H\,$\alpha$ transitions from 30~-~44.6~GHz,
only H(56)$\alpha$ is known to be blended.  Of the remaining six H\,$\alpha$ transitions, four required a 2-component fit to the line emission from
40~-~80~\kms (Figures \ref{alphaK4} and \ref{fig:histRecombs}).  The other two H\,$\alpha$ transitions (namely H(57) and H(55)$\alpha$)
did not meet the criteria for a 2-component fit, but 
a 2-component fit appears more appropriate and the reported fit is adjusted from the output of the line fitter.  The line shape appears to vary substantially,
suggesting a complex kinematic structure resulting in a line shape that is sensitive to the beam size and precise placement of the elliptical region.  Indeed Figure \ref{fig:H53A}
shows a steep velocity gradient across K4.  

Although most of the H\,$\alpha$ transitions required a 2-component fit (primary) and a weaker component in the high velocity wing (secondary), 
the composite line profile is adequately fit by 1-component primary Gaussian fit at 65~\kms and a weaker wing component (secondary) on the high
velocity side at 100~\kmss, as reported in Table \ref{alphaFits}.  Using fits to H\,$\beta$ and H\,$\gamma$ lines determined by the line fitter,
we find that a 1-component treatment of the recombination lines towards K4 has a velocity center of 66.2\,$\pm$\,0.5~\kms and width of 32.1\,$\pm$\,1.4~\kmss.  
The values we determine are in mild disagreement with those reported by \cite{dePree95}, who
report a velocity of 63.9\,$\pm$\,0.8~\kms and a width 36.0\,$\pm$\,1.9~\kmss.  This may be because we are looking at different regions of K4.  We determined
the shape and placement of the elliptical region towards K4 based on the continuum structure as sampled at 40~GHz by the ATCA, which is offset by
$\sim$1 arcsec \,SE of the continuum peak at 22~GHz observed by \cite{dePree95}. 
In the ATCA data, most of the recombination line emission associated with K4 arises 2-3 arcsec southeast of
the 22~GHz continuum peak, and higher velocity line emission is observed to the SE (Figure \ref{fig:H53A}).
Some of the higher velocity material is included in the elliptical region placed on K4.
\subsection[]{Error Estimation}
\label{sec:error}

While many sources of error exist, calibration errors (including absolute flux calibration and 
amplitude error from gain calibration) are small and fairly independent of frequency and position.
The main sources of error with which we are
concerned include:
\begin{enumerate}\item Spectral lines that are barely, or in some cases not, Nyquist sampled have poorly constrained Gaussian line parameters. 
While the precision of line center and width measurements are
obviously limited, this introduces a $\sim$10\,-\,15\% uncertainty to the line height.  We adopt a 12.5\% error to the line height for unblended lines.
It additionally introduces substantial covariance in the fit parameters of 2-component lines that are not well resolved.  This applies to the 2-component
recombination lines towards K6 and K4 and to molecular lines that are broad due to saturation or blended hyperfine components.
\item The baseline uncertainty, typically of order 0.05 mJy arcsec$^{-2}$, can increase the number of false detections and affect the
    best-fit Gaussian shapes.
While we minimized the baseline residual using a Hodrick-Prescott filter, it is impossible to obtain a fully residual-free baseline, 
and the remaining baseline can affect low signal-to-noise transitions significantly.
This contributes uncertainty in line heights and can artificially broaden the best-fit Gaussian.  As discussed in
\S \ref{sec:codeMethod}, we minimize the latter effect by fitting only the region within 0.8 $\times$ FWHM of line center.
\item Frequency-dependent variations in noise level in data from a single spectral tuning can generate false detections.
Noise can also contribute significantly to error for low signal-to-noise detections, particularly because the data is not Nyquist sampled.
\item The spectrum extracted from a region can contain flux generated at a different spatial position in low spatial resolution data.
    This applies primarily to the low spatial resolution data (from 44.6 - 50.2~GHz) in the spectrum of K6, which contains 
    flux arising in the LMH and ``h''.  This effect is discussed further in Appendix B.
\item Towards the three regions that we characterise in this work, we observe molecular gas with two primary kinematic components separated by
    $\sim$20~\kmss. For a
    few low signal-to-noise transitions, the data does not significantly disagree with a single Gaussian profile,
    so the line fitting routine returns a single fit at a velocity that is intermediate between the two
velocity components.  We indicate fits for which we suspect
this has occured with ``$BC$'' for blended components.  In these cases, the Gaussian parameters are unreliable, but
the qualitative statement of transition detection stands.  As a primary aim of the study is to demonstrate the functionality of the 
line fitting code, we correct few instances of this.  However, researchers interested in the separate contributions from the two gas cloud components
should re-fit these with the velocities and widths fixed to the mean values obtained in this work.
\end{enumerate}

Because there are many sources of uncertainty and a large number of lines detected in the survey, the simplest method for evaluating the
error is to derive it from
the fits themselves.  To do so, we assume that line radiation from all thermal molecular line transitions originates at a consistent set of
central velocities for each
targeted spatial region, setting an upper limit on the typical error of fits.
With very few high signal-to-noise exceptions, this assumption is supported by the data, as typical differences
from the mean center velocity are significantly smaller than the channel widths of 6~-~10~\kmss.

Using molecular line transitions with signal-to-noise ratios (determined as the ratio of the Gaussian fit height to the noise level in the
extracted spectrum) in the 25th - 90th percentile range, we determine the standard deviation of the line velocity centers and widths.
The 25th - 90th percentile range excludes the strongest lines which typically have wing broadening
and the weakest quartile (with a signal-to-noise ratio $\lesssim$ 5.4) which is preferentially affected 
by baseline issues and noise.  The distributions of velocity centers and line widths towards all three regions is shown in Figure \ref{fig:histKinematics}.
Using 129 unblended transitions towards K6 that fall in our target 
signal-to-noise window  (including 51 1-component transitions and 39 2-component transitions) we determine 
the standard deviation of the center 
velocity to be $\sigma_{v}$\,=\,1.1~\kms and the standard deviation of the velocity width to be $\sigma_{\Delta v}$\,=\,2.3~\kmss.
The standard deviations of both parameters are
significantly lower than the channel resolution of 6 - 10~\kmss, indicating that the method performs rather well given the limits of the dataset.
We adopt $\sigma_{v}$\,=\,1.1~\kms and $\sigma_{\Delta v}$\,=\,2.3~\kms as 1$\sigma$ errors on the fit parameters of 1-component lines 
and spectrally resolved 2-component lines with a signal-to-noise ratio \textgreater 5.4.  This includes 
unblended 1- and 2-component molecular lines and 1-component recombination lines.  These values are 
consistent with the standard deviations of the parameters fit to high signal-to-noise recombination lines 
towards L (Figure \ref{fig:histRecombs}\,b), further justifying their validity.  

For 2-component recombination lines in which the components are not spectrally resolved, we determine 
$\sigma_{v}$\,=\,3.6~\kms and $\sigma_{\Delta v}$\,=\,4.9~\kms from 17 H\,$\alpha$ and $\beta$ 
lines towards K6, and line height errors of each component are estimated at $\sim$25\%.  
Compared to molecular lines, 2-component recombination lines have larger errors because the two velocity components are poorly 
resolved.  As the line widths of the 2-component recombination line components are larger than the velocity separation
between the two components $\left(\frac{\Delta v_1+\Delta v_2}{2} \textgreater v_2-v_1 \right)$, the components are unresolved.
While a 1-component fit does a poor job of matching these profiles, a 2-component fit contains significant covariance between 
the six fit parameters.
Errors of the velocity center and width are approximately 40\% higher for 2-component H$\gamma$ and $\delta$ lines,
and the line heights have a $\sim$40\% estimated error.

Table \ref{tab:errors} summarizes the errors of the fit parameters that are adopted in this work and 
recommended for further use of the results. 

\section{Molecular Composition}
\label{sec:molComp}

\subsection{Molecular Gas Components Associated with Sgr B2}
\label{sec:molSpec}

Nearly all of the molecular lines detected towards K6, L, and K4 are in absorption against the free-free
continuum emission.  
The following molecular species were detected towards one or more of these regions.  
Tentative detections are marked with a ``(?)''.

\indent \textbf{CS Family}: CS, C$^{34}$S, C$^{33}$S, $^{13}$CS, $^{13}$C$^{34}$S, HCS$^+$, H$_2$CS, CCS, CC$^{34}$S, OCS, CCCS\\
\indent \textbf{Silicon Species}: SiO, $^{29}$SiO, $^{30}$SiO, Si$^{18}$O, SiN, SiS, SiC$_2$\\
\indent \textbf{Other Inorganic Molecules}:  SO, PN, Na$^{37}$Cl(?)\\
\indent \textbf{Hydrocarbon Chains}: {\textit l-}C$_3$H, {\it l-}C$_3$H$^+$, C$_4$H\\
\indent \textbf{Ring Species}: {\it c-}C$_3$H$_2$, {\it c-}H$^{13}$CCCH, {\it c-}HCC$^{13}$CH, {\it c-}H$_2$C$_3$O, {\it c-}H$_2$COCH$_2$\\
\indent \textbf{Amines}: NH$_3$, NH$_2$D, CH$_3$NH$_2$(?), NH$_2$CN, CH$_3$CONH$_2$, H$_2$NCO$^+$(?) \\
\indent \textbf{Isonitriles}: CH$_3$NC, HCCNC \\
\indent \textbf{Nitriles}: CCCN, HSCN, NaCN/NaNC, ∙CH$_2$CN, CH$_3$CN, CH$_3^{13}$CN, $^{13}$CH$_3$CN, CH$_2$CHCN,
CH$_3$CH$_2$CN, HC$_3$N, HCC$^{13}$CN, HC$^{13}$CCN, H$^{13}$CCCN, HC$_5$N, CH$_3$C$_3$N \\
\indent \textbf{Imines}: CH$_2$NH, {\it E-}CH$_3$CHNH, {\it Z-}CH$_3$CHNH, HNCHCN \\
\indent \textbf{Aldehydes}: H$_2$CO, H$_2$COH$^+$, CH$_3$CHO, NH$_2$CHO, NH$_2^{13}$CHO, CH$_3$OCHO, {\it cis-}CH$_2$OHCHO, 
{\it t-}CH$_2$CHCHO, CH$_3$CH$_2$CHO\\
\indent \textbf{Alcohols}: CH$_3$OH, $^{13}$CH$_3$OH, {\it t-}CH$_3$CH$_2$OH, {\it g'Ga-}(CH$_2$OH)$_2$\\
\indent \textbf{Other Oxynated Species}: HNCO, {\it t-}HCOOH, H$_2$CCO, CH$_3$OCH$_3$, (CH$_3$)$_2$CO\\
\indent \textbf{Other Organic Species}: CH$_3$SH\\
The molecular inventory of the gas includes that reported in the more extended envelope of Sgr B2 and 
in the north cloud, an extended cloud of material that peaks 1 arcmin ~north of the Sgr B2(N) core \citep{Jones08,Jones11}.  
In a survey conducted with the Mopra telescope over the same frequency range on a 6 arcmin $\times$~6 arcmin field of view, the species CS, HCS$^+$, 
CCS, OCS, SiO, SO, HNCO, HOCO$^+$, HC$_3$N, HC$_5$N, CH$_3$OH, CH$_3$CN, CH$_3$CHO, NH$_2$CHO, CH$_2$CN, and CH$_2$NH were 
detected in extended material around the North and Main cores \citep{Jones11}.  All of these molecules except HOCO$^+$ 
were detected in the significantly more sensitive ATCA survey, along with an additional $\sim$40 molecular species.
We note that the depletion of HOCO$^+$ in the absorbing material is curious and refrain from further interpreting this.  
In a 3-mm suvey conducted with Mopra over a similar field of view as the 7 mm survey, multiple other species detected here were observed 
to have an extended distribution, including NH$_2$CN, CH$_3$OCH$_3$, CH$_3$OCHO, {\it t-}CH$_3$CH$_2$OH, {\it c-}C$_3$H$_2$, CH$_2$CHCN,
CH$_3$CH$_2$CN, H$_2$CO, H$_2$CS, H$_2$COH$^+$, and CH$_3$SH \citep{Jones08}.  

Towards all three regions, molecular lines were primarily detected at two velocity components separated by 
$\sim$20~\kmss.
Figure \ref{fig:histKinematics} shows a histogram of the 25th - 90th percentile Gaussian fit 
parameters of identified molecular lines, and Table \ref{meanMolKin} presents the mean fit parameters, characterising the 
kinematic structure of molecular gas towards these positions.  K6 and K4 have velocity components centered at
64 and 82~\kms and 62 and 82~\kms respectively, whereas L has velocity components at 56 and 76~\kmss.  
While we do not include a full characterisation of the spectra towards the LMH and ``h'', the spectra of these sources 
show that line emission from the LMH hot core is at $\sim$64~\kms and emission from ``h'' is at 73~\kmss, in agreement 
with previous observations \citep{Hollis03, Belloche2013}.
Towards K6, L, and K4, both primary velocity components have median line widths \textgreater10~\kmss, indicating highly
supersonic turbulence.                                                                                                                     

A larger number of transitions are detected in the low velocity component gas towards all three regions, 
and the low velocity component has a higher total integrated line intensity in the detected transitions.  
Towards K6, the total integrated flux absorbed by the 64~\kms component is a factor of 3.5 times greater than
the total flux absorbed by the 82~\kms component.  Towards L and K4, the low and high velocity components 
typically have more similar line intensities, with total integrated flux ratios of 
$\frac{\int(T_L)|_{Low}}{\int(T_L)|_{High}}$ = 1.7 and 1.1 respectively. A larger number of
distinct molecules were detected towards the low velocity component gas in K6 than towards 
any other position or velocity component.

While nearly all molecular lines appear in absorption, a few species exhibit weakly masing transitions that 
appear in emission.  Previous work has demonstrated that the unusual environment in the Galactic Center 
produces unique excitation conditions \citep{Menten04}. For some large molecules, the 
conditions cause inversion of the energy levels in some low-$J$ lines and anti-inversion in others, leading to weak maser emission 
in the first scenario and enhanced absorption in the latter (see \S1.2. of Menten 2004 and references therein).
In the spectra towards K6 and K4, two transitions of CH$_2$NH are detected, including the 1$_{11}$\,--\,2$_{02}$ transition 
in absorption and the weakly masing 3$_{03}$\,--\,2$_{12}$ transition in emission.  Both transitions are detected both in the high 
and low velocity gas components towards K6, with similar morphologies over the K5 and K6 shells (Steber et al. in prep).   
The maser emission is detected in the low velocity gas towards K4 as well.
Additionally, the 3$_{12}$\,--\,3$_{03}$ transition of {\it t-}CH$_3$CH$_2$OH is weakly masing in the high 
velocity gas component (at 82~\kmss) towards K6 and K4, and H$_2$COH$^+$ 3$_{03}$-2$_{12}$ appears in emission towards K4. 
Finally, the 1$_{01}$\,--\,0$_{00}$ transition of H$_2$CS may show weak masing at an offset velocity in the low velocity gas in K4 
(see Figure 
\ref{fig:NLprofiles}c), 
but further investigation would be required to confirm this.  No maser emission is detected towards L.

With the exception of ammonia inversion transitions, methanol masing transitions, one 
transition of H$_2$CO, and three transitions of {\it c-}C$_3$H$_2$, all detected molecular transitions have
lower state energies below 20~K, with most being below 10~K.  For many of the detected molecules,
the highest quantum line strength transitions that fall in the 30-50~GHz range are between low energy states; however,
a number of stronger lines between higher energy states (40\,-\,80~K) are missing.  The gas is therefore rotationally cold.
Given the typical temperatures of molecular gas in the Galactic Center 
(\textgreater~80~K) (Ao et al. 2013, Mart{\'{\i}}n et al. 2008 and references therein) and the observed supersonic turbulence,  
excitation temperatures of T$_{Ex}\sim$10~K are highly subthermal. This result agrees with the conclusions of 
\cite{HuttemeisterNH3}, which determined that T$_{Ex}$ \textless ~20~K $\ll$ T$_{kin}$ in the absorbing gas components,
and with multiple recent characterisations of species detected in absorption in Sgr B2(N), which have determined that T$_{Ex}$\,$\sim$\,9~K
\citep{Loomis2013,Zaleski2013, McGuire2013}. 

Additionally, ammonia inversion transitions with lower state energies ranging from 1200 - 3500~K are detected in 
both velocity components towards K6, L, and K4.  Most of these transitions have been reported toward Sgr B2(N) 
previously with single dish observations \citep{HuttemeisterNH3, FlowerNH3}.
Although ammonia inversion lines are useful diagnostics of gas temperature \citep{NH3_1, NH3_2}
the very high energy transitions do not likely indicate the kinetic temperature of the 
molecular gas, as the energy states may be populated due to formation pumping 
(Lis et al. 2012; E. Bergin, private communication).  In work on Sgr B2 using Herschel data from the HEXOS program, 
\cite{FormationPumping} proposed that H$_3$O$^+$ and NH$_3$,
which have homologous structures with inversion transitions between metastable energy states, may be excited to 
very high energy levels by excess energy available from formation in the presence of X-ray radiation.  While they
favor an X-ray dominated environment for producing observed abundance ratios of H$_3$O$^+$ and H$_2$O, they do not rule out cosmic ray or 
shock driven chemistry.  Regardless of the exact scenario, the high energy ammonia transitions can only be explained with 
significant energy available in the gas (via X-rays, CRs, or shocks).  As such, the ATCA data are decidedly not probing cold 
material in the two primary velocity components towards K6, K4, and L, providing additional evidence that the gas
is extremely subthermally excited.

The presence of two velocity components at multiple positions across the field, the spatial 
extent of the absorbing gas over the full continuum structure (Figure \ref{fig:distrs}a), the similar excitation conditions
across the spatial field, 
and the similarity with the molecular inventory observed in more extended material by \cite{Jones11, Jones08} 
suggests we may be peering through two extended sheets of gas, perhaps with a kinematic gradient across the field.  
Previous work has suggested that absorbing gas at $\sim$65 and 82~\kms is associated with the extended envelope of 
Sgr B2 \citep{HuttemeisterNH3} and is not internal to the core.
If the components are located in the envelope, they would be anywhere from 3 - 10 pc from the cores 
\citep{HuttemeisterNH3}, 
a distance that is significantly larger than the projected distance between the cores, as L and K6 are 
separated by $\sim$ 1.3~pc and K4 and K6 are separated by 0.4~pc in the plane of the sky.
In this scenario, the extended molecular gas components need not interact directly with the local structure of the core 
(including expanding H{\sc ii} regions),
and it is unclear whether the higher or lower velocity component is closer to the star forming core.

\subsubsection{Normalised Spectral Line Ratios:  Determining Chemical or Excitation Differences In Absorbing Gas}
\label{sec:NLRs}

To determine if the system consists of two extended clouds or if each
kinematic component towards each region should be treated as a separate cloud (in which case we would be probing 3 spatial regions 
$\times$ 2 velocity components = 6 clouds), we inspect the line radiation in each component.  If the gas is located in the envelope 
with significant distance between the cores and the absorbing gas,  
we expect the gas phase molecular abundances and excitation to be fairly self-consistent in low velocity gas across the spatial field 
and in high velocity gas across the field.  The low velocity gas could be quite distinct from the high velocity 
gas however, as they are clearly separate clouds.  

If two components of gas have identical gas phase abundances and excitation, then the 
line optical depth ratios of the two gas components should be equivalent for any two lines, such that 
\begin{equation} \label{eq:tauEquiv}
\small{ \left( \frac{\tau_{C2}}{\tau_{C1}} \right)_A = \left( \frac{\tau_{C2}}{\tau_{C1}} \right)_B \pm \text{Error}}
\end{equation}  
for two gas components C1 and C2 and for spectral lines A and B.  The optical depth of an absorption line is related to the measured line and 
continuum brightness levels by 
\begin{equation} \label{eq:tauGen}
\small{ \tau = -ln\left(1+\frac{T_L}{T_C-[f(T_{Ex}) - f(T_{CMB})]}\right); \\
 f(T) = \frac{h\nu/k}{e^{h\nu/kT}-1}}
\end{equation}  
\begin{equation} \label{eq:tauThin}
\small{\tau \approx \frac{-T_L}{T_C - [f(T_{Ex})-f(T_{CMB})]} }
\end{equation}  
\begin{equation} \label{eq:tauCont}
\small{\tau \approx \frac{-T_L}{T_C}}
\end{equation}  
Equation 2 provides the general equation for absorption line optical depth, while Equation 3 is
appropriate for optically thin lines, and Equation 4 is appropriate for optically thin lines when
the continuum temperature dominates the excitation and cosmic microwave background temperatures. 
Towards K6, due to the strong continuum and low excitation temperature, Equation 4
is appropriate for optically thin lines.  It follows that the line-to-continuum ratio can be 
read as the optical depth towards K6.  The continuum is somewhat weaker towards L and K4, 
and as a result, the excitation temperature may contribute non-trivially to the denominator in Equation 3
resulting in a factor offset compared to Equation 4. While the factor varies with frequency
for a frequency-dependent continuum temperature, the change is small with frequency in the range of 30-44 GHz.  
For optically thin lines in all cases therefore, Equation 1 can be rewritten as 
\begin{equation} \label{eq:tauEquiv}
\small{ \left( {\frac{(T_L/T_C)_{C2}}{(T_L/T_C)_{C1}}} \right)_A = \left( {\frac{(T_L/T_C)_{C2}}{(T_L/T_C)_{C1}}} \right)_B \pm \text{Error}}.
\end{equation}  
We refer to the quantity on the left as the normalised line ratio (NLR) of spectral line A with respect to C1
and use the line height 
generated by the automated line fitter for $T_L$.  In the range of 29.8 to 44.6~GHz, we determine the continuum 
levels using the continuum images with central frequencies of 33.4 and 40.8~GHz and 7.5~GHz bandwidths.  
From the two images, we derived a power-law fit to the continuum strength towards K6
and linear fits to the continuum ratios $\frac{\left(T_C\right)}{\left(T_C\right)_{K6}}$ for L and K4.  
We estimate that uncertainty in the continuum ratios contributes 10\% error to an NLR.

Figure \ref{fig:NLprofiles} 
shows the profiles of selected transitions of CH$_3$CHO, CH$_3$CH$_2$CN, CH$_2$NH, and H$_2$CS.
Inspecting the profiles towards K6 
(Figure \ref{fig:NLprofiles}a), it is clear that the low (64~\kmss) and high (82~\kmss) velocity components do 
not show identical gas phase abundances and excitation. The line profiles of CH$_3$CHO and CH$_3$CH$_2$CN appear
quite similar, with the peak line strengths in the high velocity gas being $\sim$ 25\% that in the low velocity gas.
In terms of Equation 5, 
$$ \text{NLR}_{CH_3CHO} \approx \text{NLR}_{CH_3CH_2CN} \approx 0.25 $$
for the high velocity gas with respect to the low velocity gas for the two specific transitions.  
However, the NLR is clearly much higher, at $\sim$70\%, for the transition 
of CH$_2$NH and much lower for H$_2$CS.  If the line profiles of CH$_3$CHO and CH$_3$CH$_2$CN are typical towards K6, we can therefore 
say that the high velocity component towards K6 (K6:\,82~\kmss) is enhanced in CH$_2$NH 1$_{11}$\,--\,2$_{02}$ 
and depleted in H$_2$CS 1$_{01}$\,--\,0$_{00}$ relative to the low velocity gas in K6 (K6:\,64~\kmss).
Towards K6, it follows that significant differences in gas phase abundances and/or excitation exist in the two gas clouds.  
As the low and high velocity clouds are obviously distinct bodies, this result is not suprising.

More interestingly, the line profiles towards L and K4 look very distinct compared to those in K6.  We first compare only 
the low velocity gas in all three regions.  In the 
CH$_3$CHO line, the NLRs with respect to K6:\,64~\kmss
are 50\% for the low velocity gas in both L and K4 (L:\,56~\kms and K4:\,62~\kmss, respectively).  
In the line of CH$_3$CH$_2$CN, however, absorption by 
L:\,56~\kms and K4:\,62~\kms is considerably weaker, at $\sim$25 and 30\%, respectively.  Whereas the low velocity component absorption by CH$_2$NH 
is weak in L:\,56~\kms (at $\sim$22\% the strength observed in K6), it is very strong in K4.  On the other hand, the line of H$_2$CS is 
strong in the low velocity gas towards L:\,56~\kms (at $\sim$ 75\% the strength observed in K6), and weak in K4:\,62~\kms (at $\sim$25\% 
the strength observed in K6).  From these four lines alone, it is very clear that differences in excitation and/or gas phase 
molecular abundances exist on small angular scales of \textless 10 arcsec. 

Comparing the high velocity gas in the three regions produces a similar result.  With respect to K6:\,82~\kmss, the high velocity components 
in L and K4 have 40 and 80\% deeper line absorption by CH$_3$CHO.  
In the CH$_3$CH$_2$CN line, K6:\,82~\kms and L:\,76~\kms have similar line strengths, while 
the high velocity gas in K4 (K4:\,82~\kmss) shows significant enhancement.  L:\,76~\kms and K4:\,82~\kms 
have lower line strengths of CH$_2$NH 1$_{11}$-2$_{02}$, 
and while H$_2$CS is undetected in K6:\,82~\kmss, it is observed towards L:\,76~\kms and K4:\,82~\kms.  This result does not support the hypothesis
that the high velocity gas is an extended cloud located in the envelope of Sgr B2.

Utilizing the bandwidth offered by the survey, we compute the NLRs of unblended,
confidently assigned absorption lines from multiple distinct molecular families
in the segment of the spectrum extracted from high spatial resolution data (from 29.8 - 44.6~GHz).
Figure
\ref{fig:HeightPlots} 
provides NLR values with respect to K6:\,64~\kms for spectral lines in each velocity component of each spatial region, 
plotted against the line-to-continuum ratios of the lines in K6:\,64~\kmss.
As follows from Equations 2\,-\,4, the abscissa provides a proxy for the line optical depth in K6:\,64~\kmss, while 
the ordinate is proportional to the line optical depth ratio with respect to K6:\,64~\kms in the indicated gas component. 
The median NLR
is shown by the dashed line with $\pm$3$\sigma$ errors shaded around the median. 
The shaded errors are estimated by 12.5\% error on each 
line height added in quadrature with 10\% error introduced by uncertainty in the continuum ratios, resulting in a
$\sim$20\% total 1$\sigma$ error.  Finally, we note that the apparent slope at the low optical depth side relates to the detection limit
in the spectra. Because K6:\,64\,\kms has the strongest continuum level, making the spectrum towrads K6 
more sensitive to low optical depth lines
and also typically has the highest optical depth, 
lines that are weakly detected towards K6:\,64\,\kms are only detected towards other components if they have enhanced line
optical depths compared to the median.  This does not imply that these points are unreliable, but implies that the lower left corners 
of the plot cannot be filled in. 

We consider the line radiation with respect to the median ratio, as differences in the median NLR 
(such that median $\neq$ 1) do not necessarily indicate differences 
in the relative chemical abundances or excitation.  Instead the median NLR may correspond to the relative optical depth of molecular gas, and it  
also holds the factor offset
introduced by [$f(T_{Ex})-f(T_{CMB})$] in Equation 3 towards L and K4.  However, statistically significant
differences with respect to the median NLR indicate either (1) excitation differences, with distinct excitation of some molecules, or 
(2) chemical differences, with different relative gas phase molecular abundances present in distinct gas components.
If statistically significant scatter is observed in lines of a single molecule (e.g. differences in the NLR of different lines of CH$_3$CHO),
excitation differences are implicated.  Although low excitation conditions are common in the gas components discussed here, 
differences in the excitation conditions can impact the excitation of different K-states, and weak masing effects may selectively 
enhance or deplete the populations of certain states \citep{Menten04, F+14}.
Where line excitation is not implicated, the differences are attributable to chemical abundance differences.

\textbf{Trends in Figure \ref{fig:HeightPlots}:}
Transitions that appear to vary the most between different panels of Figure \ref{fig:HeightPlots} 
include CH$_2$NH 1$_{11}$\,--\,2$_{02}$,
CCS 3$_2$\,--\,2$_1$, HCS$^+$ J\,=\,(1-0), H$_2$CS 1$_{01}$\,--\,0$_{00}$, and weaker transitions of 
large aldehyde and O-bearing species.  Of these species, CH$_2$NH is known to 
be weakly masing at 1~GHz 
\citep{CH2NH} 
and is masing in this survey as well; H$_2$CS has been reported to exhibit non-thermal 
behavior at cm wavelengths 
\citep{H2CS}; 
and transitions of the O-bearing species {\it t-}CH$_3$CH$_2$OH and H$_2$COH$^+$ exhibit 
weak masing in this work.  While the lines included in
Figure \ref{fig:HeightPlots} 
appear in absorption 
and are therefore not themselves masing, the energy state populations may be very sensitive to the physical conditions. 
Significant effort will be required to determine the excitation conditions responsible for the observed line strengths,
providing powerful constraints on the physical conditions in the gas.
 
While aware of the presence of non-LTE excitation in these lines, we expect that a few of the trends indicate 
true gas phase relative abundance differences.  In this discussion, relative abundance differences are relative 
to the median abundance ratio of molecules sampled in two components of gas.
In most panels, transitions of CH$_3$CH$_2$CN appear consistently below the median value, 
indicating that K6:\,64~\kms likely has a high gas phase abundance of the species compared to other gas components. Similarly, 
K6:\,64~\kms likely has enhanced relative abundances of most S-bearing species, with only L:\,56~\kms having NLR values 
that are similar to the median.  As many of the O-bearing species have multiple lines with high NLRs, we expect that these do reflect
true abundance differences. 

We expected to observe similarity between the low velocity gas towards all three regions and the high velocity gas in the 
regions, with a distinction apparent between the two sets of spectra.  Instead a different trend is present.  
K6:\,64\,\kmss, L:\,56\,\kmss, and L:\,76\,\kms exhibit very similar excitation.  This can be seen by inspecting transitions of 
CH$_3$CHO, NH$_2$CHO, and CH$_2$CHCN in particular.  Each of these three molecules has multiple transitions sampled in Figure 
\ref{fig:HeightPlots},  
and for each of these species, the NLR values of all transitions show little variation.  On the other hand, 
K6:\,82\,\kmss, K4:\,62\,\kmss, and K4:\,82\,\kms show a different pattern of excitation and are more similar 
to each other than they are to the first set of components.  Evidence for this follows 
from the significant amount of scatter observed in lines of 
CH$_3$CHO (K6:\,82\,\kmss), NH$_2$CHO (K6:\,82\,\kms and K4:\,62\,\kmss), CH$_2$CHCN (K4:\,82\,\kmss), 
CH$_3$CH$_2$CN (K4:\,62\,\kmss), HC$_5$N (K4:\,62\,\kms and K4:\,82\,\kmss), and {\it cis-}CH$_2$OHCHO (K4:\,62\,\kmss).

In the first set of components, all lines of NH$_2$CHO and CH$_3$CHO, and most transitions of small molecules are consistent
with the median NLRs, although NH$_3$ transitions show some variability.  Observed lines of the S-bearing molecules 
SO, HCS$^+$ and H$_2$CS are comparatively depleted in L:\,76\,\kmss.  Lines of CH$_3$CH$_2$CN
are depleted in both L:\,56\,\kms and L:\,76\,\kms relative to K6:\,64~\kmss, and lines of CH$_2$CHCN may be slightly depleted in L:\,76.\,\kmss.
On the other hand, multiple lines of O-bearing species, including {\it c-}H$_2$C$_3$O (L:\,56\,\kms and L:\,76\,\kmss), H$_2$COH$^+$ (L:\,76\,\kmss), 
CH$_3$OCHO (L:\,56\,\kmss), and {\it t-}CH$_3$CH$_2$OH (L:\,56\,\kmss), are enhanced in both components in L.  
While further analysis is required, this likely indicates higher gas phase abundances of the specified O-bearing species, particularly 
where multiple lines consistently demonstrate enhancement (i.e. for {\it c-}H$_2$C$_3$O and {\it t-}CH$_3$CH$_2$OH in L:\,56\,\kmss).  Particularly
because no maser emission is detected at either component towards L, we do not anticipate that the energy state populations are significantly
enhanced or depleted by these processes, further supporting the interpretation of the results as indicating true differences in relative abundances.

In the second set of components, inclduing K6:\,82\,\kmss, K4:\,62\,\kmss, and K4:\,82\,\kmss,
we note low line ratios for some S-bearing species including H$_2$CS, CCS (K6:\,82\,\kms and K4:\,62\,\kmss), and HCS$^+$ (K6:\,82\,\kmss).  
Additionally, CH$_3$CH$_2$CN lines are lower than the median NLR towards K6:\,82\,\kms and SiO isotopologues have 
somewhat low values.  Towards all three regions, multiple O-bearing 
species exhibit high NLRs.  In K6:\,82\,\kmss, H$_2$COH$^+$, {\it cis-}CH$_2$OHCHO, {\it t-}CH$_2$CHCHO, {\it t-}CH$_3$CH$_2$OH, and the nitrile species C$_3$N,
exhibit enhanced line radiation relative to the median.  In K4:\,62\,\kmss, {\it cis-}CH$_2$OHCHO and H$_2$COH$^+$ have high NLRs, and in K4:\,82\,\kmss, lines of 
H$_2$CCO and {\it c-}H$_2$C$_3$O have high NLRs.  With respect to one another, K6:\,82\,\kms and K4:\,82\,\kms exhibit similar excitation (evidenced by 
NLR values for molecules with multiple line transitions), but very different line strengths for some S-bearing species.  
With respect to one another, they also show differences in the relative abundances of
some aldehyde and nitrile species, with NLR values that are consistently offset from the median.
They additionally show differences in which weak O-bearing transitions are detected, possibly indicating relative abundance differences.  
K4:\,64\,\kms has a very similar molecular inventory to that of K6:\,82\,\kmss, but the excitation is somewhat distinct.  Besides the relative depletion 
of SiO isotopologues in K6:\,82\,\kmss, the relative abundances are quite consistent 
with those in K6:\,64\,\kmss, indicating that these two regions have similar chemical conditions.

As is clear from this discussion and in agreement with the above discussion of
Figure \ref{fig:NLprofiles}, Figure \ref{fig:HeightPlots} 
reveals statistically significant differences from the median values for
both velocity components towards all three regions. In fact, by comparing the panels to one another, 
it is apparent that no two components have consistent line radiation, revealing variability in 
excitation and relative abundances on small spatial scales.  As a result, the gas phase molecular 
abundances and excitation of the 3 spatial regions $\times$ 2 velocity components should be treated 
separately.  This does not imply that each of the 6 clouds is an entirely, localized body, but that 
sufficient variability is observed so that signals from different positions cannot be combined,
particularly for transitions that exhibit greater degrees of variability.

\textbf{Implications for the Structure of Sgr B2:} 
If the structure of Sgr B2 included two clouds in the envelope that are extended over the continuum structure, we would observe that the molecular 
abundance ratios and excitation should be self consistent in the low velocity material and in the high velocity material.  The low velocity material
could show a very distinct pattern from the high velocity gas however.  If this were the case, the panels corresponding to low velocity material in 
Figure \ref{fig:HeightPlots} (panels b and d) should have very few points outside of the shaded 3$\sigma$ errors.  While the high velocity material (panels a, c, and e) 
could have points outside of this, the same points should appear outside of the shaded region in all three panels.
As is evident from Figures \ref{fig:HeightPlots} and from the discussion above,
this is not observed. The differences in line radiation thus indicate that we
 cannot adopt the most simple treatment of the molecular material, namely as two extended sheets of gas that are not interacting
 directly with the structure of the core.  Instead, we observe significant 
 variation on size scales of \textless 30~arcsec, indicating that both the low velocity component and the high velocity component are likely 
 located proximate to the star forming core and are not envelope components. 

As K6:\,64\,\kms and K4:\,62\,\kms components, separated by \textless10 arcsec, show a great deal of differentiation, 
it does not appear likely that the low velocity gas is located in the extended envelope, removed from the immediate mechanical and 
radiative environment of the H{\sc ii} regions and embedded sources.  The low velocity gas, varying in both excitation and 
in relative molecular abundance values towards the regions, is instead likely located proximate to the H{\sc ii} regions.  The low velocity
gas towards K6 and K4 may be part of the same structure, however we argue that the gas is located close to the core and is therefore 
responding to localised conditions.  Other work has provided evidence that the molecular material is indeed
part of an extended structure, as material has been observed at a similar velocity in the foreground of Sgr B2(M), $\sim$50 arcsec south of K6
\citep[][and references therein]{HuttemeisterNH3}.
The low velocity cloud may be an extended structure ($\gtrsim$\,1~arcmin) located in the immediate viscinity of 
the cores, creating locally variable gas phase abundances and excitation due to radiative and/or mechanical 
interactions with H{\sc ii} regions. While 
Figure \ref{fig:HeightPlots} 
demonstrates similarity in excitation between K6:\,64\,\kms and L:\,56\,\kmss, the 
ATCA data cannot provide direct observational evidence that the L:\,56\,\kms gas is part of the same structure as K6:\,64\,\kmss.
Other work has provided some evidence that they are, however  \citep[see e.g.][and references therein]{Sato00} 

While K6:\,82\,\kms and K4:\,82\,\kms are more similar to one another, differences nontheless persist. The chemistry and 
excitation may be localised to a lesser degree than the low velocity gas.  While this gas does not appear to be located in 
the extended envelope, it may be located further from the cores than the low velocity gas in Sgr B2(N).  As L:\,76\,\kms shows 
a distinct pattern of excitation and gas phase abundance, it is not clear from our observations that it is related 
to the high velocity gas observed towards K6 and K4.  However, the work of
\cite{Hasegawa94} and \cite{Sato00}
provides evidence connecting the gas clouds, indicating that the high velocity material may also be part of an extended structure 
(\textgreater\,40~arcsec).  It is plausible however that a very clumpy structure in the envelope, as proposed by \cite{Envelope},
could account for the differences between the high velocity gas in K6 and K4 while allowing the gas to be located in the envelope. 

If we adopt this structure, in which the low velocity material is located closest to the H{\sc ii} regions, then the expanding 
K6 ionization shell would interact mechanically and radiatively with the 64~\kms molecular gas given the H{\sc ii} region kinematics 
compiles in Tables \ref{alphaFits} and \ref{meanRecombFits}, driving shocks into the 
system and contributing near-UV radiation to generate a PDR. These conditions could produce the more highly varying line ratios 
for molecules towards K6.  Towards L, the low velocity molecular material would not be expected to mechanically interact 
with the H{\sc ii} regions given the recombination line kinematics measured in \S\ref{sec:recombs}, although it very likely 
sets the radiative environment.
 
Furthermore, if the high velocity components are indeed further from the H{\sc ii} regions than the low velocity components, we may be directly observing 
the colliding molecular clouds that are triggering star formation in Sgr B2, consistent with the scenario first proposed by \cite{Hasegawa94}.  
In this scenario, an extended high velocity cloud (at 75\,\textless\,v$_{LSR}$\,\textless\,90~\kmss) collides with the body of Sgr B2 
(at 40\,\textless\,v$_{LSR}$\,\textless\,50~\kmss), generating star formation in processed material at an intermediate velocity
of 55\,\textless\,v$_{LSR}$\,\textless\,70~\kmss. The observed 2-component molecular gas thus may be a direct observation of remaining material in the high 
velocity cloud continuing to collide with the processed cloud, a process that may trigger yet more star formation.  
This scenario is consistent with the observation of molecular line material at $\sim$65~\kms towards Sgr B2(M) \citep{HuttemeisterNH3}.
However, significant pieces of evidence for this scenario remain missing.  
For example, there should be observational signatures of a shock at the interface of the colliding gas clouds at a velocity 
intermediate to the high and low velocity gas components.  While the line profile of $^{28}$SiO is wing-broadened, the optically thin 
isotopologues of SiO do not show such a signature. Additionally, to elucidate the interaction between the 
H{\sc ii} region and the low velocity component molecular gas, high spatial and spectral resolution observations 
of C{\sc i}, C{\sc ii}, N{\sc i}, N{\sc ii}, and O{\sc i} are required to provide information on the PDR at the surface of the 
H{\sc ii} region.  Notably, however, carbon recombination lines are not observed in this material.
 
The evidence that the low velocity material is closer to the H{\sc ii} regions remains tenuous however, 
and additional evidence suggests the situation may be yet more complex.  Recombination lines towards K6 have 2-component 
profiles centered at velocities of 60 and 84 \kmss, while molecular gas is observed at 64 and 82 
\kmss.  If the K6 shell is an expanding H{\sc ii} region that only interacts mechanically with the 
low velocity molecular gas, then it is a coincidence that the high velocity molecular gas is at 
nearly the same velocity as the far side of the expanding shell.  Likewise, this scenario requires that the   
L:\,56 \kms molecular gas component is more closely associated with the H{\sc ii} region (at~
77~\kmss) than the L: 76~\kms molecular gas component.  The velocity correspondence between the H{\sc ii} 
region and the L:\,76~\kms molecular gas may be merely coincidental, however, a deeper look at structure is warranted.

\subsection{Molecular Absorption by Clouds Along the Line-of-Sight to Sgr B2}

The line profiles of $\sim$2 dozen molecular lines show absorption by additional velocity components associated with 
intervening clouds in the line-of-sight to Sgr B2 (Appendix B). These clouds have velocities of v\,\textless\,50\kmss.
Line-of-sight absorption by gas that is associated with (i.) spiral arm 
clouds or (ii.) material internal to the Galactic bar or Galactic Center is extensively reported in 
the literature \citep{GreavesNyman96, WirstromNH3, Gerin2010, Menten11}.  The 
clouds are typically translucent (with 10$^2$\,\textless\,n\,\textless\,10$^4$ cm$^{-3}$ and 2\,\textless\,A$_V$\,\textless\,5) and are highly interesting for 
understanding the environments and chemistry in regions that can only be 
studied through line-of-sight absorption techniques.  To date, authors have reported 
primarily di- and tri-atomic molecules in the foreground of Sgr B2 and other Galactic sources. 
Furthermore, it is typically presumed that  
the absorbing gas is extended and homogenous over the field of view \citep{Qin2010}.  
The ATCA survey provides the first sub-10" resolution view of line-of-sight molecular absorption 
towards any region in the Galaxy.

Molecules detected in line-of-sight absorption components in this survey include CS, SiO, SO, NH$_3$, 
CCS, HCS$^+$, {\it c-}C$_3$H$_2$, {\it l-}C$_3$H, {\it l-}C$_3$H$^+$, C$_4$H, CH$_3$OH, CH$_3$CN, 
HC$_3$N, CH$_3$CHO, NH$_2$CHO, and {\it t-}CH$_3$CH$_2$OH(?), with a tentative detection marked with a (?). 
\cite{Turner99} reported most of the complex organic molecules detected here in translucent clouds, however
 this is the first detection of most of the organic species in the line-of-sight to Sgr B2.  Further, 
to our knowledge, this is the first reported detection of NH$_2$CHO in translucent cloud material.  
NH$_2$CHO is of interest as it is difficult 
to produce via the gas-phase chemical reaction networks believed to dominate translucent cloud chemistry \citep{Turner00}.
The tentatively detected species {\it t-}CH$_3$CH$_2$OH is also previously undetected in translucent material.  

Towards K6, components at 6, -30, -73, and -106 \kms dominate the line-of-sight absorption spectra. 
Towards L, components at similar velocities of 35, 20, -5, -35, -75, and -106 \kms  
are detected.  The absorption profiles vary between the two regions, with relative line strengths differing. 
Towards K4, only a single velocity component at $\sim$2 \kms was detected.  The 2 \kms 
component has a highly enhanced optical depth towards K4 as compared to the 6 \kms gas 
towards K6 and the -5 \kms gas towards L.  The data thus:
\begin{enumerate}
 \item  reveal a greater molecular complexity in the line-of-sight absorbing gas towards Sgr B2(N)
 than previously known
 \item indicate that the $\sim$0~\kms component observed towards Sgr B2 cannot be treated as homogenous in the plane of 
the sky over the field of view of a single dish telescope.
\end{enumerate}

Additionally, the $\sim$2~\kms component is observed to produce maser emission by the Class I methanol 
maser at 36~GHz towards K6 and K4 (see Appendix B).
Maser emission has not previously been associated with a line-of-sight absorption cloud, indicating that this
component is likely quite distinct from other line-of-sight clouds.  Previous authors have 
proposed that the $\sim$0~\kms component consists of ejecta of Sgr B2 \citep{WirstromNH3}.  Recently, however, 
a molecular gas component at $\sim$0~\kms has been observed to be widespread across the Galactic Center 
\citep{Jones12} 
and diffuse recombination line emission has been observed at a similar velocity throughout the Galactic Center \citep{MarcCar}.

\subsection{Selected Molecular Distributions in Sgr B2} 
\label{sec:molDistrs}

In this and the following subsection (\S\ref{sec:integrative}), we confine our discussion to Sgr B2, ignoring the line-of-sight absorption components.
Most transitions detected towards K6, L, and K4 exhibit spatial distributions similar to
the absorption line distributions in Figure \ref{fig:distrs}a.  However, low energy transitions from a handful of detected
molecules show simultaneous absorption towards the continuum and emission from one or both hot cores.
Transitions of NH$_3$, HCS$^+$, CH$_3$CN, CH$_2$CHCN, CH$_3$CH$_2$CN, HC$_3$N, NH$_2$CHO, CH$_3$OCHO, and CH$_3$OCH$_3$
are prominent in absorption in the foreground of the H{\sc ii} regions and
in emission from the LMH and usually ``h'' (notably, NH$_2$CHO is not detected in ``h'').
Additionally, detected transitions of OCS, HNCO, and NH$_2$D produce strong emission towards the LMH and ``h'' and 
very weak absorption against the radio continuum.  The data also contain 
multiple instances in which molecules whose spatial distributions might be expected to be correlated 
exhibit very distinct morphologies.

The molecules SiO and HNCO are widely regarded as excellent shock tracers and therefore might be expected to exhibit 
similar spatial distributions. Enhanced gas phase abundances of SiO are achieved 
by grain sputtering upon passage of a shock \citep{Schilke97, Caselli97, PineauDF97}, and HNCO is observed to be enhanced 
in multiple regions that have known shocks \citep{RodriguezFernandez10, Minh_HNCO, Martin08}.  
The spatial distribution of HNCO has been observed to follow that of SiO in multiple sources via observations with single dish 
telescopes \citep{Zin00}.  However, Figure \ref{shocker} shows that the distributions of transitions between low-energy states of SiO and HNCO
in the core of Sgr B2(N) are anti-correlated.   The HNCO 2$_{02}$-1$_{01}$ transition produces very strong line emission 
in both hot cores, whereas only very weak absorption is detected towards K6 (the absorption contour of HNCO is at 2.5\% of the peak emission line strength). 
On the other hand, for the J = (1-0) transitions of all isotopologues of SiO, 
the images include absorption against the continuum structure and a negative ``bowl'' indicating extended emission that 
could not be recovered with the interferometer.  No compact emission by SiO isotopologues is observed 
towards the ``h'' hot core.  We note the presence of a weak emission feature on the high velocity wing of $^{28}$SiO 
eminating from the northwest edge of the LMH.  Due to the line density towards the LMH 
and the absence of an emission feature on the low velocity wing, which would be more consistent with the velocity of the LMH, 
we cannot firmly identify the noted feature as an SiO line.  

The distributions indicate chemical differentiation in Sgr B2(N).
HNCO is highly abundant in the LMH and ``h'' hot cores and depleted in the molecular gas observed against the H{\sc ii} regions;
on the other hand, even if the noted wing feature is from SiO, the species has a significantly lower column density
in the LMH than in the absorbing gas.
Any emission generated by SiO in the LMH is entirely overwhelmed by the absorbing gas components, including
in the optically thin transitions of the isotopologues.  Although the higher energy states (100~K\,\textless\,E$_{L}$\,\textless\,300~K) 
of SiO should be preferentially populated in the LMH and ``h'', \cite{HEXOSsgr} reported the non-detection of SiO in HEXOS data
sampling SiO transitions with T \textgreater 160~K.  SiO thus appears to be depleted from the gas phase in the LMH and 
``h'' hot cores, but highly abundant in molecular material in the low velocity (50\,-\,65~\kmss) and high velocity 
(70\,-\,85~\kmss) gas observed in absorption towards the H{\sc ii} regions, and in extended material around the core. 
A similar distribution of gas phase SiO has been observed in the source G34.26+0.15 (G34.26),
which contains a hot core, UC H{\sc ii} regions, 
collimated outflows, and molecular gas interacting with stellar winds \citep{Hatchell01,Watt_G34}.  
Towards G34.26, SiO was observed to be widespread and highly abundant in the material interacting with stellar winds; it was
not detected in the hot core, however \citep{Hatchell01}.

The spatial distributions of transitions from species in the CS-family present yet another 
puzzle; whereas transitions of CS, CCS, CCCS, and H$_2$CS are detected 
in the absorbing gas with no apparent emission from the hot cores, line radiation from the J = (3-2) transition of OCS shows a similar behavior to that 
of HNCO (Figure \ref{CSfam}).  
Very weak line absorption by OCS (with the lowest contour at 4.5\% of the peak emission) is detected towards K6 and K4, but nearly all of the 
OCS radiation is emission produced in the hot cores.  OCS, like HNCO, is believed to form on 
grain mantles and has been linked to shock enhancement \citep{Ren11, Garozzo10}.  
Distinct from other CS-containing species, HCS$^+$ appears both in absorption against the continuum and in emission from the hot cores. 

\subsection{Physical Environment in the Absorbing Material:  An Integrative View}
\label{sec:integrative}
The detection of SiO, CS, and other CS-bearing species and the depletion of HNCO and OCS in the absorbing molecular material
provides insight into the physical environment in the absorbing gas. The strong absorption by SiO likely results from the presence of shocks
and supersonic turbulence in the gas, although the enhancement of SiO may instead be attributed to the high X-ray flux in this 
X-ray reflection nebula
\citep{MartinP00, AmoB09}.  Other CS-bearing species
found in the absorbing gas, particularly CCS and CCCS may be formed by chemistry driven by the radical CH 
\citep{Petrie96, Yamada02, Sakai07}, and multiple larger radical species are directly detected in the spectra towards K6, L, and K4 (Appendix B).
The presence of radicals typically indicates X-ray or UV irradiation typical of XDRs or PDRs \citep{Danks83, BogerS05, Meijerink07}.  

Further, the ratio of HNCO to CS may be an excellent diagnostic of the radiative environment.
Using a limited sample of Galactic Center clouds, in which shocks are believed to be ubiquitous, \cite{Martin08} demonstrated 
that the HNCO:CS abundance ratio is high ($\sim$70) in non-PDR shocked regions and an order of magnitude lower in PDRs.  
While both species can be photodissociated, CS is less easily destroyed by UV radiation and can be replenished by a gas phase 
reaction of S$^+$, an abundant ion in PDRs \citep{Drdla,Sternberg95, Martin08}. In the Horsehead Nebula, CS is observed to be nearly 
equally abundant in the PDR and in the dense UV-shielded core \citep{Goi_CS}.
On the other hand, HNCO is believed to form efficiently on grain mantles \citep{Hasegawa93} and after being released
into the gas phase, it is easily photodissociated by UV radiation \citep{Martin08, Roberge91}.   
While theoretical work contends that the HNCO:CS ratio is time-dependent 
and more indeterminate \citep{Tideswell10}, the very high ratio 
of CS:HNCO in the absorbing gas provides evidence that this material has an enhanced UV radiation field. 
Like HNCO, OCS is believed to form on grains, and the species has been directly observed on ice grain mantles
\citep{Gibb04,Garozzo10,Palumbo11, Ren11}.  Gas phase OCS is destroyed by near-UV radiation with an even higher 
photodissociation rate than HNCO \citep{Roberge91, Sternberg95, Sato95}. The nearly complete depletion OCS
from the absorbing gas is further evidence for PDR conditions. 

As Sgr B2 is known to have a strong and pervasive X-ray flux, it is possible that X-ray radiation 
instead of UV radiation could destroy HNCO and OCS in the absorbing material.  Due to the extremely high 
densities of the LMH and ``h'' hot cores \citep{Belloche08, Belloche2014Sci}, even X-ray radiation is well shielded. 
However, HNCO and OCS exhibit widespread distributions to the west of the core \citep{Jones08}.  In this material, we expect that
the X-ray field is more typical of the bulk of material in Sgr B2.  This expectation is consistent with the prevalence
of the X-ray tracer CN in the material to the west of the core \citep{Jones08}.  The widespread distributions thus suggest
that X-ray radiation is not responsible for destroying HNCO and OCS; as such, we interpret the ATCA distributions of
CS, OCS, and HNCO as suggesting the presence of a strong UV radiation field in the absorbing gas.

Additional support for a strong UV radiation field in the absorbing material follows from the results of \S\ref{sec:molSpec}. 
First, the highly subthermal 
excitation indicates that the material is sub-critically dense, with densities typical of PDRs.  Without an excitation analysis, it is possible to estimate an upper limit 
of n\,$\lesssim$~a few\,$\times$\,10$^5$~cm$^{-3}$ using the critical densities of energy states that had undetected transitions in the 30 - 50~GHz 
range with high quantum line strengths. 
\cite{HuttemeisterNH3} determined an upper limit to the density of the absorbing gas of n \textless 10$^4$~cm$^{-3}$ 
based on the fact that the HC$_3$N J = (3-2) transition is in absorption instead of emission, and independently based on an analysis of 
the (1,1) through (6,6) NH$_3$ inversion transitions.  However, the latter estimate may not hold if the H$_3$O$^+$ formation pumping excitation mechanism proposed in 
\cite{FormationPumping} proves effective for NH$_3$. 
Densities of 10$^3$\,\textless\,n\,\textless\,10$^5$ cm$^{-3}$ are typical of PDRs \citep{TeilensScience,AnRevPDRs}, 
and densities of n\,$\approx$ a few $\times$\,10$^5$ cm$^{-3}$ are reasonable in the hot, supremely dense environment at the core of Sgr B2(N). 
In section \S\ref{sec:molSpec}, we determined that 
both absorbing gas components appear to be local to the star forming cores rather than in the extended envelope material,
with the low velocity gas possibly closer to the interface of the H{\sc ii} regions.
If the absorbing gas clouds are proximate to the H{\sc ii} regions, it is not only reasonable, 
but structurally required, that they are or contain PDRs. 

Second, the species {\it l-}C$_3$H$^+$ is detected in the absorbing material, but not in the hot core.  To date, {\it l-}C$_3$H$^+$ has only been firmly detected
in the Horsehead Nebula, the Orion Bar, and Sgr B2(N),
despite a deep search towards multiple hot cores, hot corinos, and dark clouds \citep{Pety2012,McGuire2013, McGuire2014}.
Additionally, the species was not detected in molecule-rich UV-shielded core in the Horsehead Nebula \citep{Pety2012}.  The species
is clearly associated with UV-irradiated gas, and its detection in the low and high velocity gas towards K6
supports the hypothesis that both gas components are PDR sources. 

Finally, a few of the organic species detected in the absorbing clouds were recently reported to be highly abundant in a PDR environment 
\citep{Gratier13, Guzman14}.  CH$_3$CN, HC$_3$N, H$_2$CCO, and CH$_3$CHO were observed in the 
Horsehead PDR with abundances enhanced by a factor of 1 - 30 as compared to abundances in the UV-shielded core in the Horsehead. While the 
species are known to be abundant in multiple distinct environments, their detection in the absorbing material in Sgr B2(N) is 
consistent with the absorbing material being a PDR.  

The data and discussions in this work thus support a structure in which molecular gas at 50\,\textless\,v$_{LSR}$\,\textless\,70~\kms is located 
on the interface of the H{\sc ii} regions at the core of Sgr B2.  This gas component is UV-irradiated by the O-stars that power the H{\sc ii} 
regions, and may mechanically interact with the H{\sc ii} region, particularly towards K6 where the 
shell-shaped H{\sc ii} region is thought to be expanding.
The high velocity cloud (70\,\textless\,v$_{LSR}$\,\textless\,85~\kmss) is also close to the 
core, although it may not be as deeply embedded as the low velocity material.  It may be actively 
colliding with the low velocity cloud consistent with a cloud-cloud collision scenario \citep{Hasegawa94}, 
although additional evidence is required to verify this.  Based on the detection of {\it l-}C$_3$H$^+$ and the nearly complete
depletion of OCS and HNCO, it appears that both absorbing clouds are PDRs.  This hypothesis should be further tested via comparison
to high spatial and, ideally, high spectral resolution observations of neutral and ionized carbon, sulfur and oxygen.  

However, merely calling the gas clouds PDRs is an oversimplification.
The clouds contain supersonic turbulence requiring shocks, have a high X-ray flux, and likely have an enhanced cosmic ray flux as compared
to most Galactic sources.  It is quite possible that a different factor drives the chemistry of different groups of species; 
for instance, the strong X-ray flux may produce radicals that drive the chemistry of imine molecules (with an X=N--H group), whereas 
UV radiation may drive the sulfur and carbon-chain chemistry and selectively destroy certain molecules.  Meanwhile, shocks may liberate 
silicon-bearing species and dust mantle species to be processed in the soup of ionic, atomic, and molecular gas.  

Regardless of the physical parameters driving the formation of each species, 
the extracted spectra suggest that a multitude of organic molecules, including nitriles, 
aldehydes, alcohols, and imines, either form or persist in the presence of UV irradiation, X-ray radiation, and shocks. 
While we expect that some of the observed species will not prove ubiquitous in more typical Galactic PDRs,
all of the detected species can at least survive moderate UV radiation fields and are promising targets for exploring
the chemical complexity in PDRs.

\section{Conclusions}

We completed a 7 mm spectral line survey of Sgr B2(N) with the ATCA.  The data provide: 
\begin{enumerate}
 \item the largest catalog of radio recombination lines observed with an interferometer
 \item the 30 - 50~GHz spectrum of the most line-rich interstellar source in the Galaxy, namely the LMH hot core
 \item a comprehensive picture of the chemistry, excitation, and kinematic structure of clouds of molecular gas in Sgr B2
 \item new insight into the chemistry and structure of ``diffuse'' clouds observed in the line-of-sight towards Sgr B2.
\end{enumerate}
Continuum images and continuum-subtracted spectral line data cubes are available at
at http://cdsarc.u-strasbg.fr/viz-bin/qcat?J/MNRAS/.
Spectra extracted from five positions, including the LMH and ``h'' hot cores and three H{\sc ii} regions are 
made available in the online journal and at https://github.com/jfc2113/MicrowaveLineFitter. 
Upon extraction from the data cubes, the spectra have had a primary beam correction and a baseline correction applied
(Figure \ref{HPfilter}).

In this work, we developed and applied {\sc python} scripts to fit the line emision and absorption in 1-d spectra 
extracted from the data cubes at the positions of three H{\sc ii} regions in Sgr B2.  
The code provides a purely empirical output rather than comparing with a model, as this was deemed most appropriate at 
centimeter wavelengths where line radiation is often nonthermal. The code then compares with output generated by 
Splatalogue or another line catalog after minor formatting.  The code is described in \S\ref{sec:dataHandling}, released 
in Appendix A, and available at the same locations as the extracted spectra.    

The code was applied to spectra extracted from three unique spatial positions, namely K6, L, and K4, characterised by recombination line emission and 
molecular line absorption at multiple velocity components.
The performance of the code was evaluated based on the handling of 65 recombination line profiles and 90 molecular absorption lines, 
observed against the free-free radio continuum emission.  Typical errors on the fits were significantly lower than the channel 
width in this survey.  Approximate errors are summarized in Table \ref{tab:errors}.

Using the output of the line-fitting script, the molecular inventories of two primary velocity components of 
absorbing gas were determined and the approximate excitation 
conditions were characterised, showing that the gas clouds that give rise to the absorption are highly subthermally excited (T$_{Ex}\,\sim$ 10~K $\ll$\,T$_{kin}$).
Line radiation was then inspected using the relative line strengths in different gas components 
to determine whether the molecular line radiation is self-consistent at distinct spatial positions (Figure \ref{fig:HeightPlots}). 
This methodology can determine the chemical similarity between clouds at different spatial positions or velocities
under conditions of similar excitation conditions, and the method could be quantified to further
establish the degree of similarity.  In the era of broadband spectral line observing, the method 
could be highly useful for quickly determining if observed clouds are chemically distinct and demonstrating
relative enhancement of different families of molecules.  The application 
of this method toward a variety of sources is necessary to determine its utility.

We applied the method to the ATCA data in order to test whether the molecular gas radiation is consistent with an extended 
distribution that is not interacting with the local structure of the core, as would be expected if the absorbing gas is in the 
envelope.  We determined that in both primary velocity components, the three spatial regions show statistically significant 
differences in their line radiation, indicating differences in 
either line excitation or gas-phase molecular abundances.  Many of the differences appear to be 
due to relative gas-phase abundance differences. The molecular gas in the low velocity component proved more
highly variable, indicating that it may be subjected to more localized conditions and therefore located deeper within the star-forming 
core. It may be on the interface of the H{\sc II} regions, potentially interacting
mechanically and radiatively with the structure of the H{\sc II} regions.
In agreement with cloud-cloud collision scenarios, the 82\,\kms gas is likely located further from the core.  
However, required pieces of observational evidence for the proposed structure remain unavailable, particularly high spatial 
(few arcsec resolution) and spectral (10 - 25 \kmss) resolution images of atomic and ionic lines of carbon, oxygen, and sulfur.  

Additionally, the chemistry observed in translucent clouds in the line-of-sight to Sgr B2 was briefly discussed.  The data contain 
line-of-sight absorption by transitions of $\sim$15 different molecules, including transitions from NH$_2$CHO, which was previously not 
observed in translucent gas. An absorbing component at $\sim$0 \kms proved to vary significantly in its optical depth across the spatial 
field, revealing that it contains significant density structure.    

Finally, we discussed the spatial distributions of a few detected molecular species providing evidence for chemical differentiation 
between species including SiO, HNCO, OCS, CS, CCS, and additional CS-bearing molecules.  We explored the implications of the spatial distributions 
for the physical environments of the molecular gas.  The observed distributions and physical structure
indicate that both primary absorbing gas components are likely PDR environments, although shocks, high X-ray fluxes, and 
high cosmic ray fluxes may have equal influence on the observed chemistry. The data reveal that molecules including (1) silicon-bearing species, 
(2) sulfur-bearing species, (3) carbon chain molecules, (4) multiple classes of oxygenated species including aldhydes and alcohols, 
(5) nitrogen species of pre-biotic importance including nitriles and imines are abundant in the exotic PDR environments.

This work highlights only a small fraction of the scientific capacity of the dataset presented and provides a tool to enable 
researchers to accomplish additional projects.   Towards the H{\sc II} regions, follow-up work including
multi-transition analyses of molecules are facilitated by the full-spectrum line identifications provided herein.
Furthermore, the data complement 
existing single dish data covering the same frequency range, namely the PRIMOS survey \citep{PRIMOS}.
Finally, the spectra extracted from the LMH and ``h'' are very line-dense and they remain uncharacterised.  These
spectra will provide unprecedented detail on hot core chemistry in Sgr B2(N).  

\section*{Acknowledgments}
We thank the anonymous referee for the helpful comments that improved the quality of this work.
The Australia Telescope Compact Array is funded by the Commonwealth of Australia for operation
as a National Facility managed by CSIRO.  
J.F.C. gratefully acknowledges funding by the Grote Reber Doctoral Research Program of the National
Radio Astronomy Observatory (NRAO) and by the East Asian and Pacific Summer Institude of the National 
Science Foundation (NSF).  The work was supported by the NSF grant OISE-1310963. The NRAO is a facility of
the National Science Foundation operated under cooperative agreement by Associated Universities,
Inc.  NL's postdoctoral fellowship is supported by CONICYT/FONDECYT postdoctorado, under project number 3130540.
LB gratefully acknowledges support by CONICYT Grant PFB-06.

\clearpage

\begin{table}
\caption{Log of ATCA observations showing the arrays and CABB tunings.}
\begin{tabular}{cccccc}
\hline
UT Date      & Array & Freq. A1 & Freq. A2 & Freq. B1 & Freq B2 \\
             &       & (GHz)      & (GHz)      & (GHz)      & (GHz)     \\
\hline
2011 Oct 21  & H75   & 47.40    & 49.25    & 43.70    & 45.55   \\
2013 Apr 3   & H214  & 30.75    & 32.60    & 34.45    & 36.30   \\
2013 Apr 4   & H214  & 38.15    & 40.00    & 41.85    & 43.70   \\
\hline
\end{tabular}
\label{obs_summary}
\end{table}

\begin{table}
\caption{Root mean squared (1$\sigma$) noise level in spectra extracted from K6, L, and K4}
\begin{tabular}{ccccc}
\hline
Center Freq& Beamsize & \multicolumn{3}{c}{$\sigma_{\text{RMS}}$ ($\mu$Jy arcsec$^{-2}$)} \\
 (GHz) & arcsec $\times$ arcsec &   K6 & L & K4 \\
\hline
30.75 & $ 6.3 \times 4.7 $  & 58  & 64 & 55\\
32.60 & $ 5.9 \times 4.4 $  & 54  & 71& 54\\
34.45 & $ 5.7 \times 4.0 $  & 61 &  74&51 \\
36.30  & $ 5.2 \times 4.0 $  &70 &  80& 62\\
38.15  & $ 5.1 \times 3.7 $  &72  & 105&66 \\
40.00  & $ 4.7 \times 3.6 $ & 100&  124&81 \\
41.85  & $ 4.6 \times 3.6 $  &  82&  131& 76\\
43.70  & $ 4.4 \times 3.4 $  &  132&  179& 108\\
45.55  & $ 13.0 \times 8.9 $  &  53&  83& \\
47.40 & $ 11.9 \times 8.7 $ &  51 &  88& \\
49.25 & $ 11.2 \times 8.5 $ &  70 &  148& \\
\hline
\end{tabular}
\label{rms_extracted}
\end{table}

\begin{table}
\caption{Coordinates of regions from which spectra were extracted.}
\begin{tabular}{c|cccc}
\hline
Region  & \multicolumn{4}{c}{Ellipse Shape} \\
& RA Center & Dec Center      & $\Delta$RA  & $\Delta$Dec \\

&    &    & (arcsec) &  (arcsec)\\
\hline
K6 & 17$^h$47$^m$20\fs58 & -28\degr22\arcmin13\farcs2 & 4.9 & 3.3 \\
L &  17$^h$47$^m$22\fs68 & -28\degr21\arcmin56\farcs1 & 6.8 & 4.8 \\
K4 & 17$^h$47$^m$20\fs07 & -28\degr22\arcmin04\farcs9 & 5.9 & 3.8 \\
LMH &  17$^h$47$^m$19\fs93 & -28\degr22\arcmin18\farcs2 & 6.8 & 4.6 \\
``h'' & 17$^h$47$^m$19\fs87 & -28\degr22\arcmin13\farcs6 & 4.1 & 3.4 \\
\hline
\end{tabular}
\label{regionsTable}
\end{table}

\onecolumn
\begin{table}
\begin{center}
\caption{Sample page of table providing line identifications and Gaussian fit parameters towards K6.  See Appendix C for full tables towards K6, L, and K4, and for a full description
of labeling conventions.}
\footnotesize{
\label{bigTable_K6}
\begin{tabular}{lllrrrr}
 \hline \hline

Line Rest & Species & Transition & Fit Velocity  & Height & Width & $\int I_v dv$ \\
  Freq \footnotesize{(MHz)} &  &  & \footnotesize{(\kmss)} & \footnotesize{(mJy\,arcsec$^{-2}$)} & \footnotesize{(\kmss)} & \footnotesize{ (mJy\,arcsec$^{-2}$ \hspace{1mm} }\\
  &&&&&&\footnotesize{ $\times$\,\kmss)}\\
 \hline
29853.31 & U & \hspace{4mm}------   & 64.0 & 0.20 & 19.7 & 4.3 \\   
29901.40 & CH$_3$OCH$_3$ & 1$_{10}$-1$_{01}$EE/AA/AE/EA & 64.2 & -0.64 & 20.2 & -13.9 \\   
29914.49 & NH$_3$ & (11,11) & 83.4 & -2.51 & 12.5 & -29.5 \\   
 &  &  & 64.6 & -4.96 & 15.9 & -83.5 \\   
 &  &  & 11.3 & -0.52 & 18.6 & -11.4 \\   
29956.11 & U &  \hspace{4mm}------   & 64.0 & -0.23 & 20.1 & -4.9 \\   
29971.48 & Hydrogen & H(75)$\beta$ & 84.6 & 1.20 & 26.2 & 33.6 \\   
 &  &  & 61.8 & 1.82 & 32.3 & 62.7 \\   
29979.38 & {\it t-}CH$_3$CH$_2$OH & 3$_{12}$-3$_{03}$ & $^{[1]}$81.2 & 0.41 & 12.5 & 5.5 \\   
 &  &  & 62.0 & -0.29 & 11.5 & -3.6 \\   
30001.55 & SO & 1$_{0}$-0$_{1}$ & 84.4 & -11.73 & 13.5 & -150.3 \\   
 &  &  & 64.2 & -27.56 & 15.4 & -451.7 \\   
 &  &  & 8.9 & -1.25 & 21.2 & -30.5 \\   
 &  &  & -32.4 & -0.27 & 22.9 & -6.5 \\   
 &  &  & -106.3 & -0.54 & 10.9 & -6.3 \\   
30180.60 & NaCN/NaNC$^{BC}$ & 2$_{12}$-1$_{11}$ & 74.4 & -0.19 & 20.4 & -4.2 \\   
30239.76 & U &  \hspace{4mm}------   & 64.0 & -0.22 & 16.1 & -3.8 \\   
30256.04 & U &  \hspace{4mm}------   & 64.0 & -0.17 & 28.6 & -5.1 \\   
30500.20 & Hydrogen & H(85)$\gamma$ & 72.9 & 0.90 & 37.3 & 35.6 \\   
 &  &  & 51.1 & 0.38 & 18.6 & 7.5 \\   
30513.96 & Hydrogen & H(106)$\zeta$ & 79.1 & 0.29 & 46.4 & 14.4 \\   
30707.38 & Hydrogen & H(93)$\delta$ & 74.3 & 0.43 & 49.7 & 22.7 \\   
30722.67 & U? &  \hspace{4mm}------   & 64.0 & 0.22 & 90.9 & 21.3 \\   
31002.30 & CH$_3$CH$_2$CHO & 3$_{21}$-2$_{12}$ & 65.9 & -0.25 & 33.6 & -8.9 \\   
31106.15 & CH$_3$OCH$_3$ & 2$_{11}$-2$_{02}$EE/AA/AE/EA & 65.1 & -1.06 & 23.2 & -26.1 \\   
31186.68 & Hydrogen & H(74)$\beta$ & 66.8 & 2.23 & 39.9 & 102.6 \\   
31199.39 & Helium & He(74)$\beta$ & 73.5 & 0.22 & 36.9 & 8.7 \\   
31223.31 & Hydrogen & H(59)$\alpha$ & 78.3 & 7.51 & 29.2 & 233.7 \\   
 &  &  & 55.7 & 5.95 & 23.1 & 146.5 \\   
31236.04 & Helium & He(59)$\alpha$ & 89.7 & 0.32 & 16.6 & 5.7 \\   
 &  &  & 65.3 & 0.71 & 34.8 & 26.2 \\   
31262.33 & NaCN/NaNC & 2$_{02}$-1$_{01}$ & 62.0 & -0.66 & 15.8 & -11.1 \\   
31424.94 & NH$_3$ & (12,12) & 84.5 & -2.50 & 13.3 & -35.5 \\   
 &  &  & 65.2 & -7.30 & 19.4 & -150.5 \\   
 &  &  & 8.2 & -0.56 & 25.6 & -15.4 \\   
31482.38 & Hydrogen & H(99)$\epsilon$ & 75.6 & 0.31 & 45.0 & 14.8 \\   
31583.40 & Hydrogen & H(84)$\gamma$ & 77.7 & 0.84 & 33.4 & 30.0 \\   
 &  &  & 51.2 & 0.49 & 23.6 & 12.2 \\   
31698.47 & Hydrogen & H(92)$\delta$ & 68.7 & 0.28 & 33.6 & 10.1 \\   
31727.07 & {\it cis-}CH$_2$OHCHO & 5$_{23}$-5$_{14}$ & 77.6 & -0.25 & 13.7 & -3.6 \\   
 &  &  & 60.9 & -0.26 & 11.7 & -3.2 \\   
31833.03 & {\it cis-}CH$_2$OHCHO & 6$_{24}$-6$_{15}$ & 61.2 & -0.22 & 20.1 & -4.8 \\   
31911.20 & CH$_3$CH$_2$CHO & 3$_{21}$-2$_{20}$ & 65.4 & -0.27 & 10.0 & -2.9 \\   
31951.78 & HC$_5$N & 12-11 & 84.8 & -0.34 & 8.6 & -3.1 \\   
 &  &  & 64.2 & -1.78 & 12.6 & -23.9 \\   
32373.67 & NaCN/NaNC & 2$_{11}$-1$_{10}$ & 62.8 & -0.29 & 28.1 & -8.6 \\   
32398.56 & $^{13}$CH$_3$OH$^{Maser}$ & 4$_{-1,4}$-3$_{03}$ & 82.3 & 0.93 & 12.6 & 12.5 \\   
 &  &  & 62.8 & 0.40 & 10.7 & 4.6 \\   
32432.32 & Hydrogen & H(98)$\epsilon$ & 74.0 & 0.26 & 44.3 & 12.1 \\   
32468.48 & Hydrogen & H(73)$\beta$ & 78.2 & 1.92 & 32.8 & 66.8 \\   
 &  &  & 55.1 & 1.36 & 23.8 & 34.4 \\   
32627.30 & {\it l-}C$_3$H$^{Adj}$ & J=$\frac{3}{2}$-$\frac{1}{2}$ $\Omega$=$\frac{1}{2}$ F=2-1f & 63.7 & -0.50 & 17.2 & -9.2 \\   
 &  &  & 14.5 & -0.16 & 12.0 & -2.1 \\   
 &  &  & $^{[2]}$-2.5 & -0.31 & 18.8 & -6.1 \\   
32640.21 & U &  \hspace{4mm}------   & $^{[3]}$64.0 & -0.31 & 26.2 & -8.7 \\   
32627.30 & {\it l-}C$_3$H & J=$\frac{3}{2}$-$\frac{1}{2}$ $\Omega$=$\frac{1}{2}$ F=2-1f & -108.5 & -0.20 & 21.1 & -4.5 \\   
32660.65 & {\it l-}C$_3$H$^{Adj}$ & J=$\frac{3}{2}$-$\frac{1}{2}$ $\Omega$=$\frac{1}{2}$ F=2-1e & 62.6 & -0.66 & 11.7 & -8.3 \\   
32663.38 & {\it l-}C$_3$H$^{Adj}$ & J=$\frac{3}{2}$-$\frac{1}{2}$ $\Omega$=$\frac{1}{2}$ F=1-0e & 61.8 & -0.19 & 10.5 & -2.1 \\   
32660.65 & {\it l-}C$_3$H$^{Adj}$ & J=$\frac{3}{2}$-$\frac{1}{2}$ $\Omega$=$\frac{1}{2}$ F=2-1e & 17.4 & -0.20 & 24.8 & -5.3 \\   
 &  &  & $^{[4]}$-2.1 & -0.33 & 14.7 & -5.2 \\

\hline \hline
\end{tabular}
}\end{center}
\end{table}

\twocolumn

\begin{table*}
\caption{Best fits to the composite average of recombination line transitions.  The errors quoted are from the diagonal terms of the covariance matrix.  However, the matrices contain significant covariance terms,
causing the error to be highly overestimated.  Fits to the composite average of H$\alpha$ and H$\beta$ transitions agree to a precision that is significantly smaller than the covariance matrix output.}
\begin{tabular}{llrrrrrrr}
\hline
Region & Species & \multicolumn{3}{c}{Gaussian Component 1} &\multicolumn{3}{c}{Gaussian Component 2}& Height Ratio\\
\hline
 & & \multicolumn{1}{c}{Height}  &  \multicolumn{1}{c}{Velocity} &  \multicolumn{1}{c}{Width}  &
 \multicolumn{1}{c}{Height} &  \multicolumn{1}{c}{Velocity} &  \multicolumn{1}{c}{Width} &   \multicolumn{1}{r}{H$_1$/H$_2$} \\
& &  \multicolumn{1}{c}{(mJy arcsec$^{-2}$)} &  \multicolumn{1}{c}{(\kmss)} & \multicolumn{1}{c}{(\kmss)} &  \multicolumn{1}{c}{(mJy arcsec$^{-2}$)} &
\multicolumn{1}{c}{(\kmss)} & \multicolumn{1}{c}{(\kmss)}\\
\hline
K6& H\,$\alpha$ & 7.1$\pm$1.0 & 59.3$\pm$3.1 & 	28.0	$\pm$3.9 & 	5.7	$\pm$1.1	 & 83.6$\pm$3.7 &	27.0$\pm$4.4& 1.37\\
& H\,$\beta$&1.9 $\pm$ 1.2& 59.7 $\pm$ 17& 30.0 $\pm$ 24& 1.4 $\pm$ 1.9& 84.2 $\pm$ 20& 27.2 $\pm$ 21& 1.25\\ 
& He\,$\alpha$&0.67 $\pm$ 0.28&60.7 $\pm$ 10 & 23.5 $\pm$ 21& 0.51 $\pm$ 0.35& 85.6 $\pm$ 12& 19.8 $\pm$ 23& 1.31\\
L & H\,$\alpha$&7.3	$\pm$0.3&	76.6	$\pm$0.4&	24.9	$\pm$1.0\\
& H\,$\beta$&2.0 $\pm$0.3& 76.5 $\pm$1.6&25.3$\pm$3.8 & \\
& He\,$\alpha$&0.63 $\pm$ 0.27&76.1 $\pm$ 4.8& 22.4 $\pm$ 11& \\
K4& H\,$\alpha$ & 2.5 $\pm$ 0.3&65.0 $\pm$ 2.5&28.4 $\pm$ 5.8 & 0.35 $\pm$ 0.26 & 100.3 $\pm$ 17 & 27.8 $\pm$ 44& 7.1\\
\hline
\end{tabular}
\label{alphaFits}
\end{table*}

\begin{table*}
\caption{Mean parameters and standard errors of fits to H$\alpha$~-~$\gamma$ transitions.  Parameters are calculated from 1-component fits towards L, 2-component fits towards K6, 
and 1-component fits towards K4.}
\begin{tabular}{|c|c|rr|rr|c}
\hline
Region & Number of Fits&\multicolumn{2}{c|}{Gaussian Component 1} &\multicolumn{2}{c|}{Gaussian Component 2}& Height Ratio \\
\hline
 & &  \multicolumn{1}{c}{Velocity} &  \multicolumn{1}{c|}{Width}   &  \multicolumn{1}{c}{Velocity} &  \multicolumn{1}{c|}{Width} &
 H$_1$/H$_2$\\
& &  \multicolumn{1}{c}{(\kmss)} & \multicolumn{1}{c|}{(\kmss)} &    \multicolumn{1}{c}{(\kmss)} & \multicolumn{1}{c|}{(\kmss)} & \\
\hline
K6 & 25& 58.9 $\pm$ 0.8 &28.7 $\pm$ 1.1&84.7 $\pm$0.8 & 27.2$\pm$1.2&1.36$\pm$0.10\\
L & 26& 76.3 $\pm$ 0.2 & 26.3 $\pm$ 0.7&&&\\
K4 & 14 & 66.2 $\pm$ 0.5&32.1 $\pm$ 1.4&&&\\
\hline
\end{tabular}
\label{meanRecombFits}
\end{table*}

\begin{table*}
\caption{Recommended errors to fits to the ATCA data output by the automated line fitter.}
\begin{tabular}{|l|l|l|r|r|r}
\hline
\multicolumn{2}{c|}{Line Type} & S:N Regime & \multicolumn{1}{c}{$\sigma_{v}$} & \multicolumn{1}{c}{$\sigma_{\Delta v}$} &  \multicolumn{1}{c}{$\sigma_{Height}$}  \\
& & & \multicolumn{1}{c}{(\kmss)} & \multicolumn{1}{c}{(\kmss)} & \multicolumn{1}{c}{(\%)}  \\ 
\hline
\multicolumn{2}{l}{2-Component Recombination Lines} \\
& Primarily H\,$\alpha$ \& H\,$\beta$& S:N \textgreater 15 & 3.6 & 4.9 & 25 \\
 & Primarily H\,$\gamma$, H\,$\delta$ \& He$\alpha$& 5\,\textless\,S:N\,\textless\,15 & 5.0 & 6.9 & 40 \\
\multicolumn{2}{l}{1-Component Recombination Lines} & S:N \textgreater 5.4 & 1.1 & 2.3 & 12.5 \\
\multicolumn{2}{l}{1- \& 2-Component Molecular Lines} & S:N \textgreater 5.4 & 1.1 & 2.3 & 12.5 \\
\hline
\end{tabular}
\label{tab:errors}
\end{table*}

\begin{table*}
\caption{Kinematic measurements of molecular gas towards K6, L, and K4 from lines in the 25th - 90th percentile
signal-to-noise ratio. }
\begin{tabular}{|c|crr|crr|c|}
\hline
Region &\multicolumn{3}{c|}{Low Velocity Component} &\multicolumn{3}{c|}{High Velocity Component}& Height Ratio$^*$ \\
\hline
 & N &  \multicolumn{1}{c}{Velocity} &  \multicolumn{1}{c|}{Width}   & N & \multicolumn{1}{c}{Velocity} &  \multicolumn{1}{c|}{Width} &
 H$_1$/H$_2$\\
& Fits &  \multicolumn{1}{c}{(\kmss)} & \multicolumn{1}{c|}{(\kmss)} &  Fits &  \multicolumn{1}{c}{(\kmss)} & \multicolumn{1}{c|}{(\kmss)} & \\
\hline
K6 & 90 & 63.5\,$\pm$\,0.1 &13.1\,$\pm$\,0.3 & 39 & 81.9\,$\pm$\,0.2 & 11.0\,$\pm$\,0.3&3.2$\pm$0.3\\
L & 47 & 56.3\,$\pm$\,0.2 & 14.1\,$\pm$\,0.4& 29 & 75.9\,$\pm$\,0.3&12.0\,$\pm$\,0.6&1.6\,$\pm$\,0.3\\
K4 & 30 & 61.8 $\pm$ 0.5&14.9\,$\pm$\,0.7& 20&82.0\,$\pm$\,0.5&15.3\,$\pm$\,0.6&1.0\,$\pm$\,0.1\\
\hline
\end{tabular}
\label{meanMolKin}
\footnotesize{\\ $^*$The height ratio is determined from lines that have both a high and low velocity component.}
\end{table*}

 \clearpage

\begin{figure}
\includegraphics[width=8.5cm]{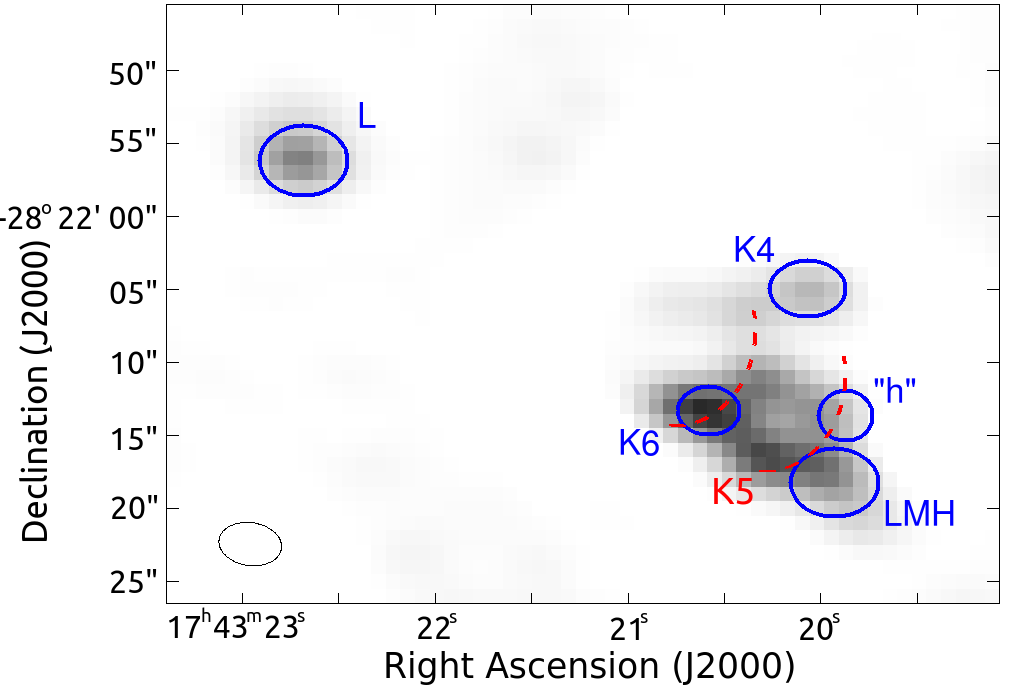}
%\plotone{RegionsOverContinuumN.png}
\caption{Continuum image of Sgr\,B2(N) and L at 40~GHz showing the elliptical regions from which spectra
were extracted. The dashed line arcs point out the K5 and K6 shell-shaped H{\sc ii} regions.  The synthesized beam shape is shown
in the lower left corner.}
\label{regionsFig}
\end{figure}

\begin{figure}
\includegraphics[width=8.5cm]{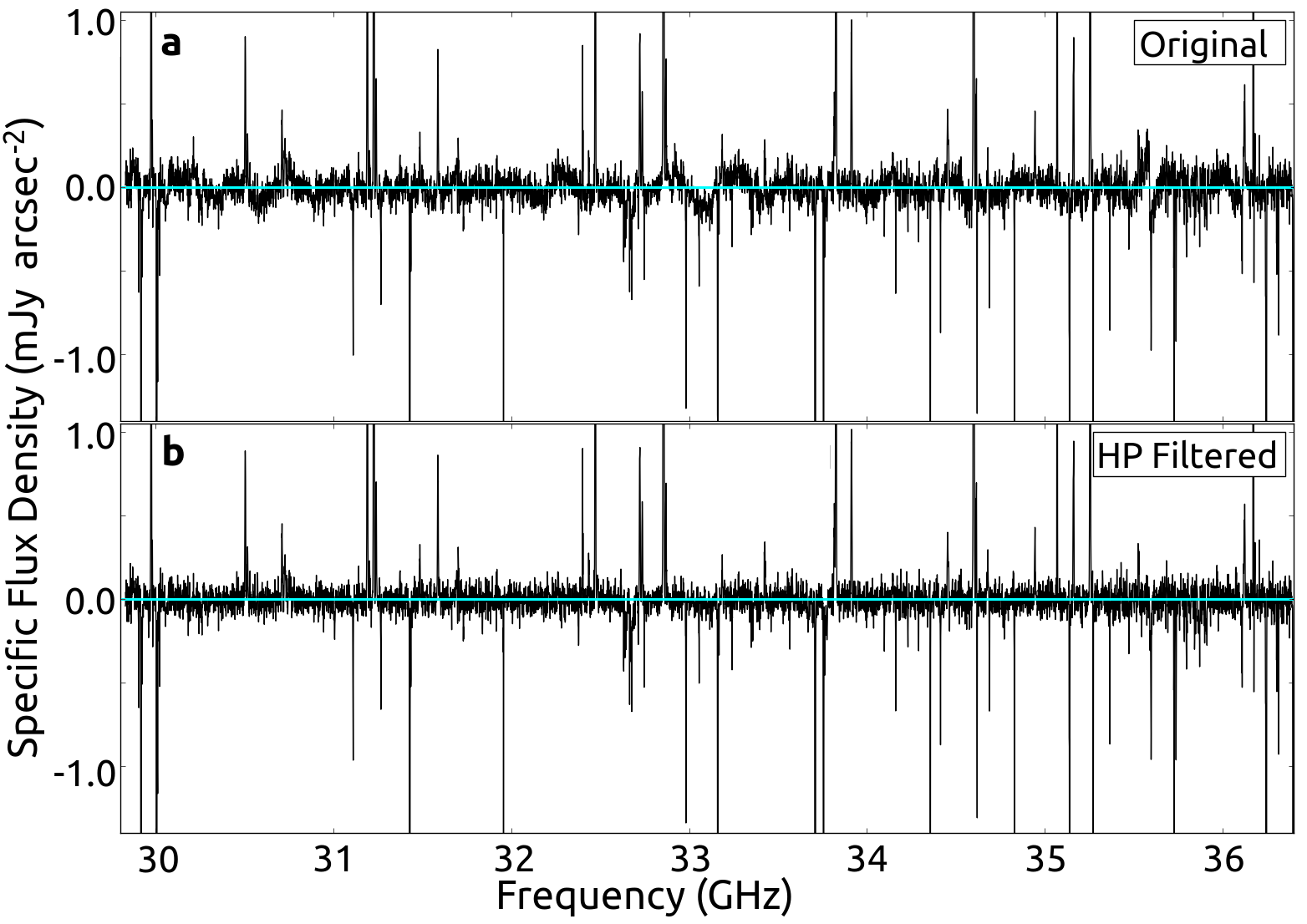}
%\plotone{RegionsOverContinuumN.png}
\caption{A segment of the full spectral coverage towards K6. \textbf{a.} shows the spectrum extracted from the data cubes
and \textbf{b.} is the spectrum after baseline removal using a Hodrick-Prescott filter.}
\label{HPfilter}
\end{figure}

\begin{figure*}
\includegraphics[width=6.5in]{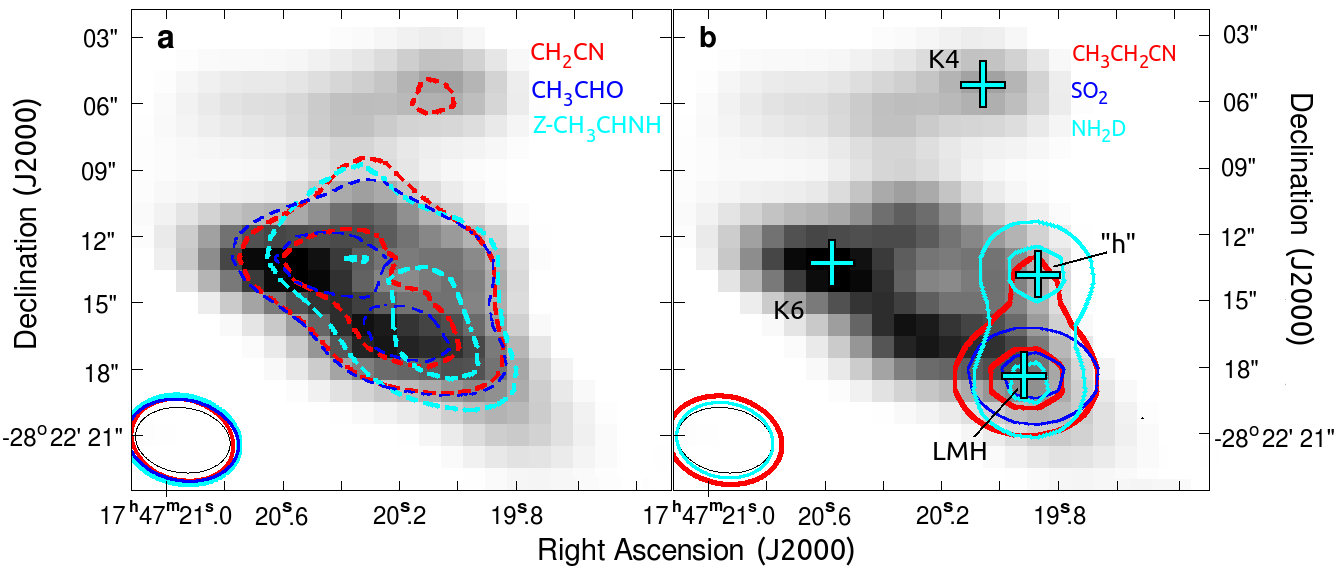}
\caption{40\% and 80\% integrated line emission contours over 40~GHz continuum illustrate the two most common distributions of molecular line radiation. 
\textbf{a.} CH$_2$CN (2 - 1), CH$_3$CHO (2$_{02}$ - 1$_{01}$),
 and {\it Z-}CH$_3$CHNH (2$_{02}$ - 1$_{01}$) are observed in absorption (dashed contours) preferentially towards the K5 and K6 shells.
 \textbf{b.} NH$_2$D (3$_{13}$ - 3$_{03}$), CH$_3$CH$_2$CN (4$_{31}$ - 3$_{30}$), and SO$_2$ (19$_{3,18}$ -  18$_{3,15}$) are detected in
 emission towards the LMH hot core, and NH$_2$D and CH$_3$CH$_2$CN are detected in emission towards ``h''.  The central positions of the elliptical 
 regions from which spectra were extracted are indicated by the crosses.
 The synthesized beam shapes for the continuum (thin black ellipse) and spectral lines (colored ellipses) are shown in the lower left corner.}
\label{fig:distrs}
\end{figure*}

\begin{figure*}
\includegraphics[width=6.5in]{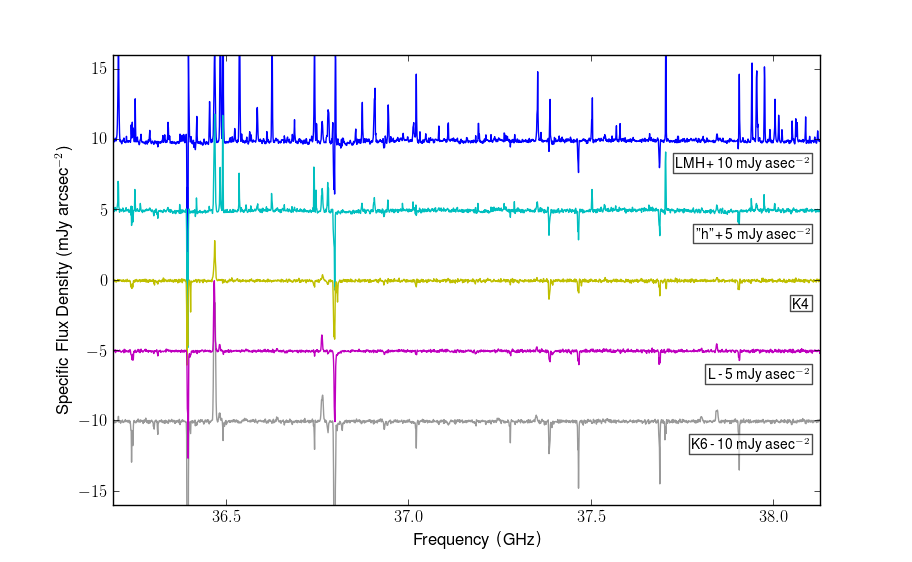}
\caption{A representative segment of spectra extracted from all five regions targeted in this study.  Whereas the LMH and ``h'' have
line-dense spectra dominated by molecular line emission,
K4, L, and K6 have lower line densities with molecular lines observed in absorption and recombination lines in emission.}
\label{segmentSpectra}
\end{figure*}

\begin{figure*}
\includegraphics[width=6.0in]{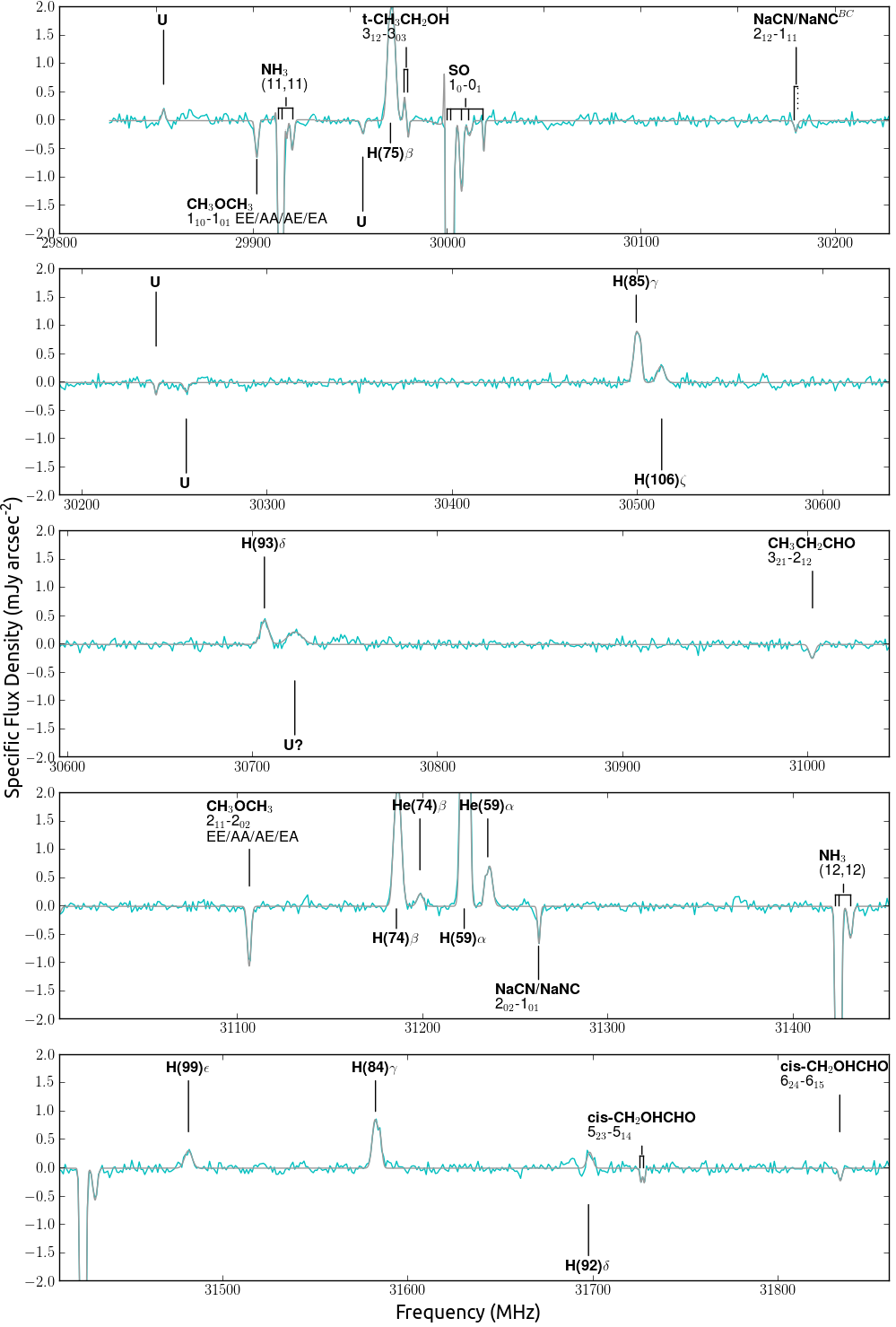}
\caption{A sample page of the full data spectrum extracted towards K6 with line identifications labeled and the output of the automated 
line-fitter overlaid.  See Appendix B for full figures for K6, L, and K4.}
\label{fig:FullSpectrum_K6}
\end{figure*}

\begin{figure*}
\includegraphics[width=6.5in]{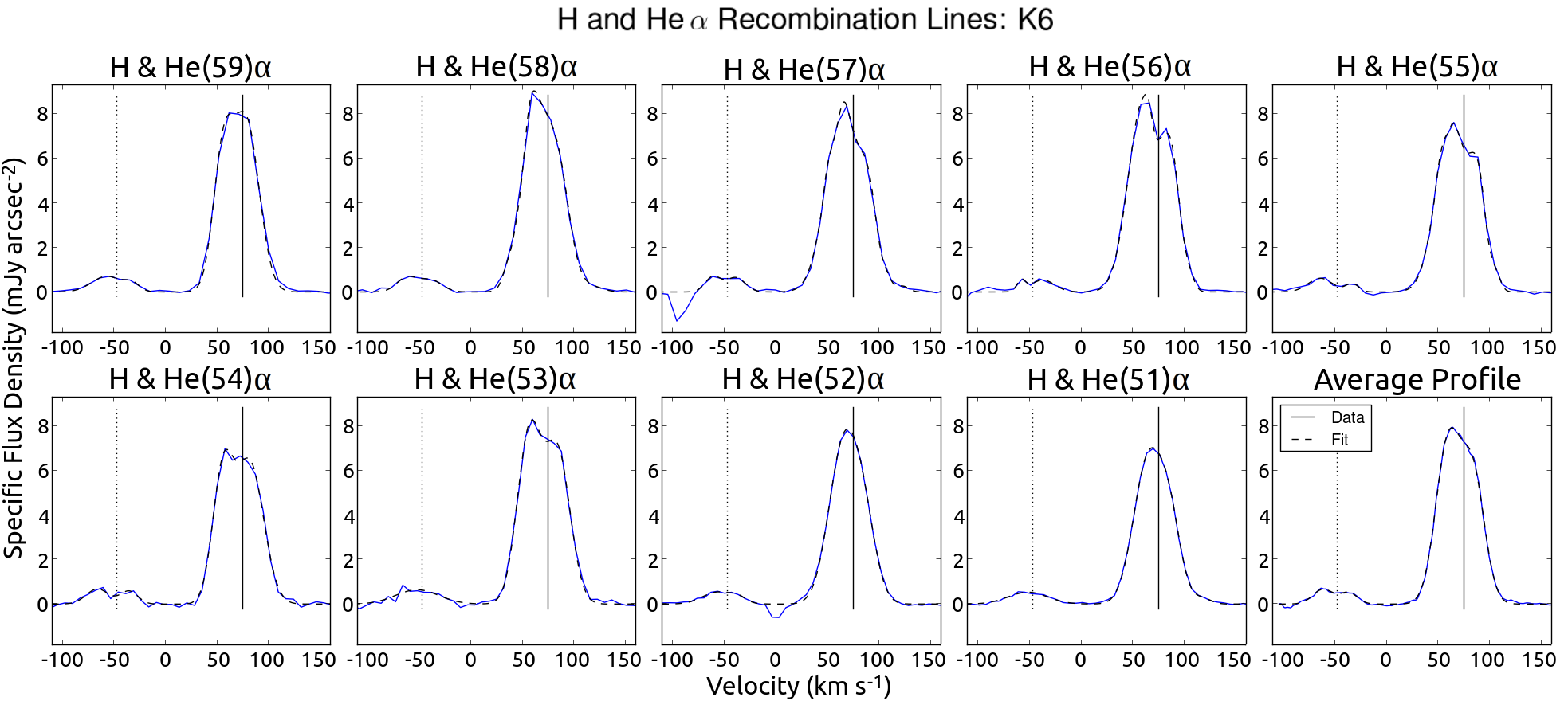}
\caption{Hydrogen and Helium\,$\alpha$ recombination lines towards K6.  Profiles of the H(59) - H(53)$\alpha$ transitions, extracted from
high spatial resolution data,
show a double peaked profile, whereas the H(52) and H(51)$\alpha$ transitions are well fit by a single component. 
The averaged profile is best fit by a Gaussian centered at 59\,\kms and a weaker component at
84\,\kmss. The solid and dashed vertical lines represent the H and He\,$\alpha$ transitions at 75\,\kms respectively.}
\label{alphaK6}
\end{figure*}

\begin{figure*}
\includegraphics[width=6.5in]{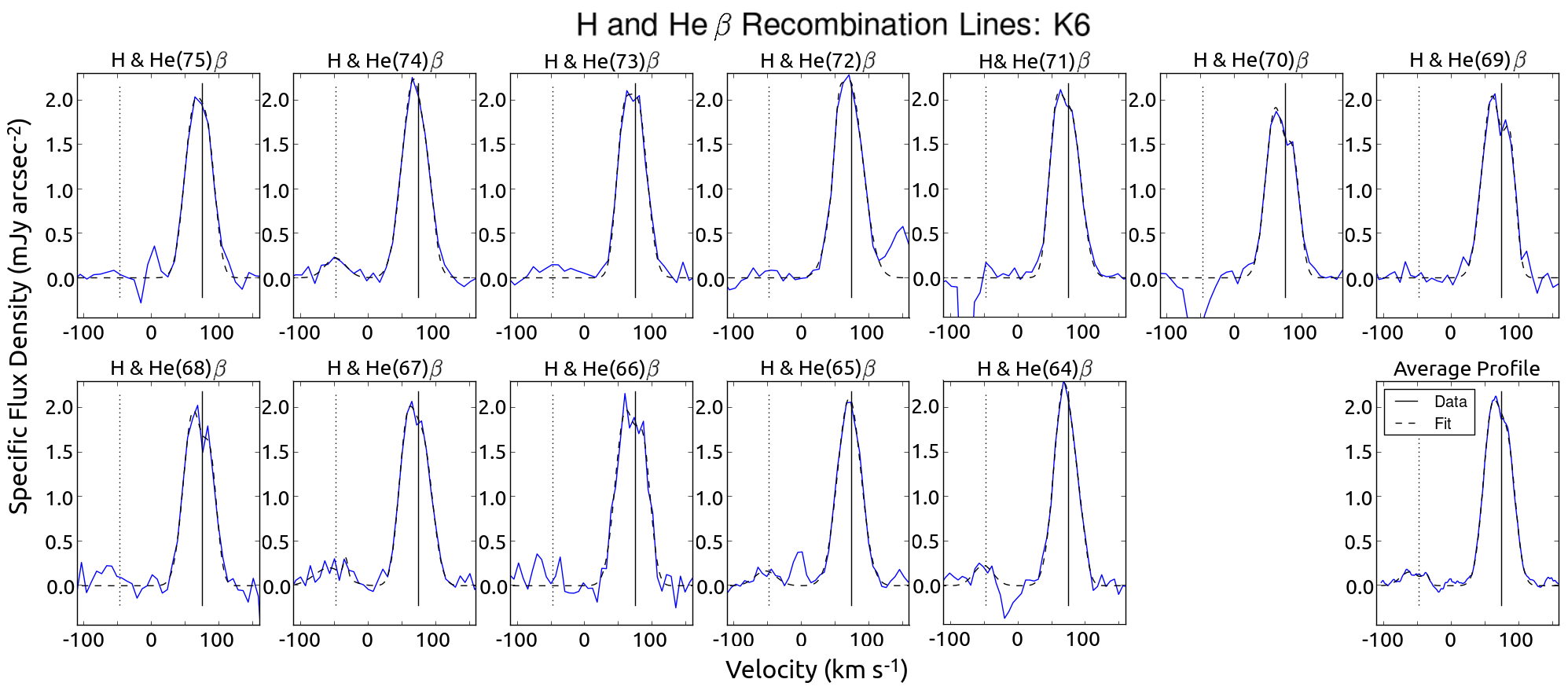}
\caption{Hydrogen and Helium\,$\beta$ recombination lines towards K6.  Profiles of the H(75) - H(66) $\beta$ transitions, extracted from high
spatial resolution data,
show a double peaked profile, whereas the H(65) and H(64)$\beta$ transitions are well fit by a single component.  The averaged profile is best
fit by a Gaussian centered at 60\,\kms and a weaker component at
84\,\kmss. The solid and dashed vertical lines represent the H and He\,$\beta$ transitions at 75\,\kms respectively.}
\label{betaK6}
\end{figure*}

\begin{figure*}
\includegraphics[width=6.5in]{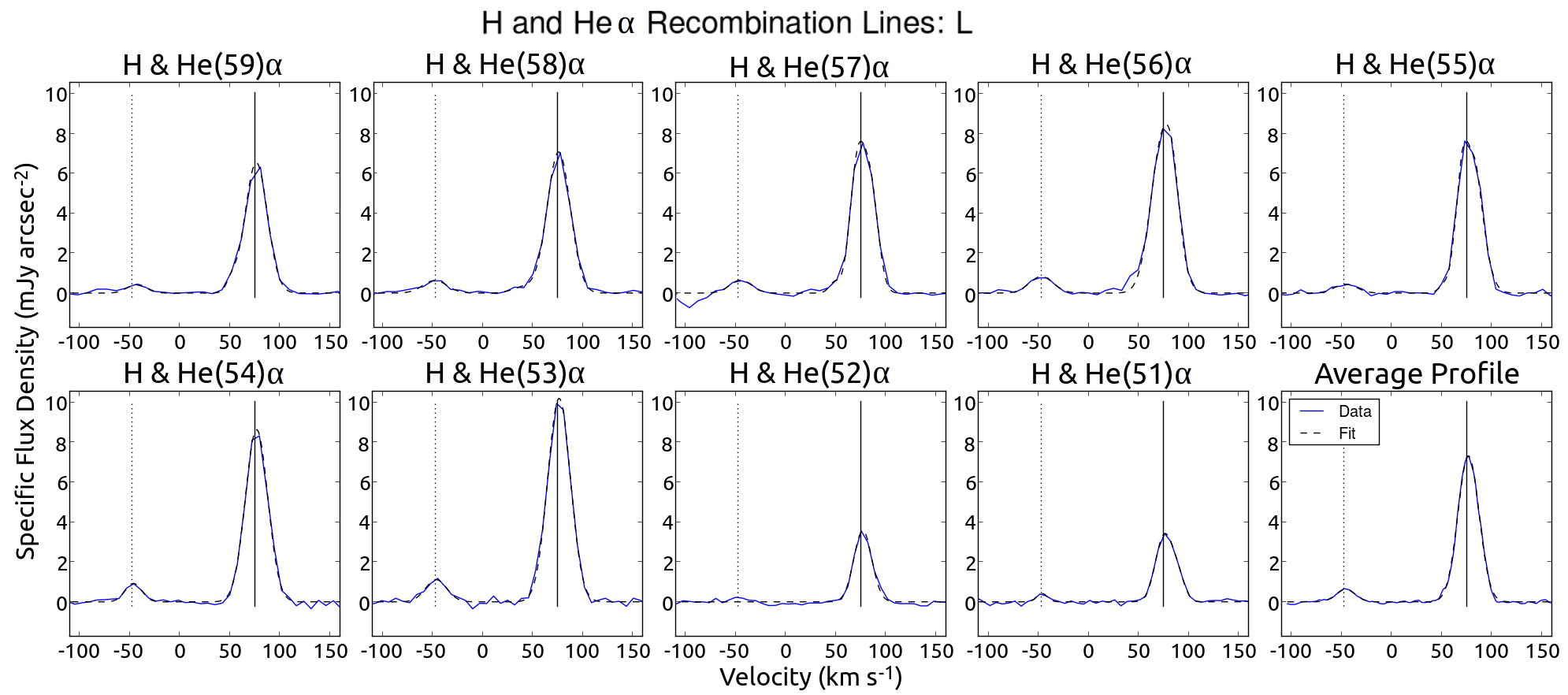}
\caption{Hydrogen and Helium\,$\alpha$ recombination lines towards L are best fit by a Gaussian at 76.5 \kmss. 
The solid and dashed vertical lines represent the H and He\,$\beta$ transitions at 75\,\kms respectively.}
\label{alphaL}
\end{figure*}

\begin{figure*}
\includegraphics[width=6.5in]{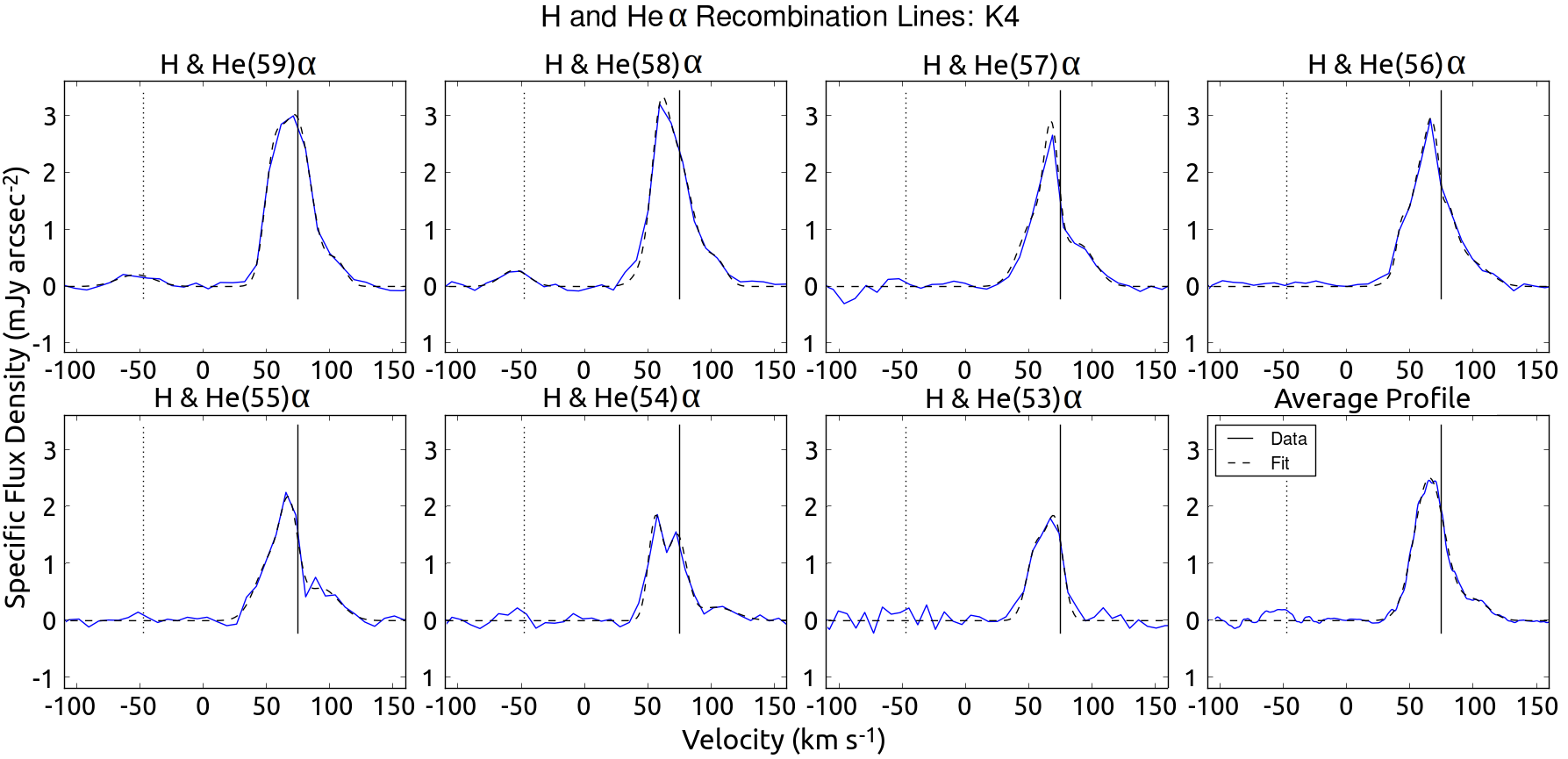}
\caption{Hydrogen and Helium\,$\alpha$ recombination lines towards K4 show unusual line shapes that 
are highly sensitive to the spatial resolution of the image from which they were extracted.
The lines are typically best fit by two primary components.  
In addition, a weaker wing component appears at $\sim$100~\kmss.
The solid and dashed vertical lines represent the H and He\,$\beta$ transitions at 75\,\kms respectively.}
\label{alphaK4}
\end{figure*}

 \begin{figure*}
\includegraphics[width=6.5in]{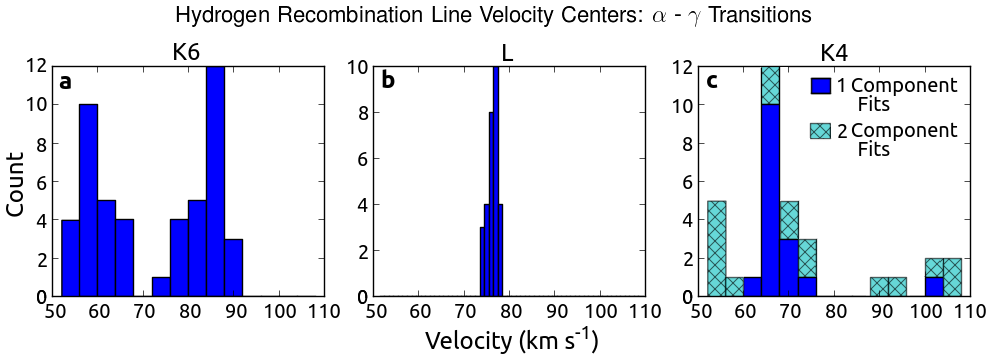}
\caption{Velocity center of H $\alpha$ - $\gamma$ transitions fit towards L, K6, and K4.  \textbf{a.} includes transitions towards K6 that were fit with 2 Gaussian components;
\textbf{b.} includes unblended transitions towards L;  \textbf{c.}
includes all transitions towards K4.  Components with $v\,>\,$80~\kms towards K4 are fitting a high velocity 
wing that is not a primary components of the line profile.}
\label{fig:histRecombs}
\end{figure*}

\begin{figure*}
\includegraphics[width=6.5in]{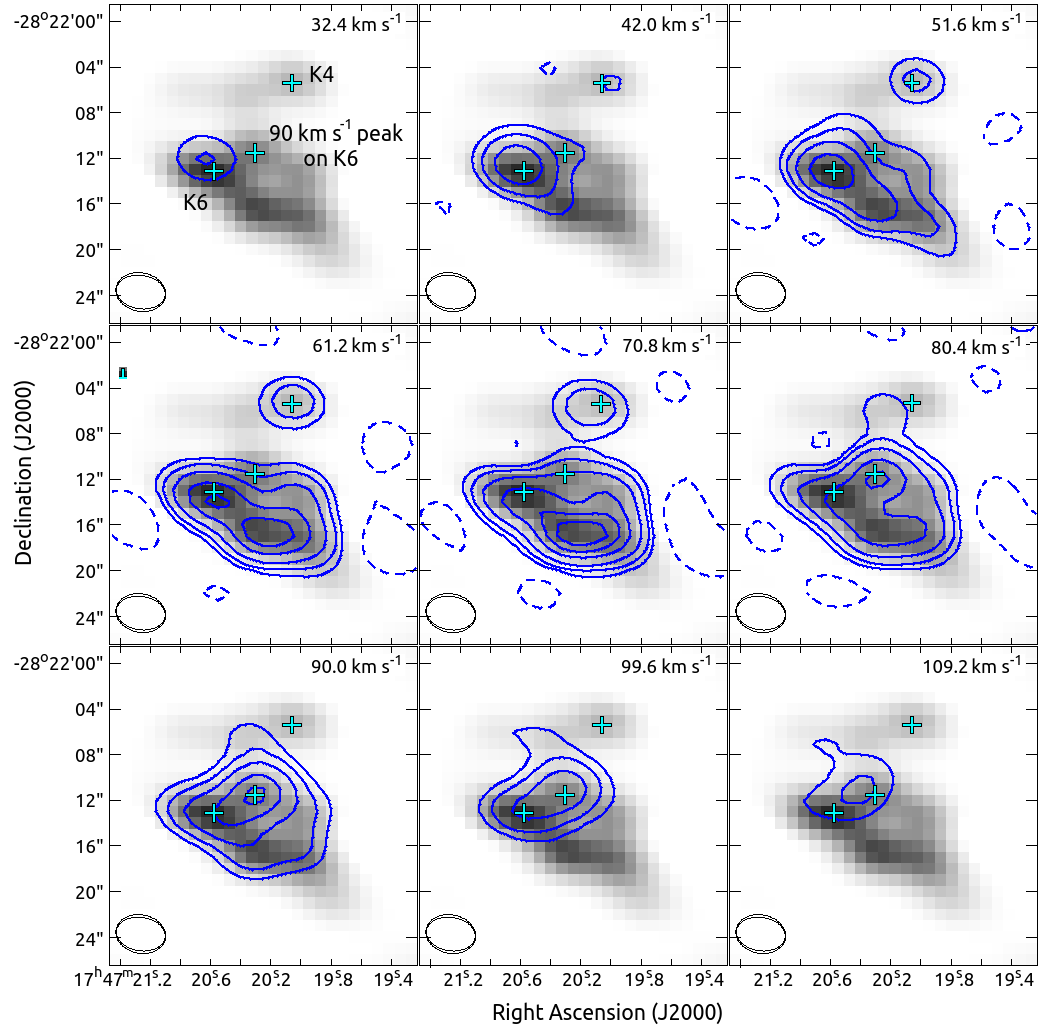}
\caption{Channel maps of the H(53)$\alpha$ transition at 42951~MHz shown in contours over the 40~GHz continuum
in greyscale.  The synthesized beams for the continuum and line images are shown in the lower left corner. The figure shows a strong velocity gradient from
the NW to the SE of K4.  The crosses mark the center positions of the regions selected for K4 and K6 (at the base of the K6 shell)
and indicate the peak position of the 90~\kms gas.  Notice the offset positions of high and low velocity gas on the K6 shell and the velocity gradient 
observed at K4.}
\label{fig:H53A}
\end{figure*}

\begin{figure*}
\includegraphics[width=6.5in]{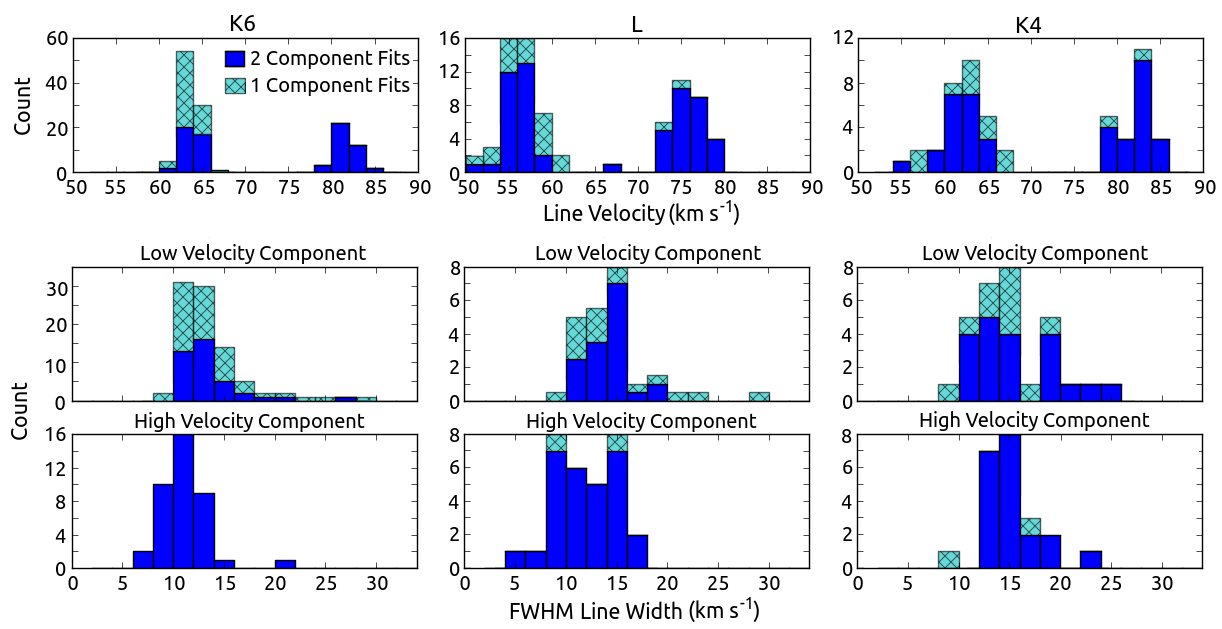}
\caption{Kinematic parameters of molecular line fits output by the automated line fitter.  
1-component fits shown include fits with a signal-to-noise ratio in the 25th to 90th percentile of all unblended 
1-component fits.  2-component fits shown include unblended transitions
in which the low velocity component has a signal-to-noise ratio in the 25th to 90th percentile 
of the low velocity components of all 2-component lines.}
\label{fig:histKinematics}
\end{figure*}

\begin{figure*}
\includegraphics[width=6.5in]{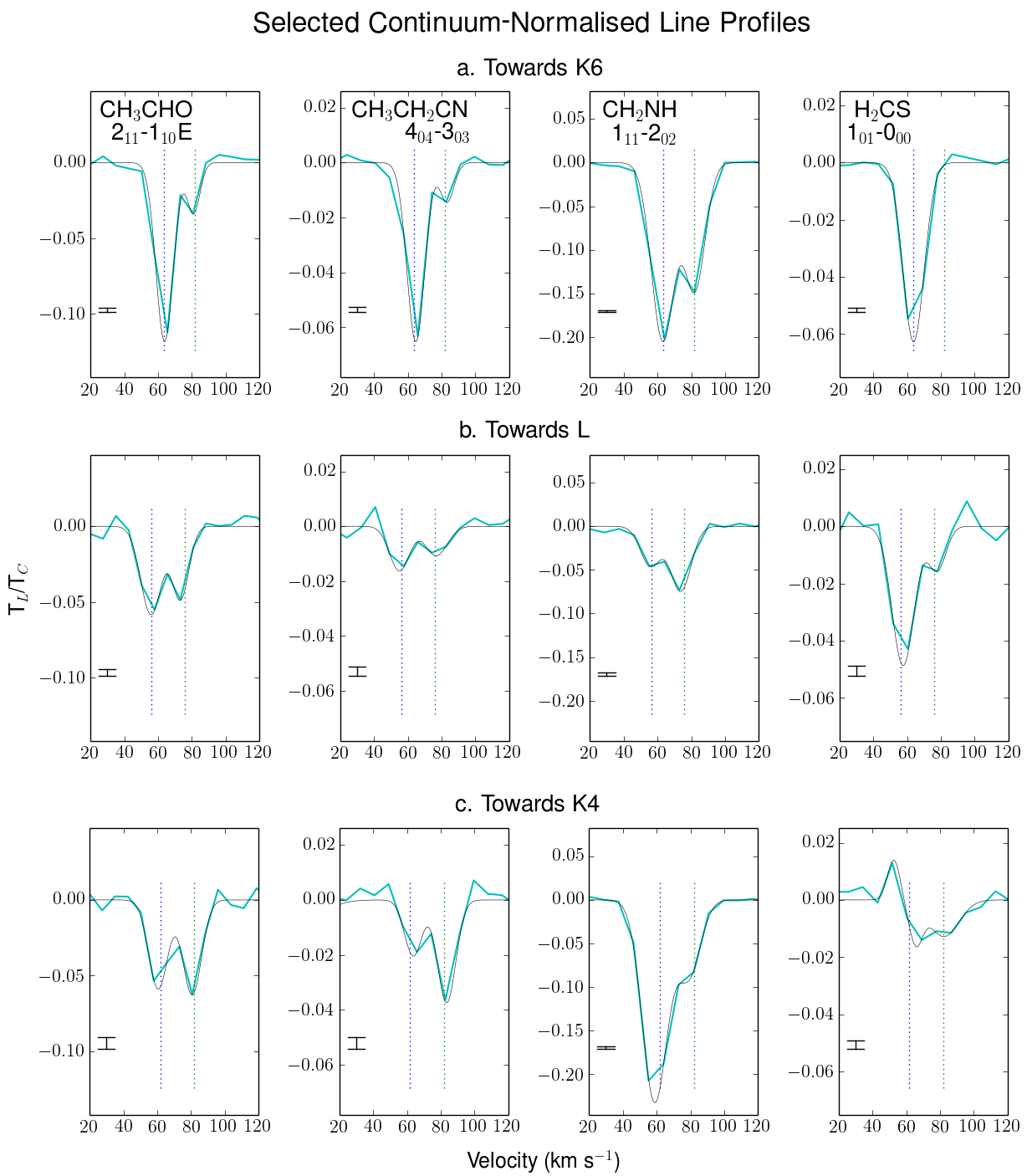}
\caption{Line profiles of CH$_3$CHO 2$_{11}$-1$_{10}$E (39362~MHz), CH$_3$CH$_2$CN 4$_{04}$-3$_{03}$ (35722~MHz),
CH$_2$NH 1$_{11}$-2$_{02}$ (33704~MHz), and H$_2$CS 1$_{01}$-0$_{00}$ (34351~MHz) reveal differences in the line optical  
depth ratios between all spatial positions and all velocity components.  
The data (cyan) are overlaid with the automated line fits (black), and vertical dotted 
lines indicate the velocities associated with the low and high velocity gas towards each spatial region as reported in Table 
\ref{meanMolKin}.
\textbf{a.} In K6, while CH$_3$CHO and CH$_3$CH$_2$CN
have similar line profiles, the selected transition of CH$_2$NH has a comparatively enhanced line strength in K6:\,82~\kmss.  Further, 
the line radiation by H$_2$CS in the K6:\,82~\kms is significantly depleted to below the detection threshold.   
Towards L (\textbf{b.}) and K4 (\textbf{c.}) the line profiles vary significantly, and line strengths do not scale linearly with 
the strengths observed towards K6.    The scale of the ordinate is the same for each line towards all three regions, and the scale is set relative 
to the normalised line height in the K6:\,62~\kms component.
The root-mean-squared values of the extracted spectra are shown in the lower left corner of each panel.
}
\label{fig:NLprofiles}
\end{figure*}

\begin{figure*}
\includegraphics[width=7in]{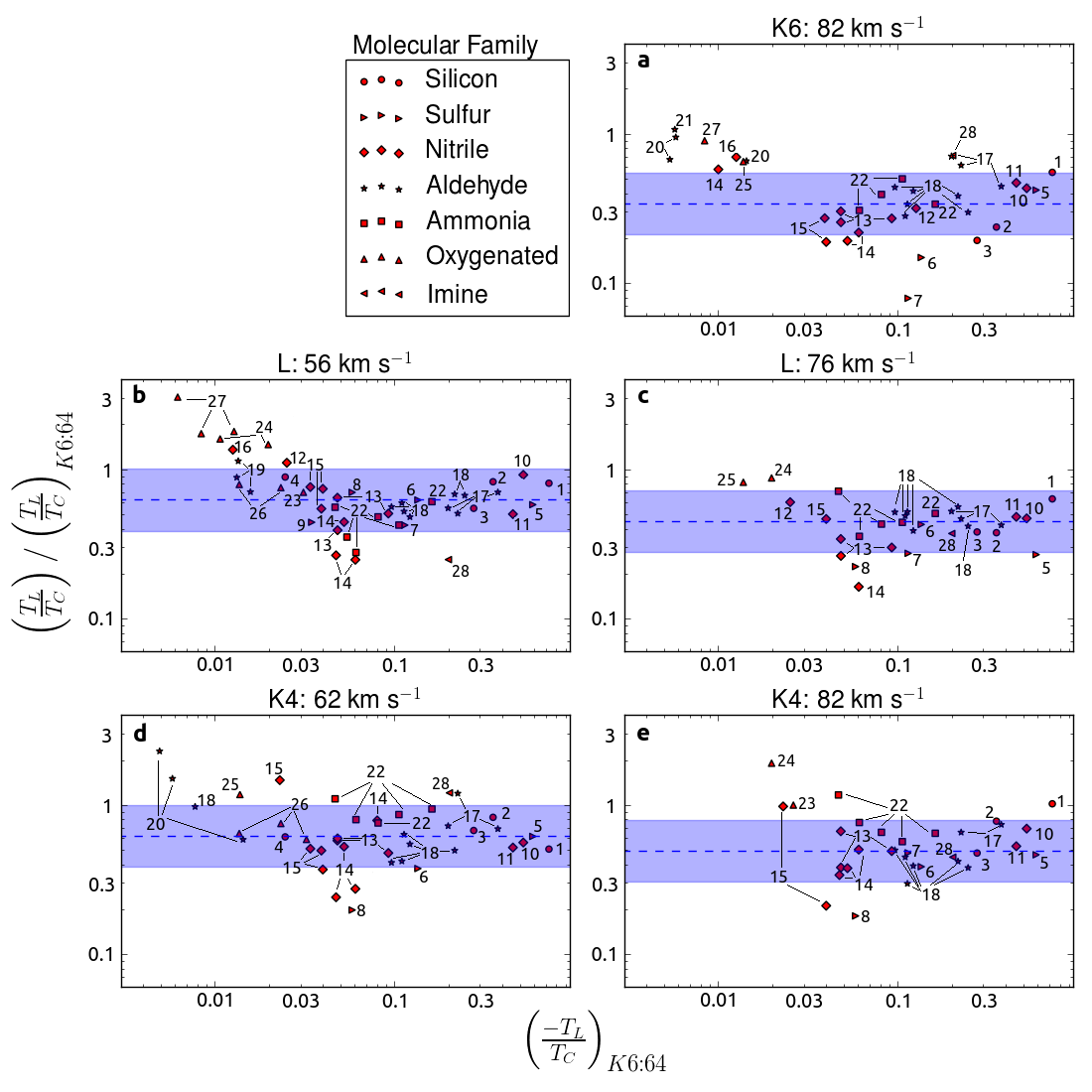}
\caption{Line height ratios with respect to the K6:\,64~\kms component indicate statistically significant differences in 
gas phase molecular abundances or excitation between different spatial positions.  In the plot, the abscissa shows the line 
 height-to-continuum ratios of lines detected in K6:\,64~\kmss.  Given the low excitation temperatures observed in this gas and the 
 high background continuum temperature,  this provides a proxy of the line optical depth towards K6:\,64~\kmss. 
 The ordinate provides the normalized line ratios (NLRs) of the lines in the relevant gas component with respect to K6:\,64\,\kmss. 
 For all but the strongest lines,
the y-axis values are proportional to the line optical depth ratio with respect to K6:\,64~\kmss.
The median ratio is represented with the dashed line, and shading indicates the  
$\pm$3$\sigma$ range.  Represented data only include lines within the high spatial resolution portion of the spectrum 
(from 30 - 44.6~GHz) for selected families of molecules. Species are numbered as follows: 1 - 4 are SiO, $^{29}$SiO, 
$^{30}$SiO, and SiS.  5 - 9 are sulfur-bearing species SO, CCS, HCS$^+$, H$_2$CS, and OCS. 
10 - 16 are nitrile species HC$_3$N, CH$_3$CN, CH$_2$CN, CH$_2$CHCN,
CH$_3$CH$_2$CN, HC$_5$N, and CCCN.
17 - 21 are aldehyde species NH$_2$CHO, CH$_3$CHO, CH$_3$OCHO, {\it cis-}CH$_2$OHCHO, and {\it t-}CH$_2$CHCHO. 
22 is NH$_3$.  23 - 27 are non-aldehyde oxygenated species H$_2$CCO, {\it c-}H$_2$C$_3$O, H$_2$COH$^+$, CH$_3$OCH$_3$, and {\it t-}CH$_3$CH$_2$OH. 
28 is the imine species CH$_2$NH.
Note that the apparent slope towards the low optical depth segment of the panels is related to the detection limit;  because K6:\,64\,\kms
has the highest optical depth, lines that are weakly detected towards K6:\,64\,\kms are detected towards other components only if they have 
higher than the median line optical depth.  This does not imply that these points are unreliable, but implies that the lower left corners of 
each panel are below the detection threshold. }
\label{fig:HeightPlots}
\end{figure*}

%\begin{figure*}
%\includegraphics[width=4in]{FIGURES/PNG_format/fig14.png}
%\caption{The Q-value between different gas components provides a measure of how distinct 
%each gas component is from each other.  As detailed in Equations 6 and 7,
%this value provides the average offset of points in Figure \ref{fig:HeightPlots} from the median in units of the 1$\sigma$ error.
%We cycled through each component as the reference component to obtain Q-values for every component pair.}
%\label{fig:Qvals}
%\end{figure*}

\begin{figure*}
\includegraphics[width=6.5in]{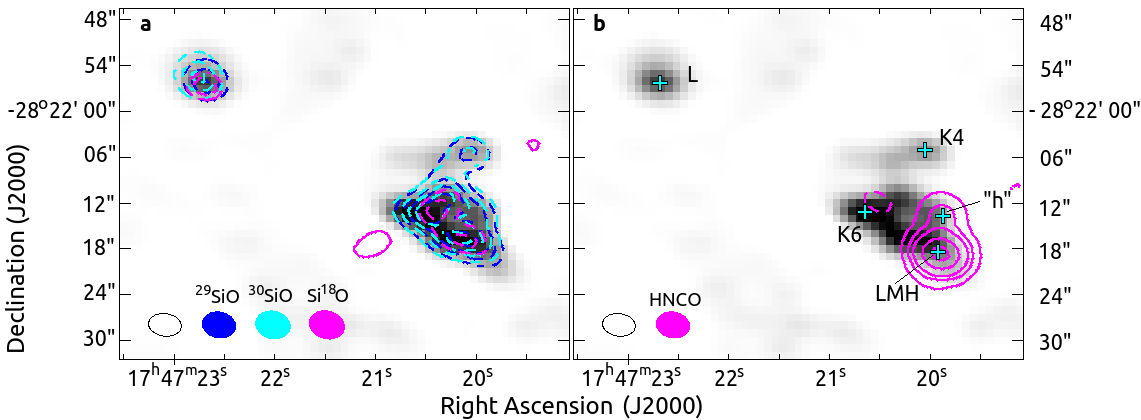}
\caption{Integrated line contours of the (1-0) transitions of SiO isotopologues are anticorrelated with
the 2$_{02}$-1$_{01}$ transition of HNCO.  \textbf{a.}  $^{29}$SiO, $^{30}$SiO, and Si$^{18}$O
are observed in absorption (dashed contours) against the continuum and exhibit no emission associated with the hot cores.    \textbf{b.} HNCO is
present in emission towards the LMH and ``h'' hot cores.  A very weak absorption feature (with a contour at 2.5\% of the peak line strength) 
is detected slightly north of K6.  The crosses indicate the center positions of the elliptical regions from which spectra were extracted.
The synthesized beams for the continuum and line images are shown in the lower left corner.}
\label{shocker}
\end{figure*}

\begin{figure*}
\includegraphics[width=6.5in]{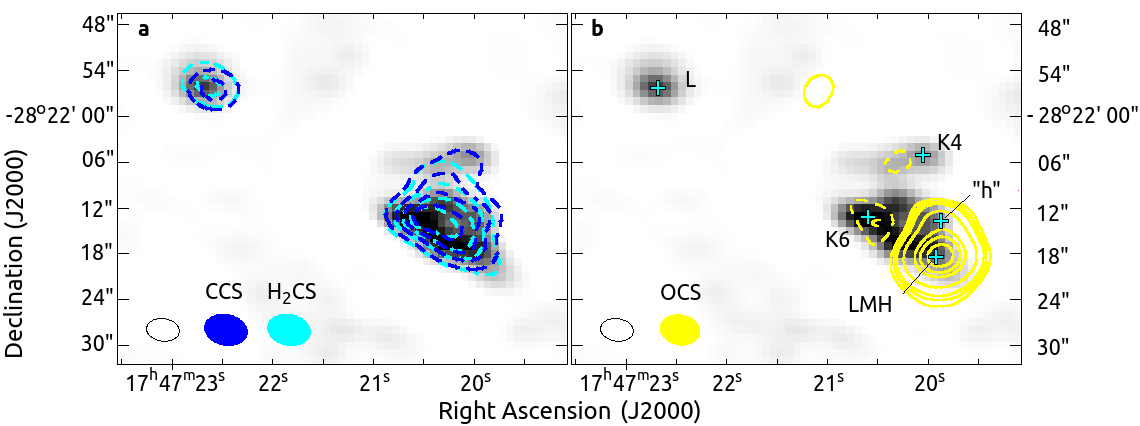}
\caption{Integrated line contours of H$_2$CS (1$_{01}$-0$_{00}$), CCS (3$_2$-3$_1$), and OCS (3-2). \textbf{a.} H$_2$CS and CCS are 
observed in absorption (dashed contours) against K4, K5, K6, and L, with no detected emission from hot core regions.  
Contour levels are at 20, 40, 60, and 80 \% of the peak line strength.
\textbf{b.}  OCS appears in emission from the hot core regions, the LMH and ``h'', with very weak absorption observed
towards K6 and near K4.  The contour levels are at 4.5, 7, 20, 40, 60, and 80 \% of the peak line strength.   
The crosses indicate the center positions of the elliptical regions from which spectra were extracted.
The synthesized beams for the continuum and line images are shown in the lower left corner.}
\label{CSfam}
\end{figure*}

% extended tables and figures for online edition
% may need B&W figures for paper edition too

\twocolumn

\bsp

\label{lastpage}

%, which typically has n~\textgreater~10$^5$ cm$^{-3}$ \citep{Lis91}

\clearpage

\section*{Appendix A\\Automated Line Fitter Code Release} \label{Ap:AppendixA}

All scripts used for automated line fitting and semi-automated line identification are provided in the online
journal and in GitHub at https://github.com/jfc2113/MicrowaveLineFitter.
We provide instructions for use including a description of the input and
output files and system requirements.

The scripts have been tested in {\sc ipython} version 0.12.1 with {\sc python} 2.7.3, 
{\sc numpy} 1.6.1, and {\sc matplotlib} 1.1.1rc. 
{\sc Python} 2.7.3 includes the {\sc math} and {\sc csv} modules that are 
used in the scripts.  

The packaged file provided in the journal includes eight scripts, spectra characterized in this work, and 
csv files holding line data.  
Two of the scripts are run directly, namely  `FullCode.py' is run before `VelocityFileMaker.py'.
After optional adjustments by the user, `VelocityFileMaker.py' uses the output of `FullCode.py' 
to generate a final csv file that contains 
Gaussian fits to the full spectrum and line identifications.  Of the remaining six scripts, 
five contain modules called in `FullCode.py'.  The modules 
and their primary purposes include:\begin{itemize}
         \item `LineFind.py' grabs segments of the spectrum that contain lines, detecting ``line-channels'' with single-channel intensity values greater 
         than $\sigma_{rms} \times $ an input threshold.  The module also selects a set number of channels on either side of  line-channels.  The number of neighboring
         channels has a default value of 5, but can be adjusted when LineFind is called.
         \item `FTinterp.py' interpolates a segment of the spectrum, outputting a spectrum with a factor of 3 improvement in the spectral resolution.  It performs the
         interpolation by Fourier-domain zero padding.  This should only be used for data with poor spectral resolution compared to the linewidths.
         \item `Gfit.py' provides the most essential functionality; through `FullCode.py', the function Gfit.GauFit is passed a single segment of the spectrum that contains one or more 
         lines.  Gfit.GauFit returns a 1- or 2-component Gaussian fit to a spectral feature within the segment.  The script selects a specific portion of the line-containing segment
         and first attempts a one-component fit.  It then follows an algorithm to adjust the section
         of the segment used for line-fitting and re-fit 1-component Gaussians.  It passes information to `G2criteria.py', which 
         evaluates the fit against a set of criteria to determine whether a 2-component Gaussian fit is justified.  The output of G2criteria 
         determines whether Gfit.GauFit fits a 2-component Gaussian, following an algorithm to adjust the section used for line-fitting and attempting 2-component Gaussian fits.
         \item `G2criteria.py' determines if a 2-component Gaussian fit is required.  Gfit.GauFit provides G2criteria with information about the data and the best 1-component 
         Gaussian fit achieved; G2criteria then determines if the data differs significantly from the 1-component fit, requiring a 2-component fit.  G2criteria takes into account the 
         signal-to-noise ratio of the line in question and the degree to which the data differs from the 1-component Gaussian fit.  The primary factors considered include
         (1) the difference between the position of the absolute maximum of the data and the position of the 1-component Gaussian center; (2) the maximum residual between the data and
         the fit, compared to the rms noise level at that frequency tuning; (3) the mean residual between the data and the fit, compared to the rms noise level; (4) the width of the
         fit compared to the typical width of features.  The criteria, mostly empirically determined, works very 
         well for this dataset.  However, we advise inspecting results obtained on other datasets carefully and likely adjusting the criteria, as they
          will change with the parameters of the data.  
         \item `LineID.py' provides semi-automated line identification of features fit over the full spectrum.  FullCode compiles the output of Gfit.py and passes it to LineID, along 
         with one or two csv files containing spectral line catalog data.  The input csv files can quickly and easily be output from the ALMA Spectral Line
      Catalogue - Splatalogue\footnote{Available at www.splatalogue.net; Remijan et al. (2007)} after selecting a set of molecules that might appear in your spectrum.
        \end{itemize}
   
FullCode manages the full data spectrum, compiles the results of Gfit.GauFit, and calls LineID to output the data.  All of the modules except G2criteria are written to 
be easily tunable, and the main parameters that users will want to adjust are specified and explained in the first section of FullCode.  These parameters include:
whether the results should be output; what plots should be produced; the names of input files; whether data interpolation is desired; 
information about the source including guidelines for appropriate linewidths and line velocities; and information about the data structure 
(e.g. how many separate data sections are in a single spectrum; in this work, the line-fitter inspected a spectrum including 11 tunings, each of which had different $\sigma_{rms}$
noise levels.  If all tunings have the same noise level in the intensity units you are using, they can be treated as a single section).

FullCode requires a spectrum input in standard ascii format with frequency and baseline subtracted intensity.
The file `K6\_fullSpec.txt', provided with the scripts, is in the correct format.  If the user chooses to perform line identification, 
the script requires files containing recombination line or molecular line data output in csv format from any spectral line catalog.  The line data should 
be formatted similarly to the files `recombLines.csv' and `molecularLines.csv', provided.  If your source has recombination lines, 
we recommend preserving the file name of `recombLines.csv', but including lines that are relevant for your frequency range.  
All csv files read into the code should be text csv files with {\sc unicode UTF-8} characters and a colon (:) as the field delimiter. 
The code outputs csv file data with this format as well.

FullCode plots the full data spectrum with the line fits overlaid, marking the locations of lines from the input catalogue 
data that are near the fit frequency centers. The code can output a csv file containing Gaussian line fits and 
possible line assignments.  This output of FullCode is named ALLFITS\_(r).csv, where (r) is the name of the source or region targeted.  
It contains:
\begin{itemize}
 \item Gaussian line fit parameters (height, center frequency, and width in frequency units) of all features.
 \item two measures of the quality of the fit.  First, the rms residual between the data and the fit; second,  for 1-component fits, the file provides
the offset between the frequency at which the data has its maximum absolute value and the Gaussian center frequency. Fits with abnormally large values may be suspect. 
 \item the `bin' indicates how the line was handled in Gfit.GauFit. Lines with bin = 0 are 1-component fits that were deemed appropriate.  Lines with bin = 1 are 2-component
 fits that were found to be reasonable.  For lines with bin = 2, the data significantly differed from the best 1-component fit obtained, however Gfit.GauFit was unable to 
 obtain a reasonable 2-component fit and therefore used the best 1-component fit obtained.  These should be inspected and possibly re-fit manually.  
 For bin = 3 lines, the residual between the data and the fit
 is significantly larger than the noise level of the spectrum.  These should be inspected and possibly re-fit manually.  Bin = 5 lines are entirely unreasonable,
 with very broad line widths or unreasonable line heights.  They are used as placeholders to enable the code to proceed, and they will need to be re-fit.  Of 617 components 
 used to fit the spectrum towards K6, 
360 have bin = 0, 242 have bin = 1, 1 has bin = 2, 12 have bin = 3, and 2 have bin = 5. 
 \item possible atomic or molecular carriers of the detected Gaussian features.  
 The output file lists any transitions from the input recombLines.csv and
 molecularLines.csv files that are within a specified velocity range of the detected features. 
 It does not choose the most probable line at this step, but does so when the
 second code is run.  These can be prioritized before running VelocityFileMaker.py to obtain
 the final line results.
\end{itemize}

Upon inspection of the results and manual updates to any bad fits, run VelocityFileMaker to generate a final output csv file containing line identifications and Gaussian fit parameters,
with parameters listed in velocity space. VelocityFileMaker generates an output csv file entitled `velocity\_ALL\_(r).csv'.  The file 
contains line rest frequencies, species, transitions, Gaussian fit velocities, heights, widths (in velocity units), 
center frequencies, and line types.

Before running VelocityFileMaker, edit parameters in `velEDIT.py'.  Specify the names of the positions from which you extracted 
spectra, csv file input names, and velocity ranges that are appropriate for different types of lines, including recombination lines, molecular lines from the primary source, 
and additional lines observed in absorption by foreground diffuse or translucent clouds.  The latter are only relevant in some lines-of-sight.

The code implements primarily kinematics-based consistency checks for line identifications, and can use one or two optional 
input csv files to prioritize line identifications. In order to minimize mis-identification, the final line-identification process 
operates on a priority system as follows:
\begin{itemize}
 \item  Priority 1: The code first inspects identified recombination lines for kinematic consistency and consistent line-intensities.  
 This step requires a sufficiently large number of recombination lines to work well (\textgreater10 high signal-to-noise lines or \textgreater20 lines recommended), 
 as it derives the mean Gaussian fit parameters and standard deviations in order to evaluate consistency.
Lines that are self-consistent with the mean and standard deviations are output as firm line identifications, and assigned a ``LineType'' of either ``H Recomb'' or ``He Recomb''.
The script is not presently equipped to handle carbon recombination lines or absorption by recombination lines, but could be easily adjusted to do so.
Features that are inconsistent with recombination lines include absorption lines, features that are significantly more broad or narrow than the mean, features 
that are too far from the mean center frequency,
and lines that are significantly too strong or weak (e.g. an H$\,\gamma$ line is as strong as most H$\,\alpha$ lines).  These may be output as tentative line identifications,
which the user should inspect for line blending or data issues.  If you do not have recombination lines, the code will simply move to Priority 2.  
 \item  Priority 2: The user has the option of specifying a file entitled `strongLines.csv'. (Of course, the file name can be changed per the users preference).  
 The file ``strongLines.csv'' contains the strongest lines, which can be firmly identified prior to identifying weaker molecular lines.  Lines that fall within the 
 velocity range set in the function vEDIT.strongVrange are output as firmly identified, and assigned ``LineType = strong''.  Verify that the first two columns of 
 your strongLines file are formatted the same as the file included in the package.
 \item  Priority 3: The user can use a file entitled `SAClines.csv', which indicates which transitions that may be observed in foreground absorption by 
 diffuse and translucent cloud material, often referred to as Spiral Arm Clouds (abbreviated SAC). The foreground absorption components are at velocities that
 are inconsistent with the targeted cloud, and must be constrained in VelocityEditFile.  Lines that fall within the 
 velocity range set in the function vEDIT.SACvRange are output as firmly identified, and assigned the ``LineType = SAC''.  Verify that the first two columns of 
 your SAClines file are formatted the same as the file included in the package.
 \item  Priority 4: The code then assigns remaining lines tentatively assigned in ALLFITS\_(r).  In cases where multiple transitions are near the Gaussian center frequency, 
 the line with the best kinematic match to the velocity, as specified in VelocityEditFile, is output.  While this typically works well
 at centimeter wavelengths, we recommend inspecting lines that have multiple line entries in ALLFITS\_(r), especially at higher frequency.  
 If the code gets the ``wrong'' answer at this point, mark the line as blended, and consider only keeping the preferred line in ALLFITS\_(r)
 \item  Priority 5: Finally, the code outputs all fits that were not firmly identified in Priorities 1-4.  These are given a Line Type of ``Unidentified''.  
 If a fit was associated with a known transition in ALLFITS\_(r), the transition parameters will be output so that the user can decide whether or not to 
 adopt the transition as identified.
\end{itemize}
While FullCode.py can fit most blended lines well, VelocityFileMaker does not handle identification of blended lines with much sophistication; 
however the prioritization system significantly improves the likelihood that most of the line
radiation can be ascribed to the identified carrier transition.  The user is responsible for inspecting the output of ALLFITS\_(r) and velocity\_ALL\_(r) 
to ensure that any lines that may be blended (which should have multiple transitions listed in ALLFITS\_(r)) are handled appropriately.

Within each priority, the script also handles wing components.  As described in \S 3.2, numerous spectral features are non-Gaussian, including recombination lines, 
strongly masing lines, and optically thick molecular lines.  In FullCode, in which the line fitting is 
conducted, after subtracting a best-fit 1- or 2-component Gaussian fits from the raw data (called the primary component(s)), 
high signal-to-noise lines may have wing components with intensities exceeding the detection threshold.  
Such wing features, which are typically weak as compared to the primary 1- or 2-component fit, are fit to the residual.
VelocityFileMaker selects these features to prevent them from incorrectly being assigned to other molecular carriers or
labelled as unidentified.  Wing features are selected based on (1) their separation from the primary line component 
as compared to the width of the primary line component and (2) their height as compared to the primary line component.  
If a feature meets either of the following criteria, it is identified as a wing of an identified line and the velocity is 
calculated with respect to the rest frequency of that line. The criteria include:
\begin{enumerate}
 \item ${\frac{H_{\text{Feature}}}{H_{\text{Primary}}}}$\,\textless\,0.22 and $ v_{\text{Feature}} - v_{\text{Primary}}$\,\textless\,$\Delta v_{\text{Primary}}$
 \item ${\frac{H_{\text{Feature}}}{H_{\text{Primary}}}}$\,\textless\,0.05 and $ v_{\text{Feature}} - v_{\text{Primary}}$\,\textless\,$1.5 \times \Delta v_{\text{Primary}}$,
\end{enumerate}
for the line heights, velocity centers, and FWHM widths $H$, $v$, and $\Delta v_{\text{Primary}}$, respectively.  These are assigned ``LineType = wing''.
Within each priority, VelocityFileMaker identifies the primary components, which must be consistent with the kinematic 
information input, and then selects any wings associated with these lines.  As such, the recombination lines and wings of recombination 
lines are identified, followed by designated ``strong lines'' and the wings of strong lines, etc.   

\clearpage

\section*{Appendix B\\Full Spectra and Line Identification Figures} \label{Ap:AppendixB}

The raw data spectra extracted towards K6, L, and K4 are provided with the output of the line fitter overlaid and line identifications labeled.  
All data are frequency shifted to the nominal source velocity of 64~\kmss. The following conventions are used in all figures.  
Tentative identifications are noted with a `?', as are unidentified lines suspected to be artifacts.  
Strong masing transitions are marked with `$Maser$', blended lines are marked with `$Blend$', 
and line fits that have been adjusted from the output of the line fitter are marked with `$Adj$'. 
In cases where the line-fitter determined a 1-component fit that we suspect to be a 2-component transition
with insufficient signal-to-noise to be identified as such by the fitter, the species is marked with `$BC$'.

For cases in which a transition has blended spectral 
structure (such as blended transitions of A and E states), the listed transition indicates what components are expected to be blended, 
with the first listed component corresponding to the state that is strongest or best matches the expected source velocity.  The 
transition listed first sets the rest frequency in Column 1 in the table provided in Appendix C.

Recombination lines are marked at 75, 76, and 64~\kms ~for K6, L, and K4, respectively.
Identified molecular line components are marked at the following velocity components:
\begin{itemize}
  \item[K6:] Primary velocity components are at 64 and 82~\kmss; translucent cloud components are at 6, -30, -73, and -106~\kmss.
  \item[L:] Primary: 56 and 76~\kmss; translucent cloud: 35, 20, -5, -35, -75, and -106~\kmss.
  \item[K4:] Primary: 62 and 82~\kmss; translucent cloud: 2~\kmss.
\end{itemize}

Tentatively identified velocity components of firmly identified transitions are marked with a dashed pointer.  
For lines identified in the high velocity gas but not in the low velocity gas, the low velocity component is marked with a dotted line.
Figure \ref{dummyFig} illustrates these conventions.

\subsection*{Notes on U-lines, Tentative U-lines, and Line Contamination}
The figures and tables in Appendices B and C include firmly identified features, contamination 
from flux originating from other locations on the image (``contam''), tentatively
identified features (marked with a ``?''), unidentified lines (marked with ``U'' in figures and tables
and referred to as U-lines), and tentative U-lines (marked with ``U?'' in figures and tables
and referred to as ``U?-lines'').

Nearly all instances of line contamination are present in the spectrum of K6 and are lines that originate in the 
LMH or ``h'' hot cores. In spectra extracted from the high spatial
resolution data (29.8 to 44.6 GHz), the hot cores are sufficiently separated from the elliptical region
placed on K6.  As a result, the line emission from the hot cores does not generate artifacts
in the spectra towards the regions characterised here for frequencies $\nu$~\textless~44.6~GHz.  In this region 
of the spectrum, we have carefully inspected the images and spectra to verify that strong lines in 
the hot cores do not have imaging artifacts including negative bowls that are detected in K6, K4, and L.
In the spectra extracted from the low resolution data (towards K6 and L), no imaging artifacts are 
observed towards L, while 30 imaging artifacts are present in the spectrum of K6.  The authors have
carefully inspected the images and spectra to verify that these are properly handled.  The instances
of contamination occur more frequently in the lowest resolution image (from 44.6 - 46.5 GHz) than in
any other image, as would be expected.  In cases where line absorption is observed towards K6 and line
emission is produced in the LMH and/or ``h'', the line profiles in the low resolution spectra will not be 
representative of the true line flux towards K6.  For this reason, we excluded low resolution 
spectra from the analysis of line ratios.  

In addition to the imaging artifacts towards K6, a single feature is labeled as contamination towards L
in the high spatial resolution portion of the spectrum.  The methanol masing transition at 36~GHz 
appears as an absorption feature in the spectrum of L. Inspecting the image however, it is clear that imaging 
artifacts are implicated, as the very strong masing transition has limited dynamic range and image 
fidelity.  We have inspected the images and spectra to ensure that the masing features identified towards
K4 and K6 appear to arise from these regions.  No instances of line contamination were observed towards K4.

Tentative U-lines (U?-lines) are features that meet the detection threshold, but upon inspecting the spectra and images, 
the authors remain unconvinced that a real line is present.  As noted in \S3.2, with the established 
threshold of 3.5$\sigma$, one channel in 2149 line-free channels should meet the detection threshold assuming 
perfect baseline subtraction and frequency-independent noise within each tuning.  With minor baseline residuals, it 
is reasonable to expect an average of $\sim$1 falsely detected feature per tuning.  In the spectra of K6 and L, composed of 11 tunings, 
we report 11 and 10 U?-lines respectively consistent with this expectation; in the spectrum of K4, composed of 7 tunings, we report 11 U?-lines.
The number in emission and absorption are similar, as would be expected for random noise and baseline variations.

As is typical of line surveys of Sgr B2, unidentified transitions are detected in the spectra towards all three positions.  
However, while a large fraction of detected lines are unidentified in most surveys of the LMH hot core 
(see e.g. Belloche et al. 2013; Nummelin et al. 1998), roughly 90\% of all detected features 
are confidently assigned towards K6, L, and K4.
Of the 496 features detected towards K6 (excluding contamination from the LMH and ``h''), 89\% are confidently assigned,
while tentative assignments, U-lines, and U?-lines compose 2, 7, and 2\% of the detected features, respectively.  Towards K6, 
three of the U-lines are observed at two velocity components, with velocities consistent with the 64 and 82~\kms components
characterised in \S3.2. An additional 23 lines are observed
in absorption at a single velocity component, and 4 are observed in emission.  The fraction of U-lines in emission is thus 
$\sim$12\%, significantly larger than the 1\% and 2.5\% observed for firmly assigned molecular lines (excluding the methanol masing transitions)
in the low and high velocity gas, respectively.

Towards L, 93\% of the 273 detected features (excluding contamination) are firmly identified. 
Tentative assignments, U-lines, and U?-lines compose 2.6, 1.1, and 3.7\% 
of the transitions, respectively.  Only three U-lines are reported towards L, and all are 
in absorption.  Two of the three U-lines in L are detected in absorption towards K6, 
and the third does not meet the detection threshold in K6, but a clear absorption 
feature is present towards K6.  Excluding recombination lines and U?-lines, all 
detected molecular lines are in absorption towards L.  
Neither strong maser transitions nor weakly masing transitions are detected towards L. 

%for some reason, it is putting a - mark at the start of this paragraph!!
Of the 192 lines detected towards K4, 90\% are firmly identified; tentative assignments, U-lines, and U?-lines compose 1.6, 2.6, and 5.8\% respectively.
Weak masing constitutes 3 and 2\% of the firmly detected molecular lines in the low and high velocity gas, respectively, 
in K4, and one of the five U-lines observed towards K4 appears in emission.  Two of the absorbing U-lines detected in K4 are also detected in 
K6, but K4 and L have no U-lines in common with one another.  It is possible that the U-line observed in emission towards K4 is produced by H$_2$CS 
at the same velocity as the methanol masers towards K4.

\begin{figure*}
\includegraphics[width=3.0in]{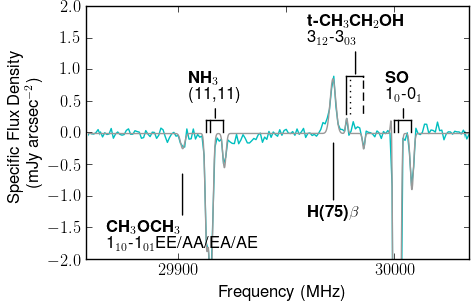}
\caption{A segment of the spectrum towards K4 illustrates the marking conventions used in this appendix.  The t-CH$_3$CH$_2$OH transition 
is firmly detected at 82~\kmss, indicated by the solid line pointer.
No emission or absorption is detected at 62~\kmss, although this component is typically the most prominent.  The dotted
line marks 62~\kms, at which we surprisingly detect no line radiation.  Additionally, a 2~\kms component is tentatively detected,
producing a dashed pointer.    The transitions of SO and NH$_3$ each have three 
firmly identified velocity components marked by pointers at 82, 62, and 2~\kmss. The position of CH$_3$OCH$_3$ at 62~\kmss is marked by 
the single pointer at 62~\kmss. Pointers for recombination lines are at 65~\kmss.}
\label{dummyFig}
\end{figure*}

\begin{figure*}
\includegraphics[width=6.0in]{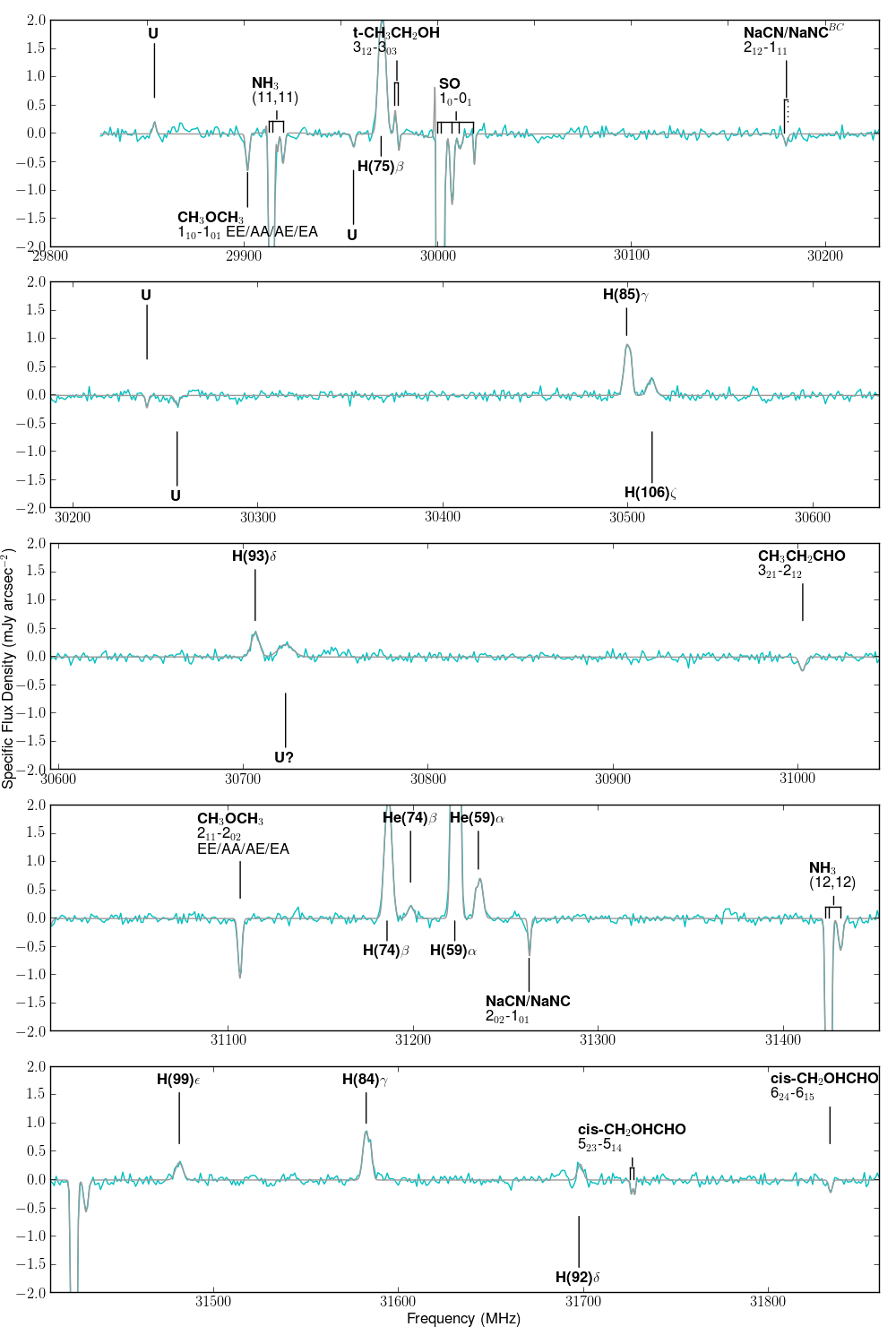}
\caption{Spectrum Towards K6.}
\label{fig:FullSpectrum_K6}
\end{figure*}

\begin{figure*}
\includegraphics[width=6.0in]{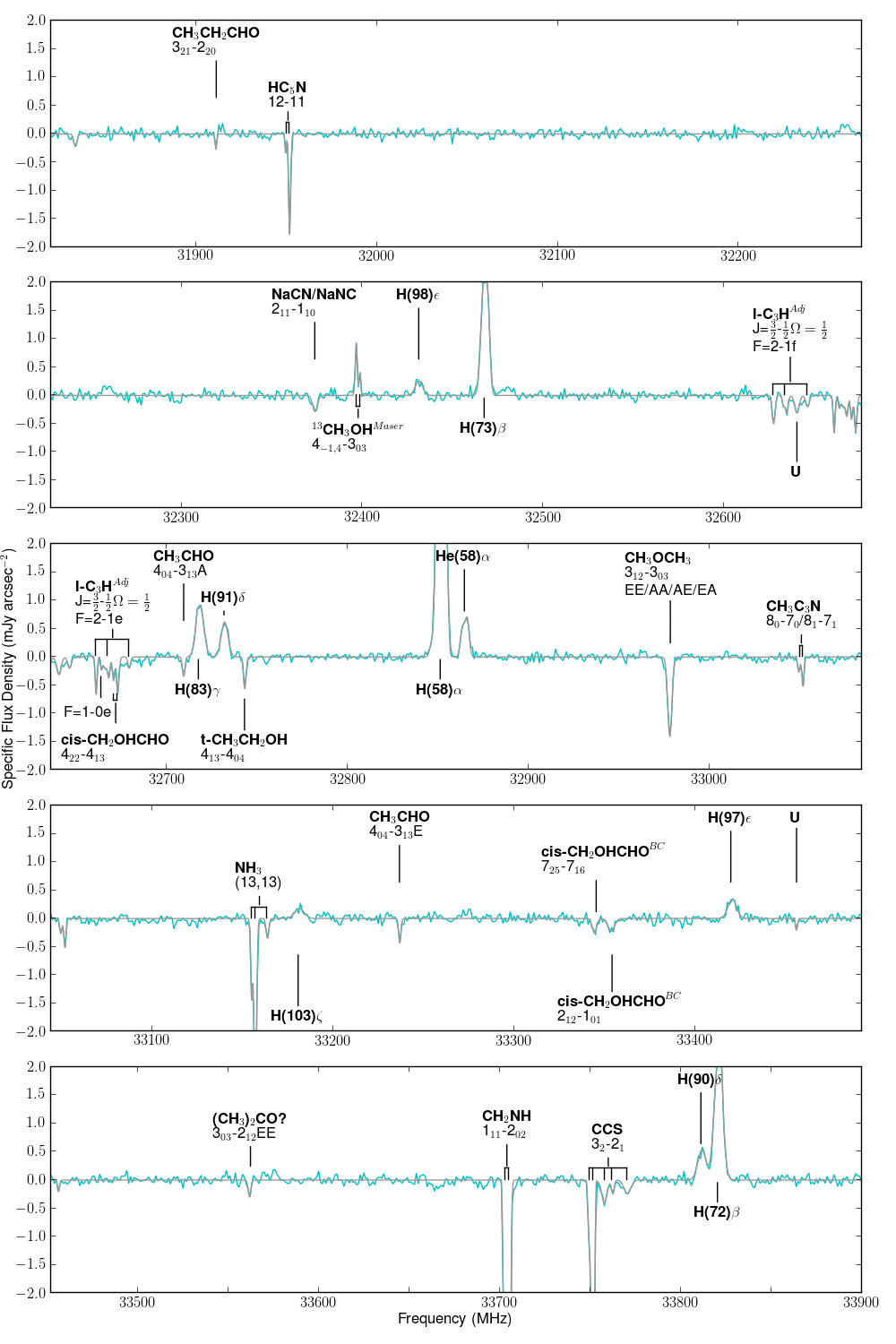}
\caption{Spectrum Towards K6.}
\label{fig:FullSpectrum_K6}
\end{figure*}

\begin{figure*}
\includegraphics[width=6.0in]{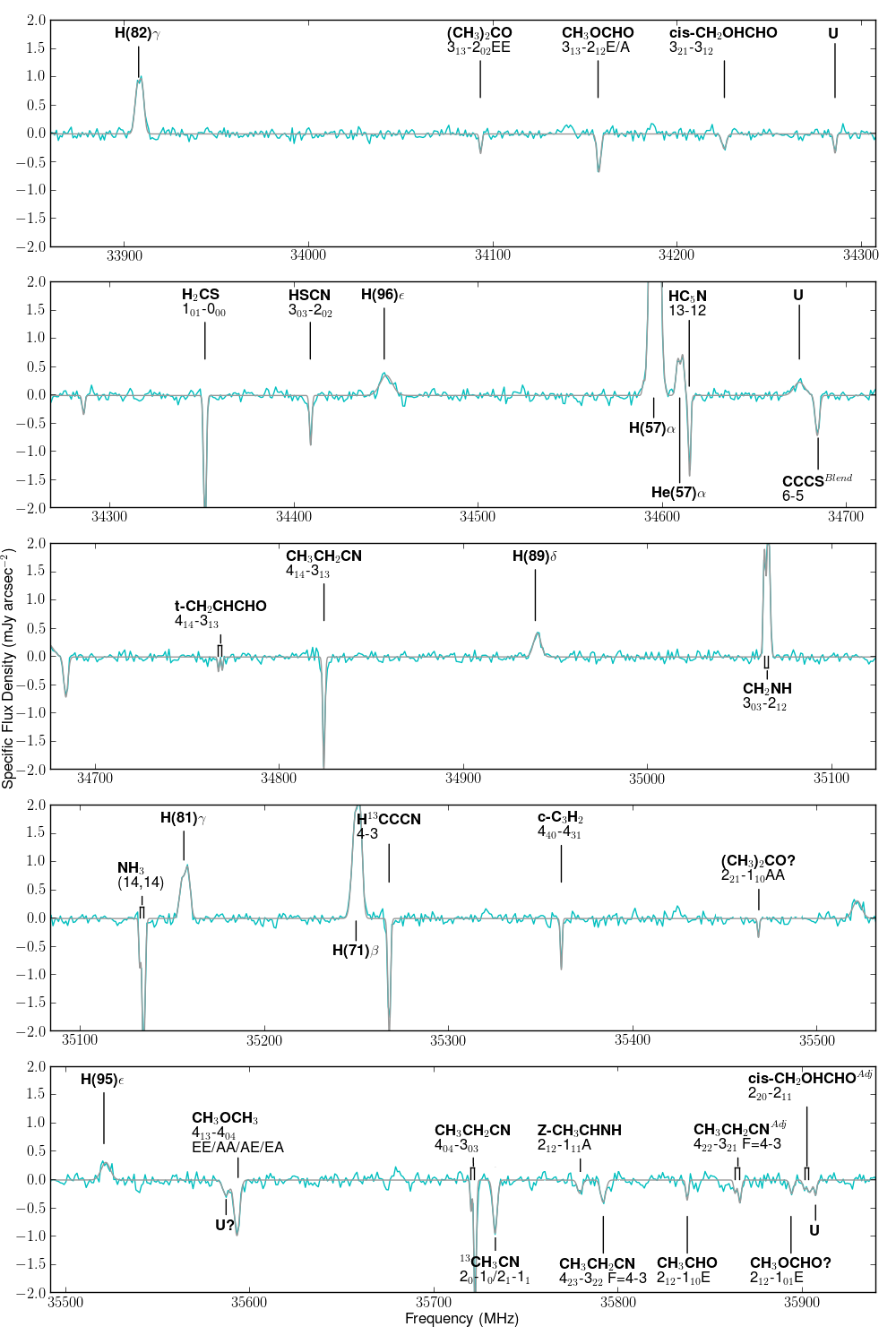}
\caption{Spectrum Towards K6.}
\label{fig:FullSpectrum_K6}
\end{figure*}

\begin{figure*}
\includegraphics[width=6.0in]{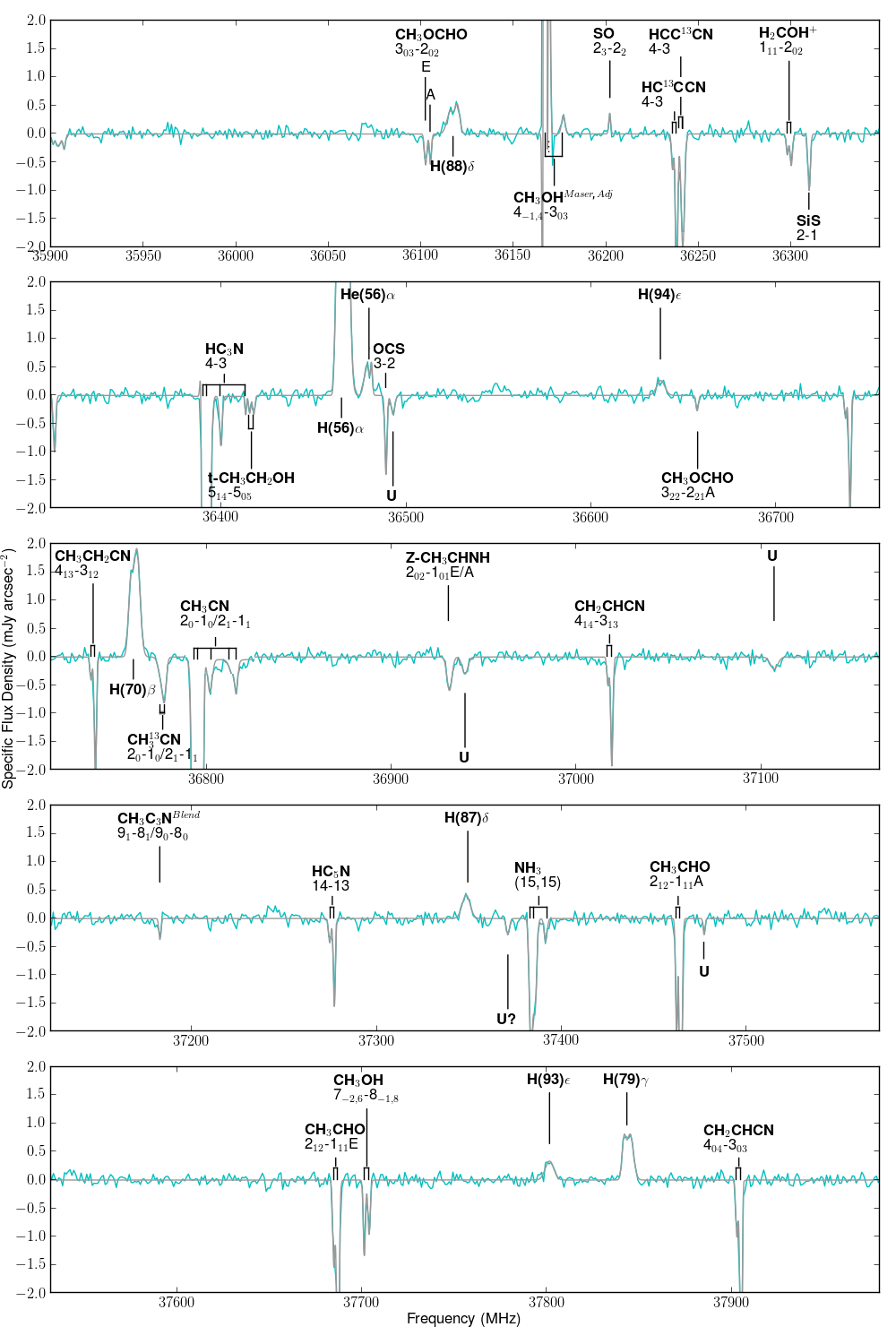}
\caption{Spectrum Towards K6.}
\label{fig:FullSpectrum_K6}
\end{figure*}

\begin{figure*}
\includegraphics[width=6.0in]{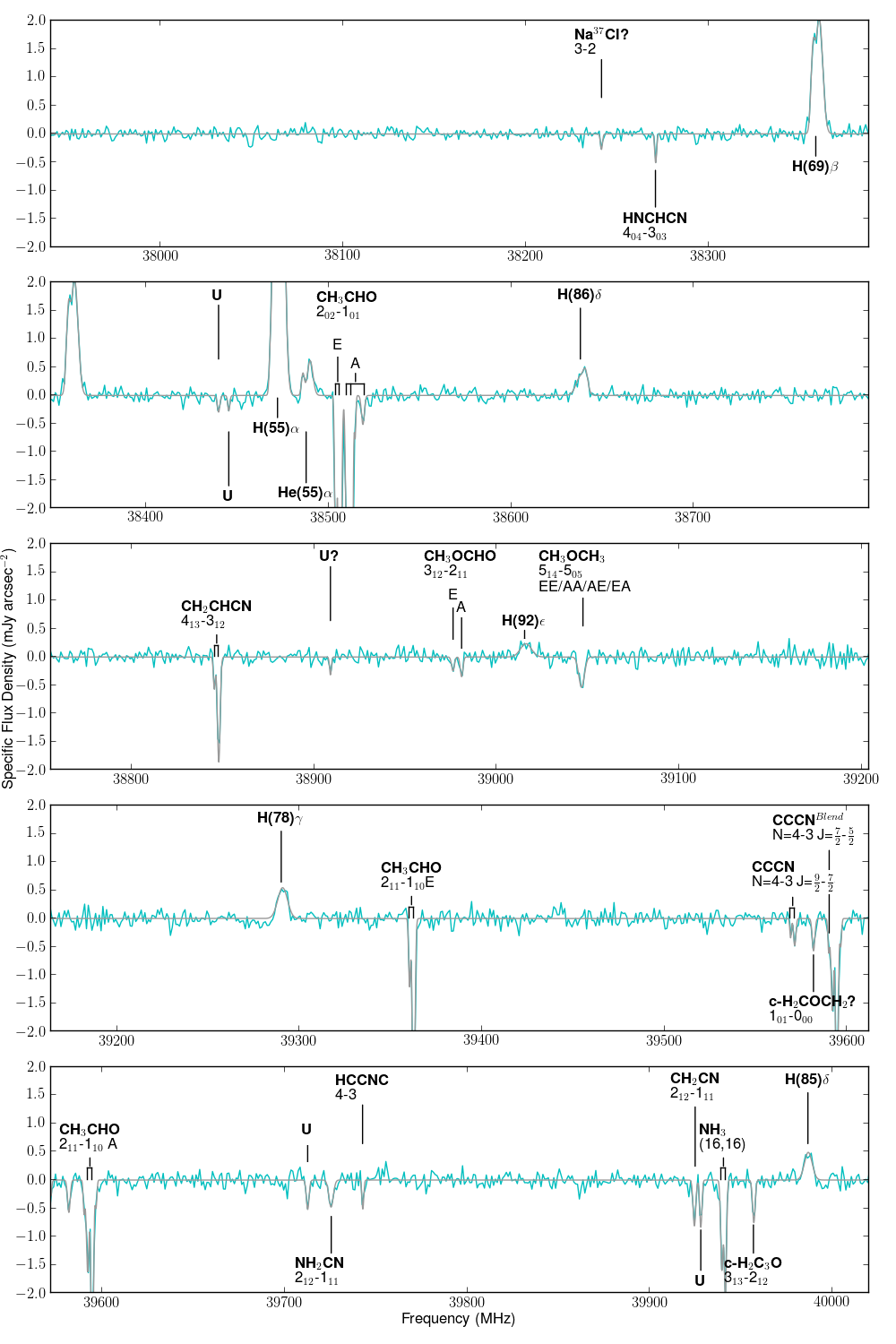}
\caption{Spectrum Towards K6.}
\label{fig:FullSpectrum_K6}
\end{figure*}

\begin{figure*}
\includegraphics[width=6.0in]{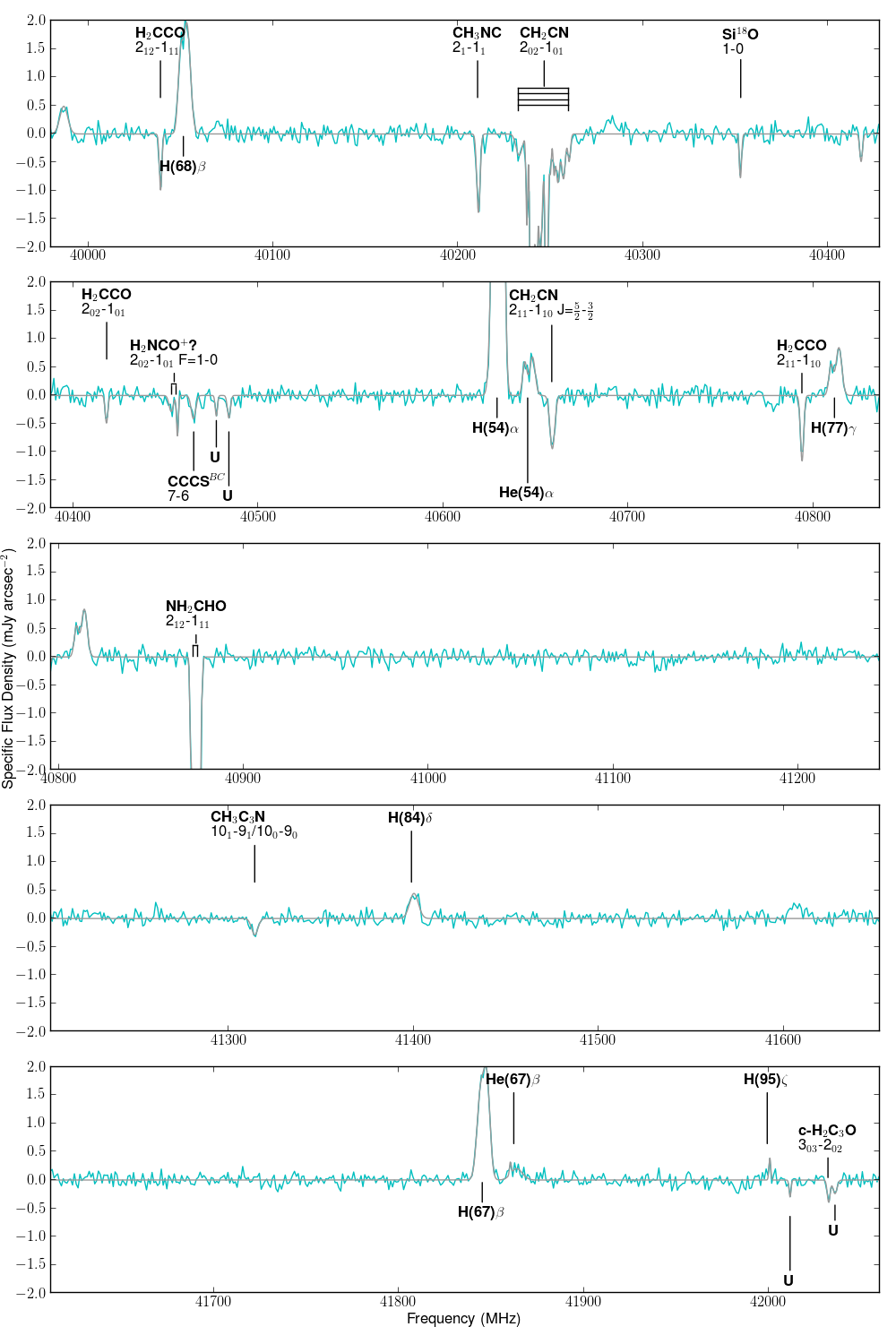}
\caption{Spectrum Towards K6.}
\label{fig:FullSpectrum_K6}
\end{figure*}

\begin{figure*}
\includegraphics[width=6.0in]{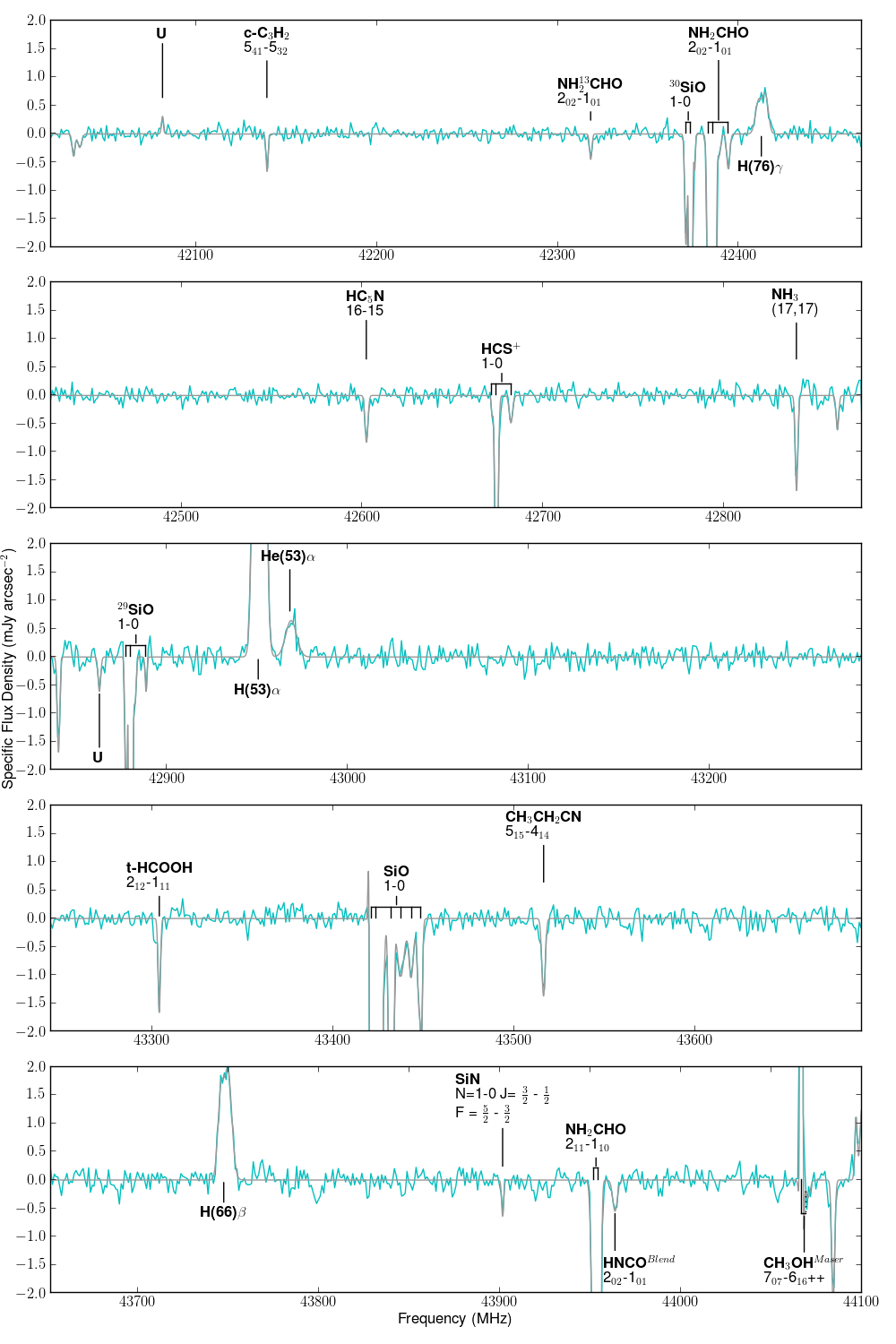}
\caption{Spectrum Towards K6.}
\label{fig:FullSpectrum_K6}
\end{figure*}

\begin{figure*}
\includegraphics[width=6.0in]{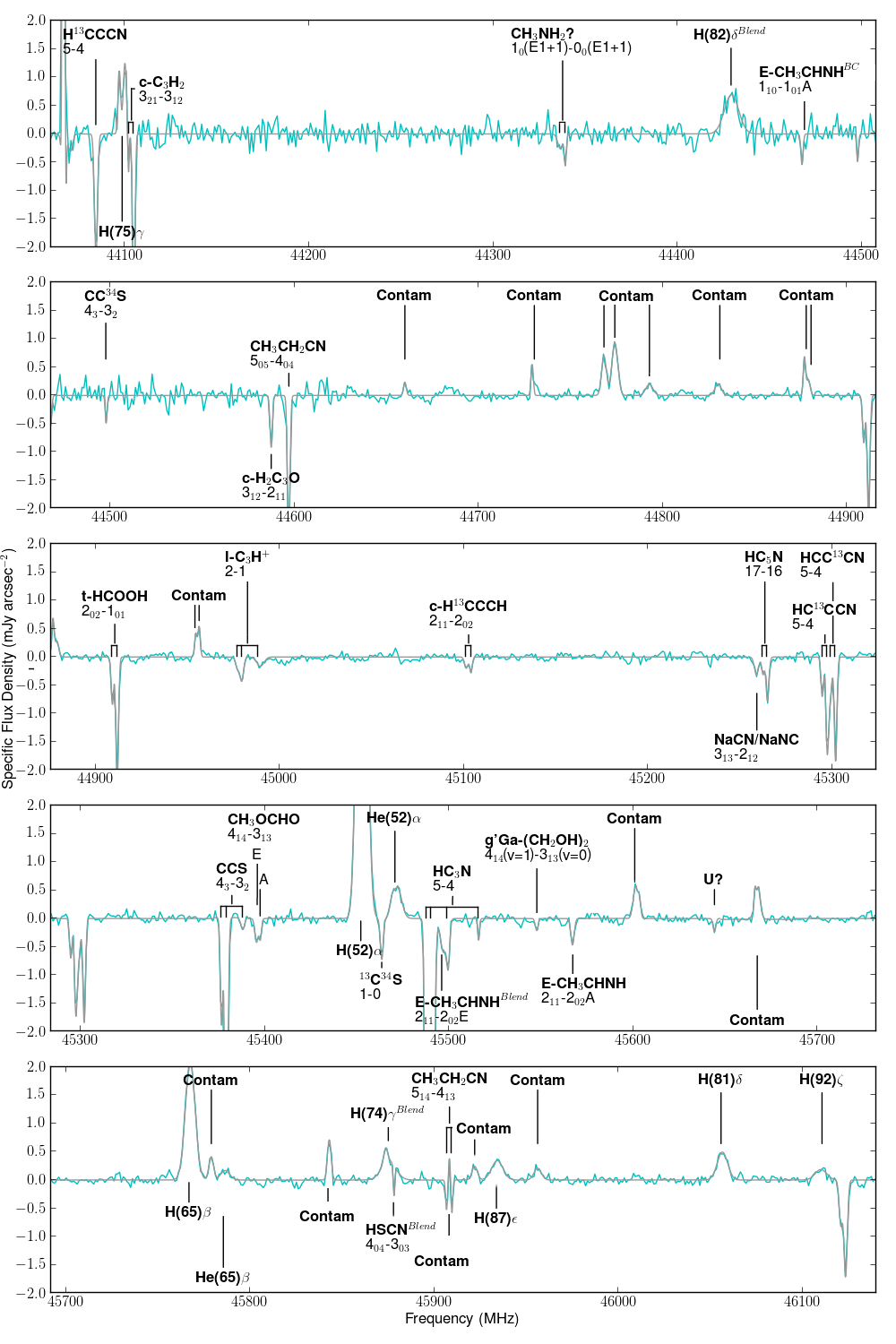}
\caption{Spectrum Towards K6.}
\label{fig:FullSpectrum_K6}
\end{figure*}

\begin{figure*}
\includegraphics[width=6.0in]{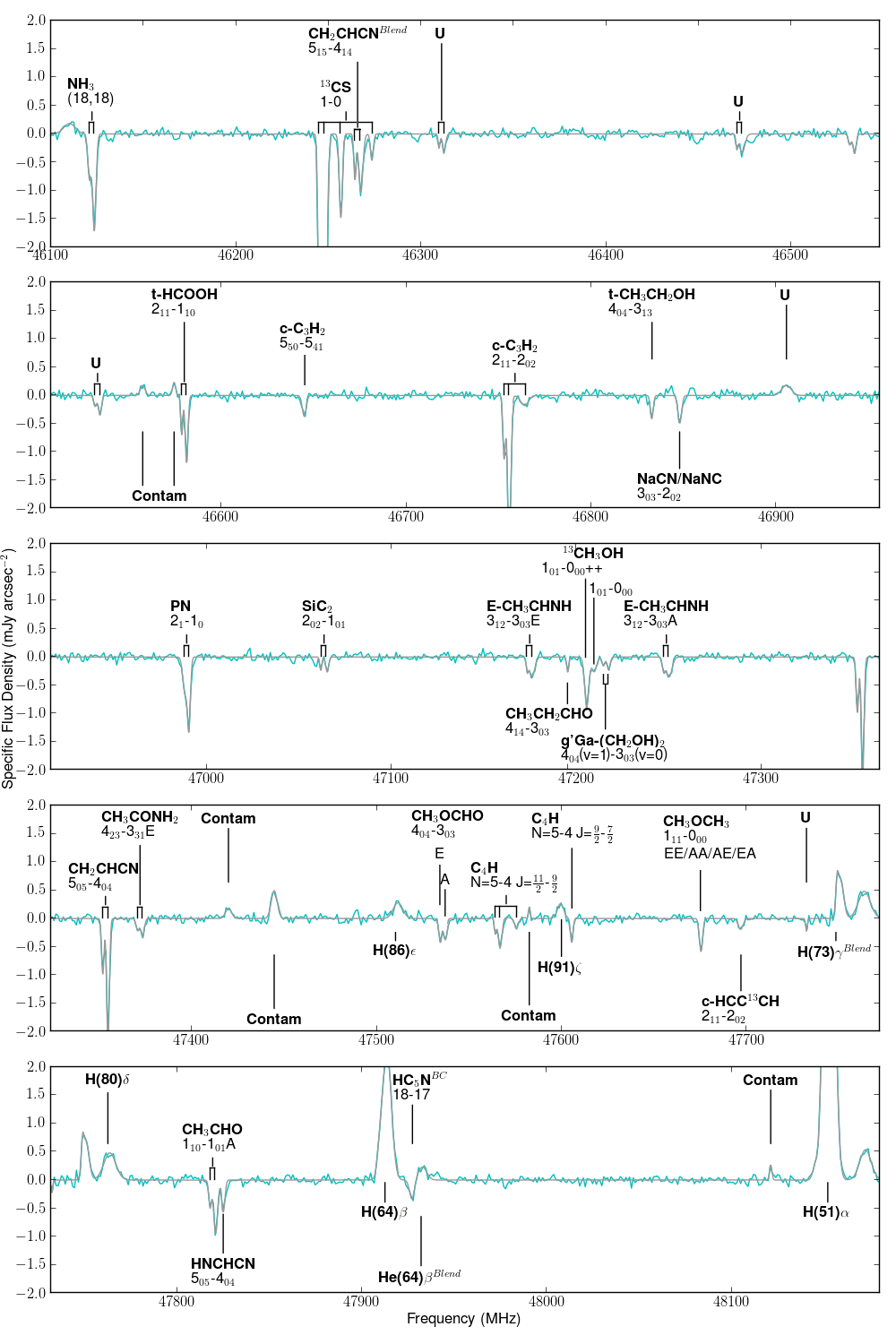}
\caption{Spectrum Towards K6.}
\label{fig:FullSpectrum_K6}
\end{figure*}

\begin{figure*}
\includegraphics[width=6.0in]{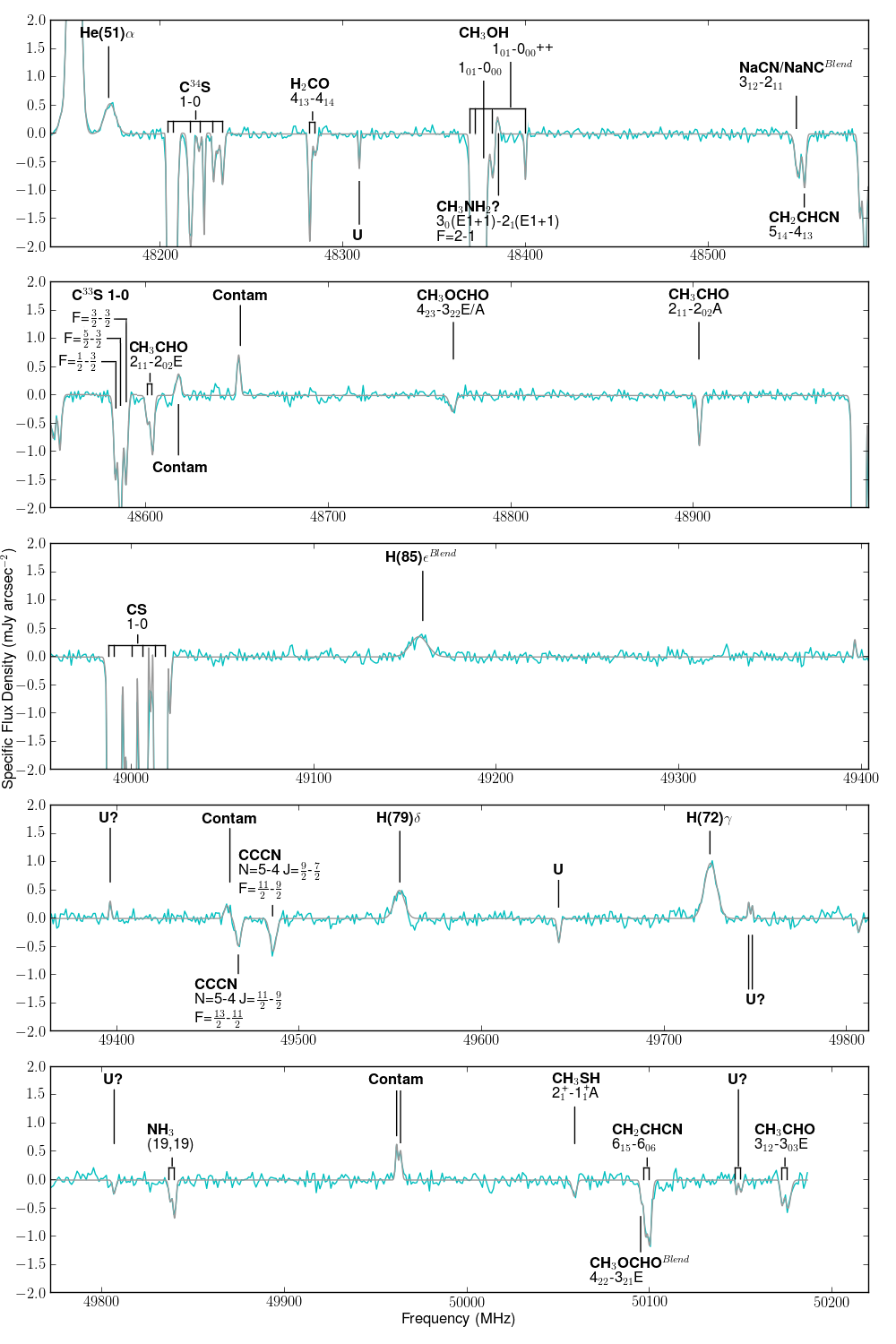}
\caption{Spectrum Towards K6.}
\label{fig:FullSpectrum_K6}
\end{figure*}

%%%%%%%%%%%%%%%%%%%%%%%%%%%%%%%%%%STARTS L.

\begin{figure*}
\includegraphics[width=6.0in]{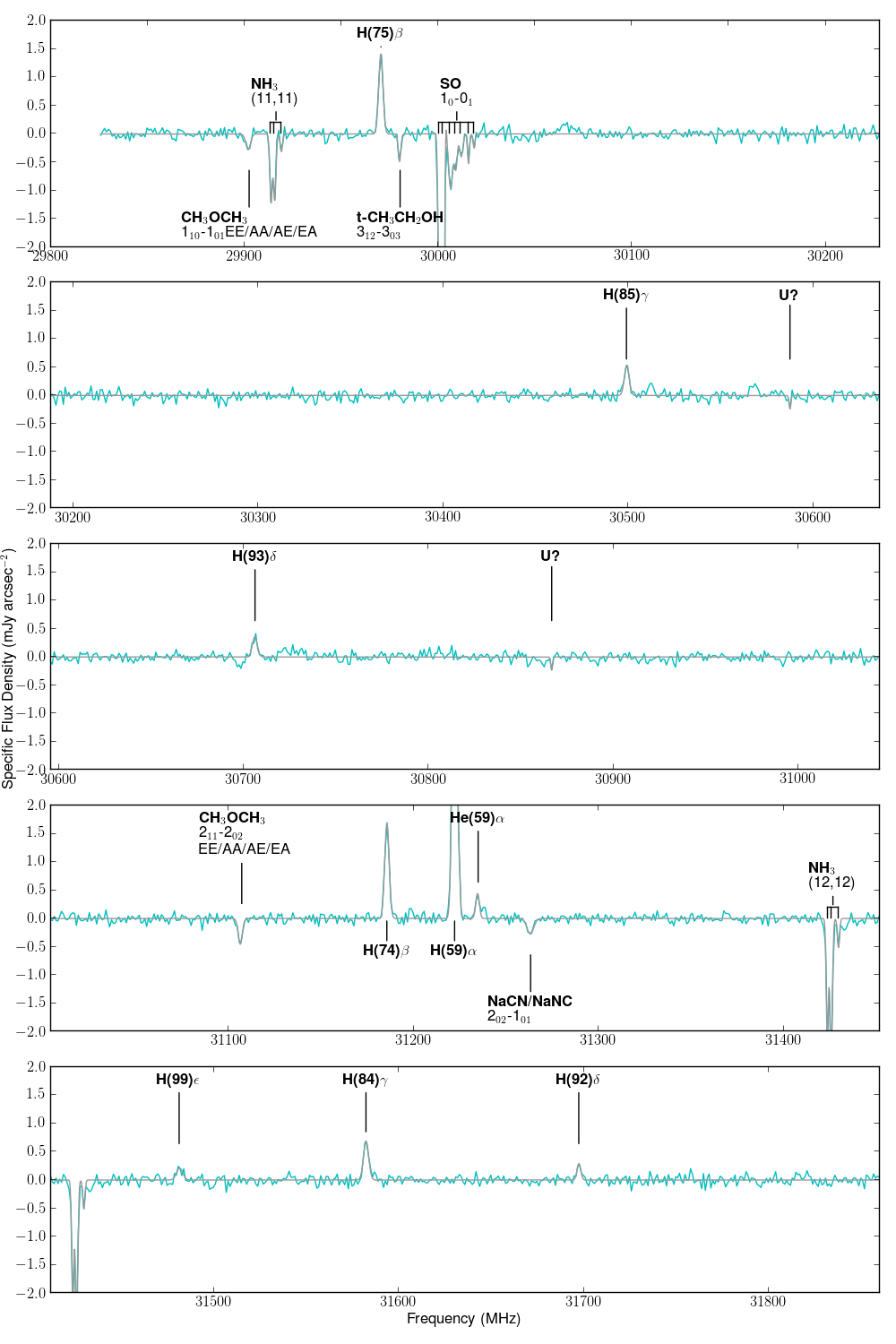}
\caption{Spectrum Towards L.}
\label{fig:FullSpectrum_K6}
\end{figure*}

\begin{figure*}
\includegraphics[width=6.0in]{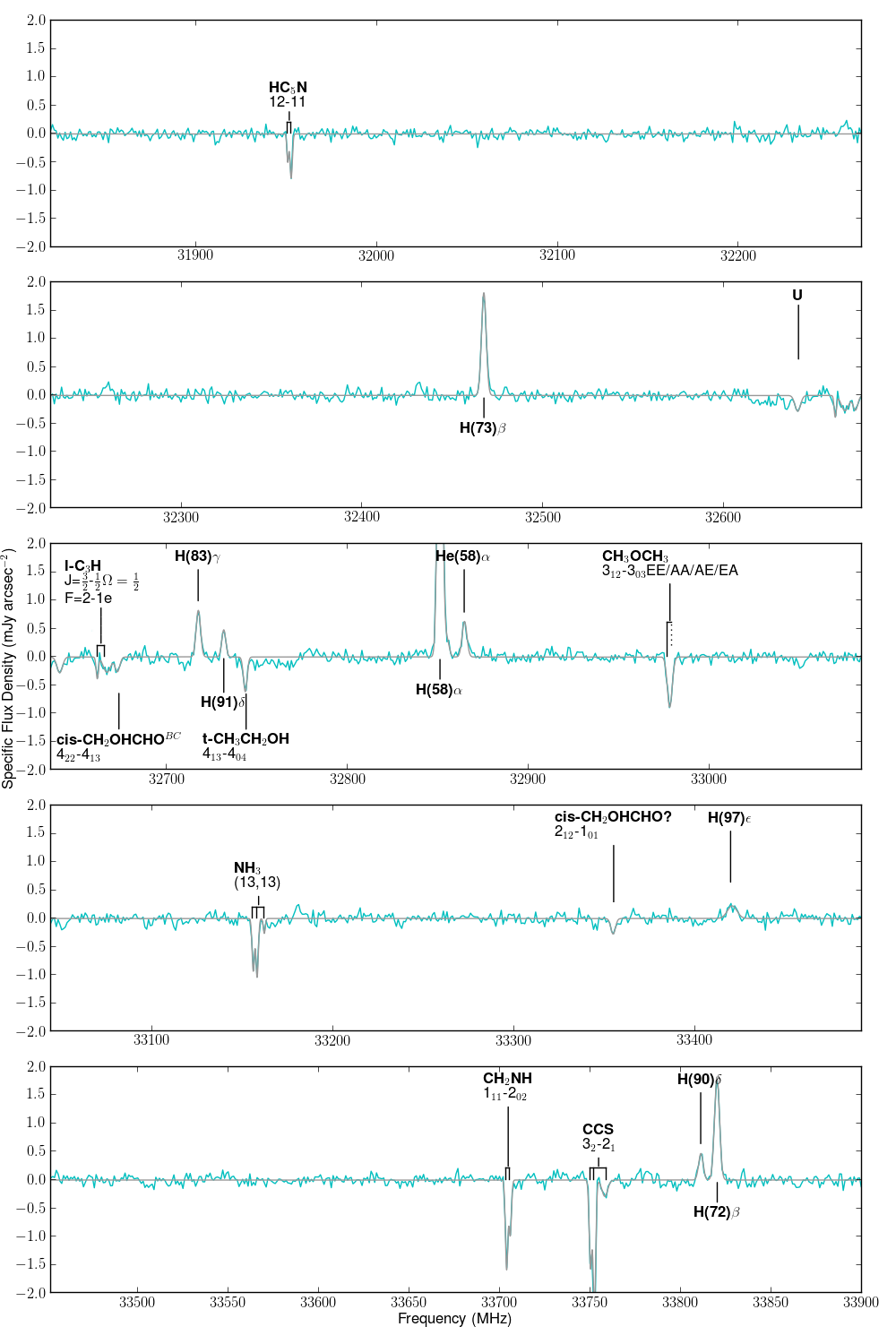}
\caption{Spectrum Towards L.}
\label{fig:FullSpectrum_K6}
\end{figure*}

\begin{figure*}
\includegraphics[width=6.0in]{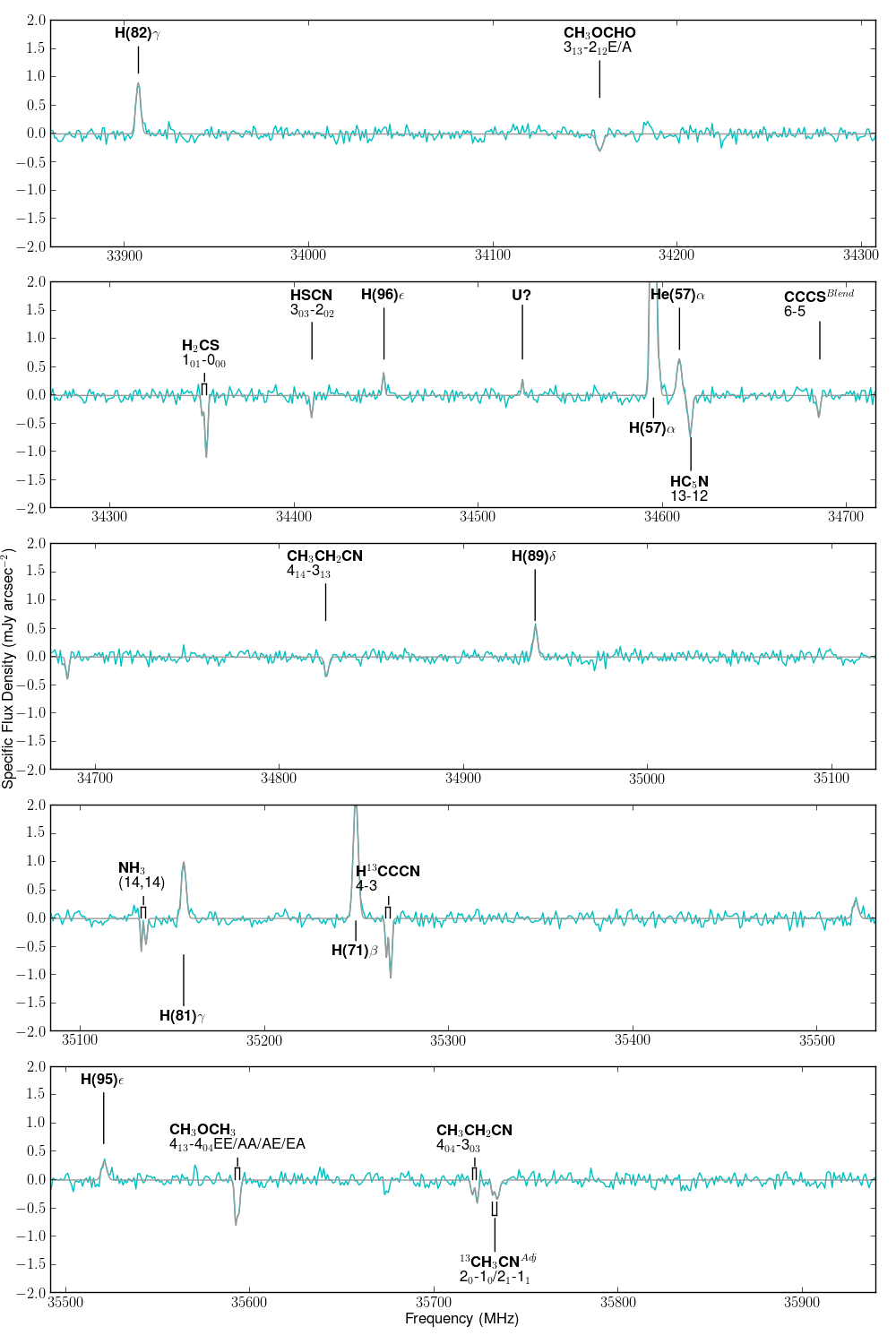}
\caption{Spectrum Towards L.}
\label{fig:FullSpectrum_K6}
\end{figure*}

\begin{figure*}
\includegraphics[width=6.0in]{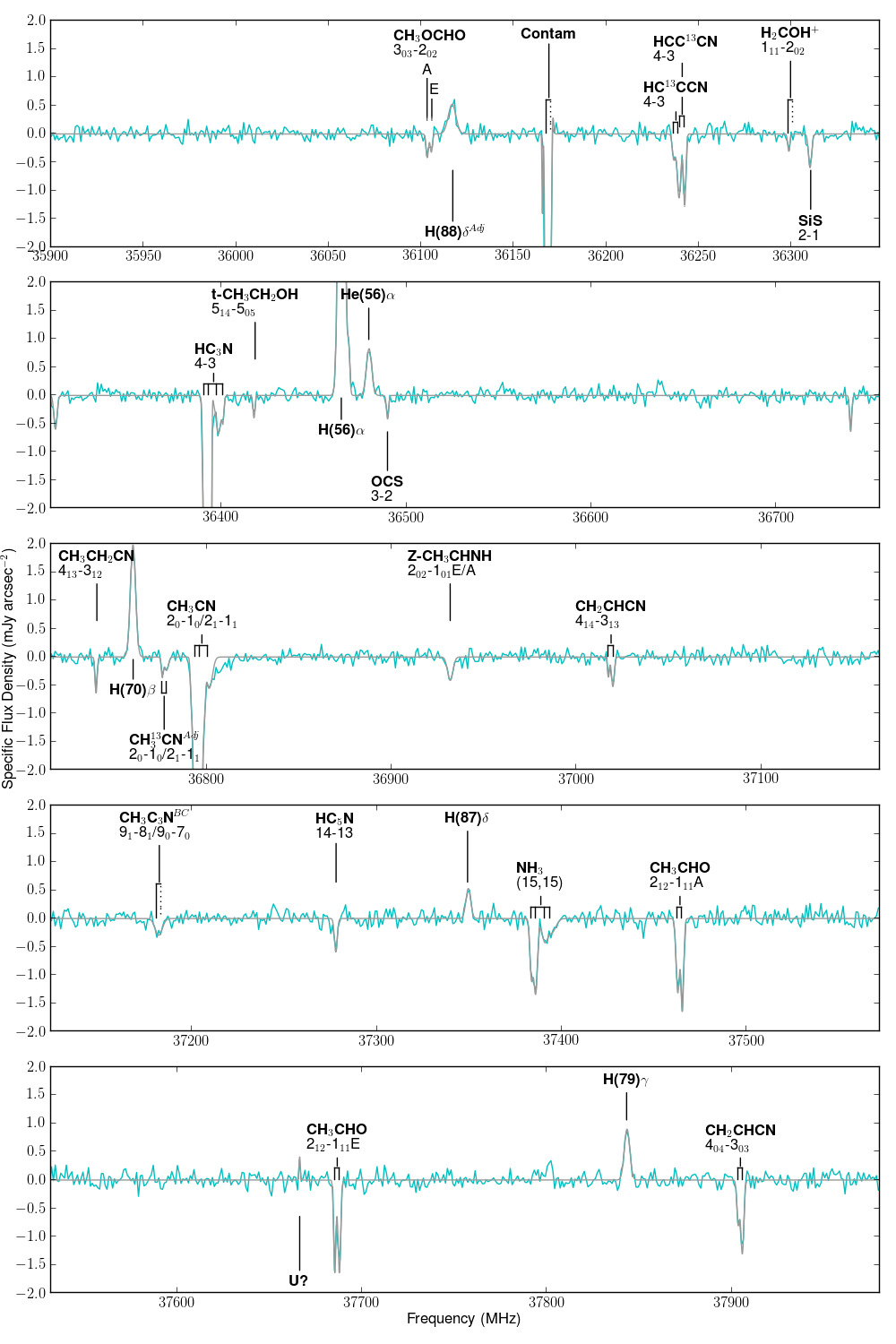}
\caption{Spectrum Towards L.}
\label{fig:FullSpectrum_K6}
\end{figure*}

\begin{figure*}
\includegraphics[width=6.0in]{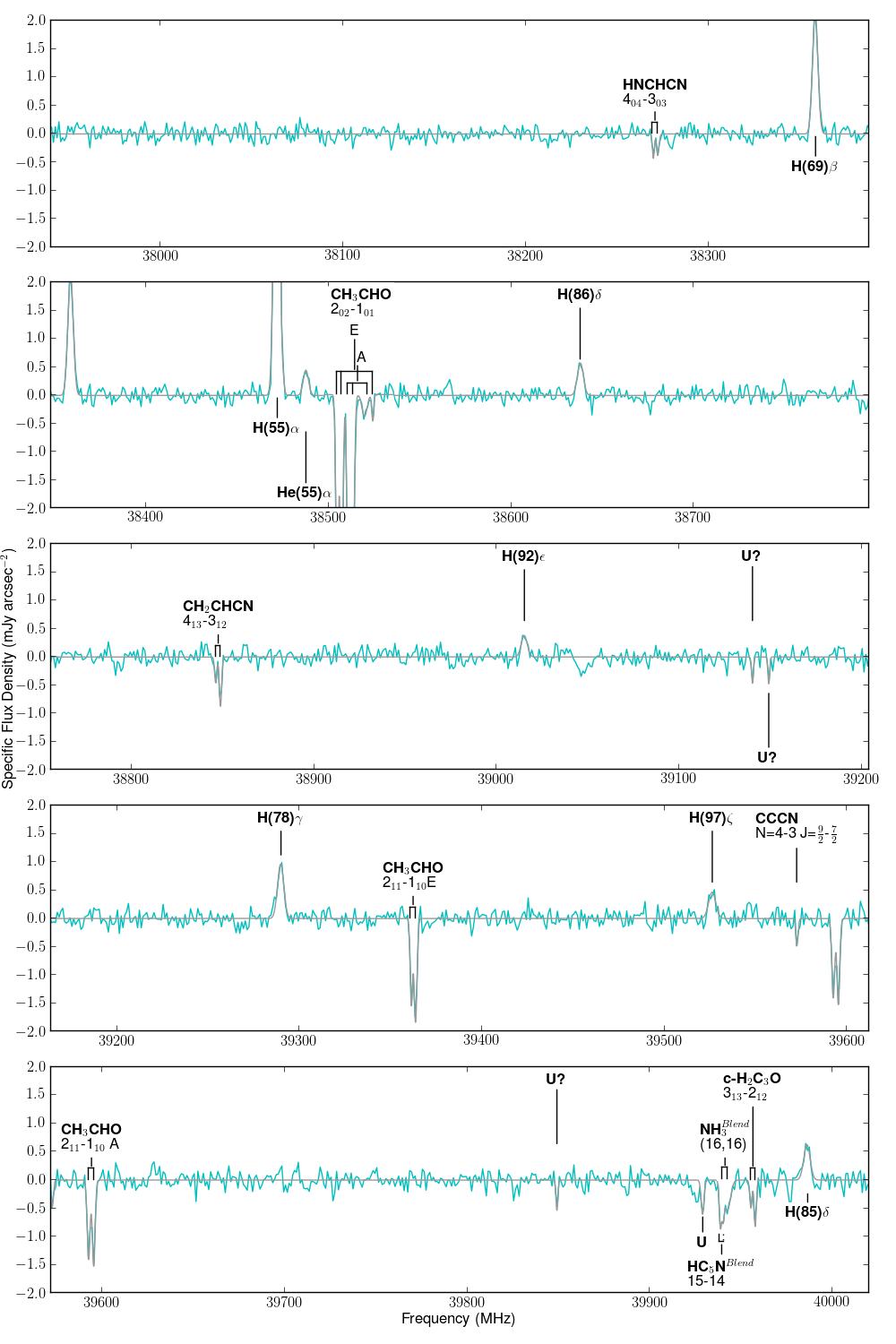}
\caption{Spectrum Towards L.}
\label{fig:FullSpectrum_K6}
\end{figure*}

\begin{figure*}
\includegraphics[width=6.0in]{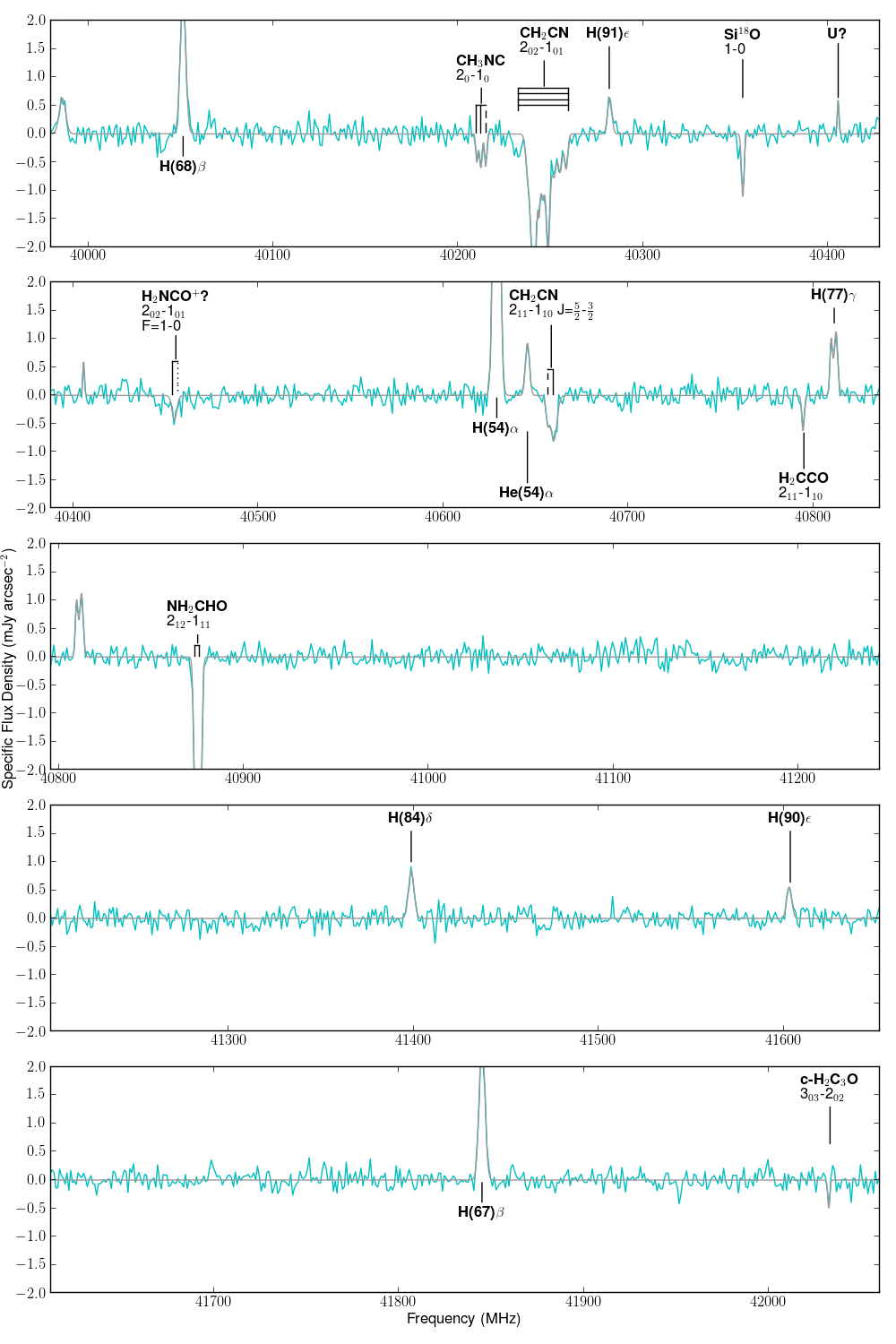}
\caption{Spectrum Towards L.}
\end{figure*}

\begin{figure*}
\includegraphics[width=6.0in]{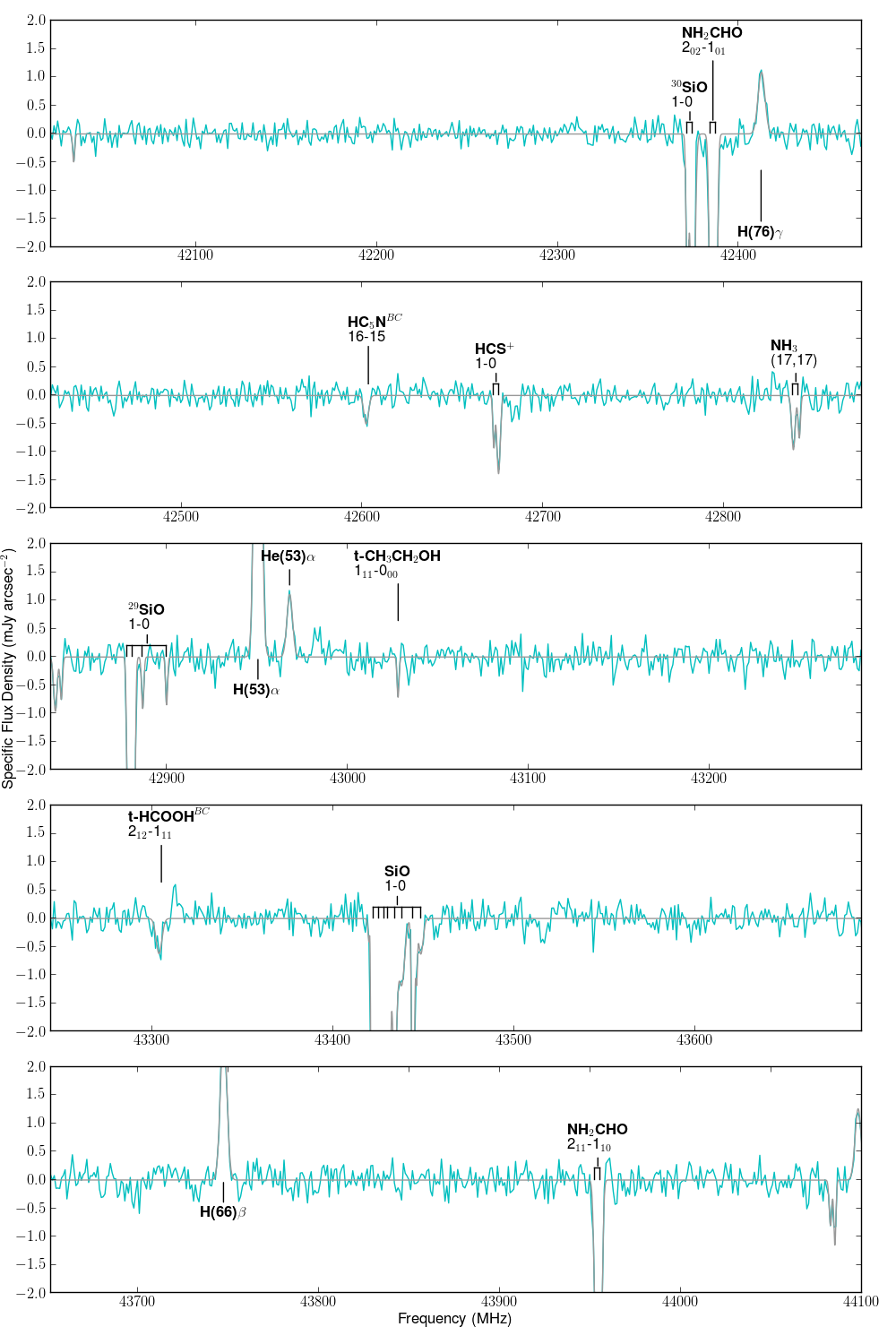}
\caption{Spectrum Towards L.}
\end{figure*}

\clearpage

\begin{figure*}
\includegraphics[width=6.0in]{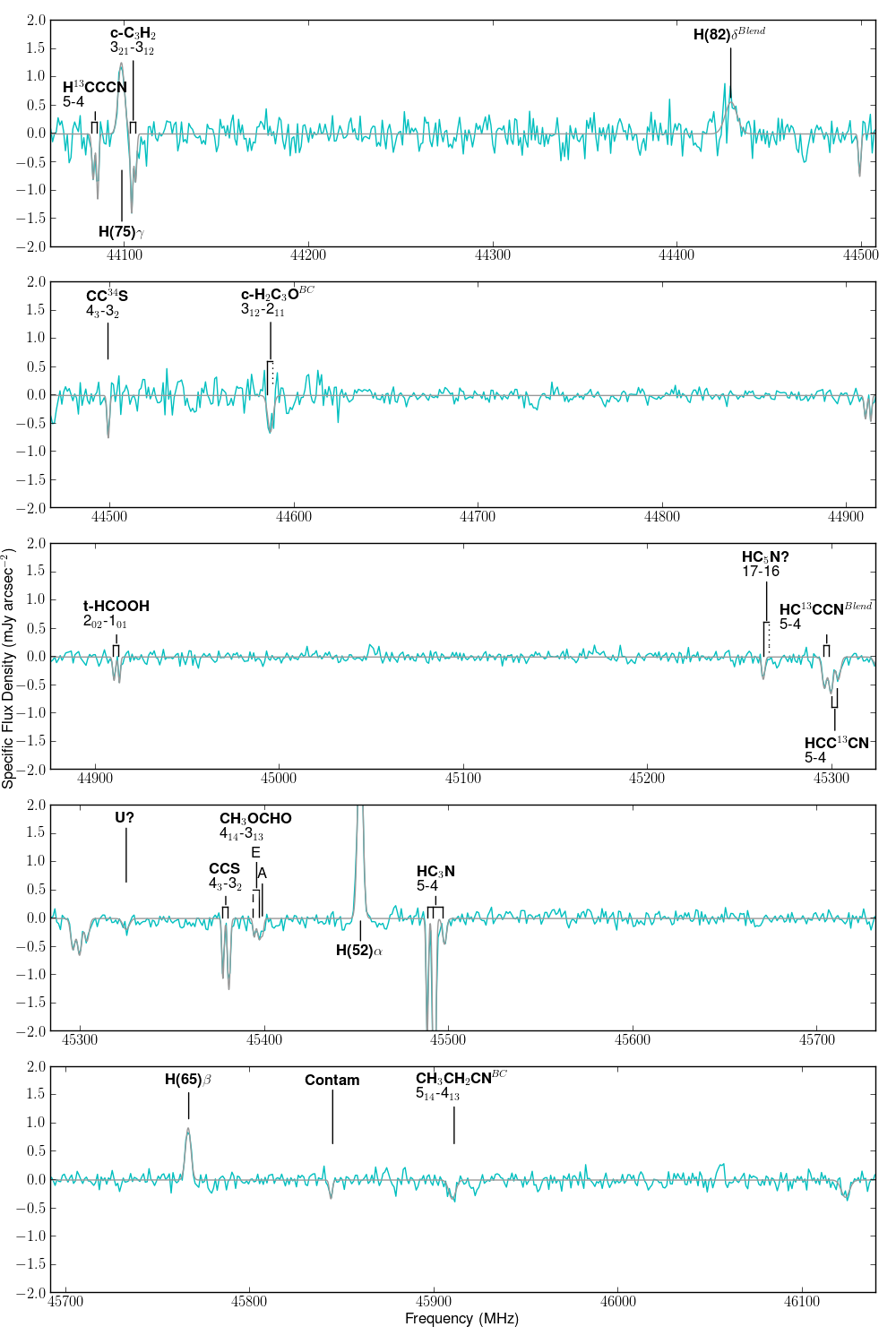}
\caption{Spectrum Towards L.}
\end{figure*}

\begin{figure*}
\includegraphics[width=6.0in]{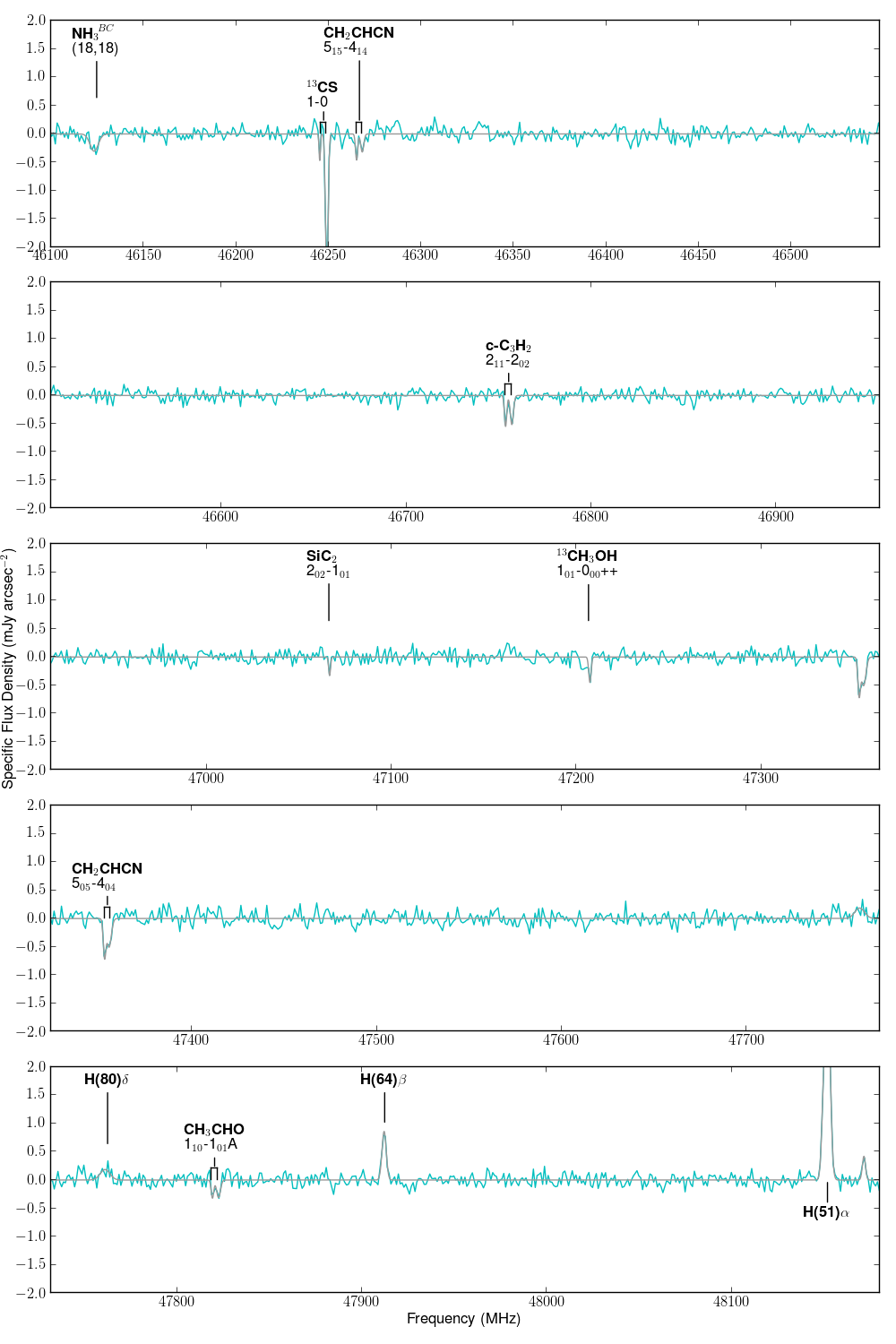}
\caption{Spectrum Towards L.}
\end{figure*}

\begin{figure*}
\includegraphics[width=6.0in]{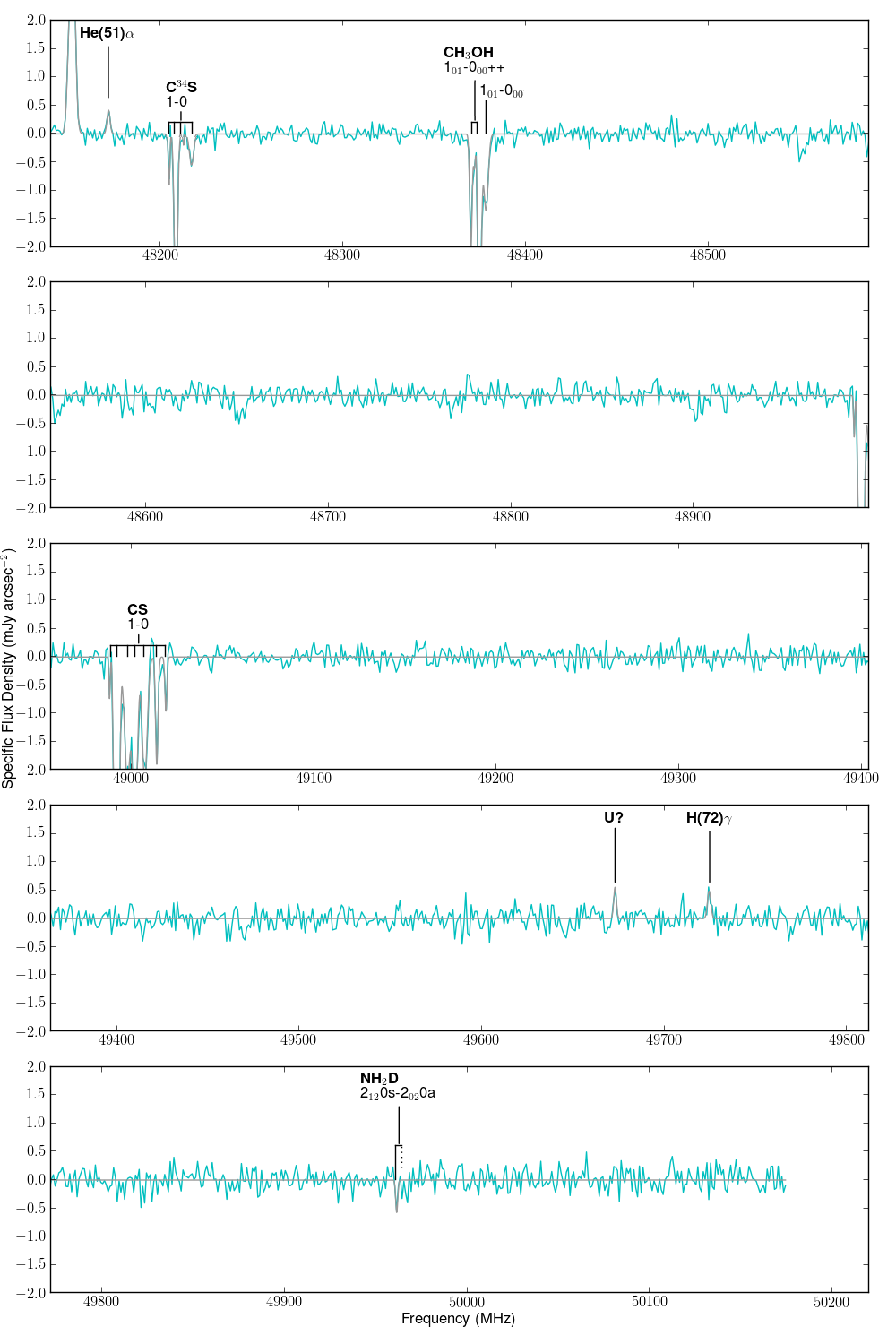}
\caption{Spectrum Towards L.}
\end{figure*}

%%%%%%%%%%%%%%%%%%%%%%%%%%%%%%%%%%STARTS K4.

\begin{figure*}
\includegraphics[width=6.0in]{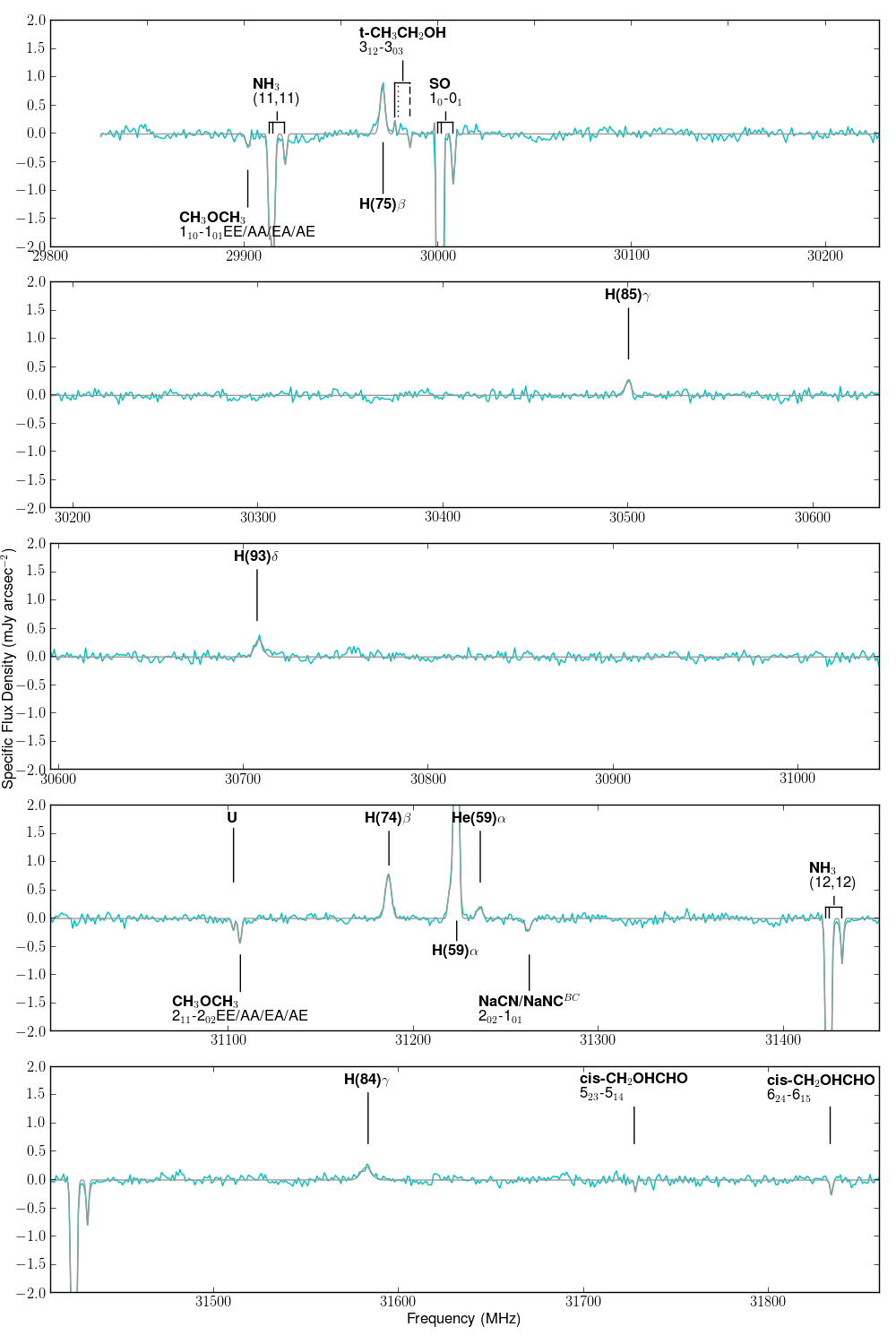}
\caption{Spectrum Towards K4.}
\end{figure*}

\begin{figure*}
\includegraphics[width=6.0in]{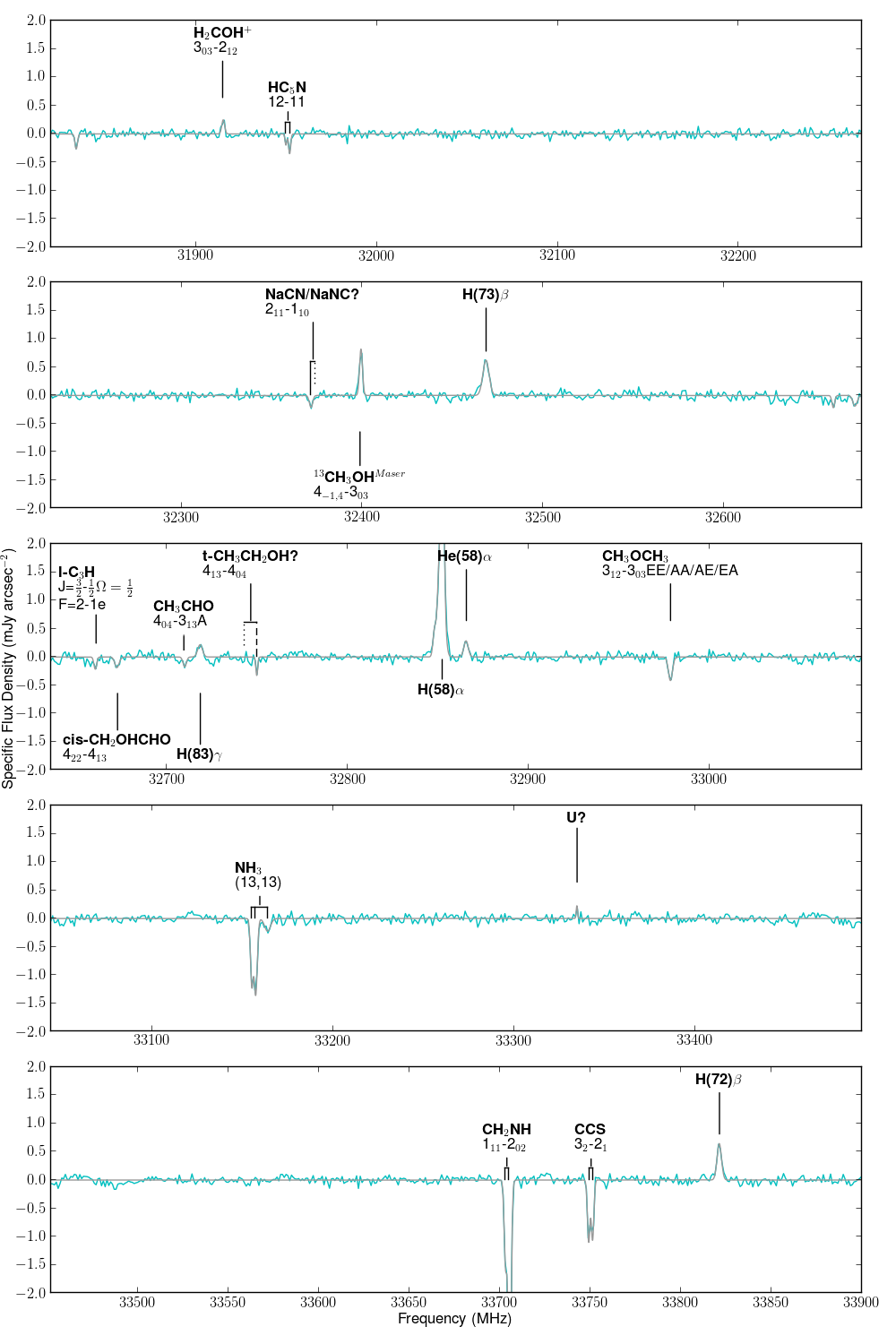}
\caption{Spectrum Towards K4.}
\end{figure*}

\begin{figure*}
\includegraphics[width=6.0in]{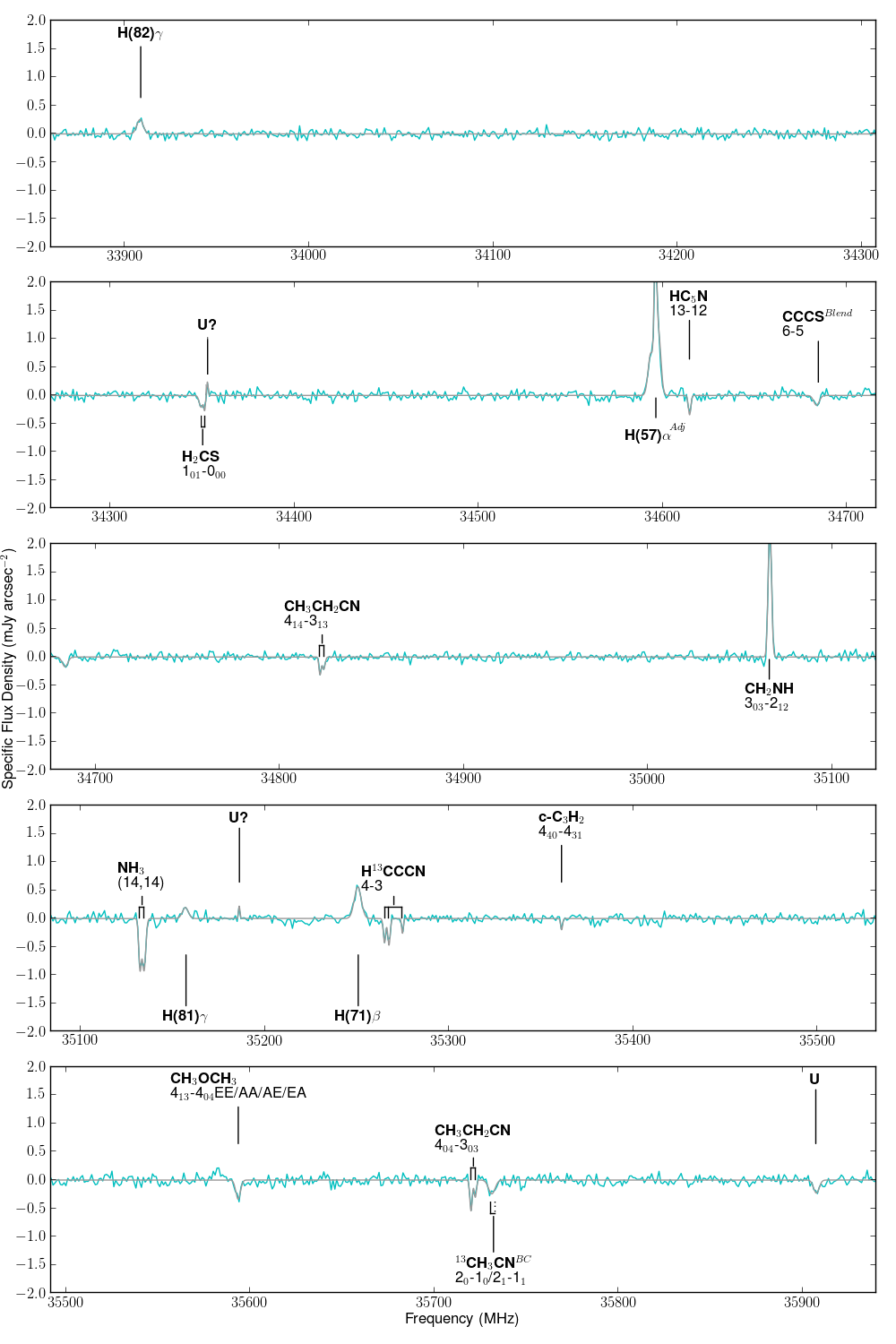}
\caption{Spectrum Towards K4.}
\end{figure*}

\begin{figure*}
\includegraphics[width=6.0in]{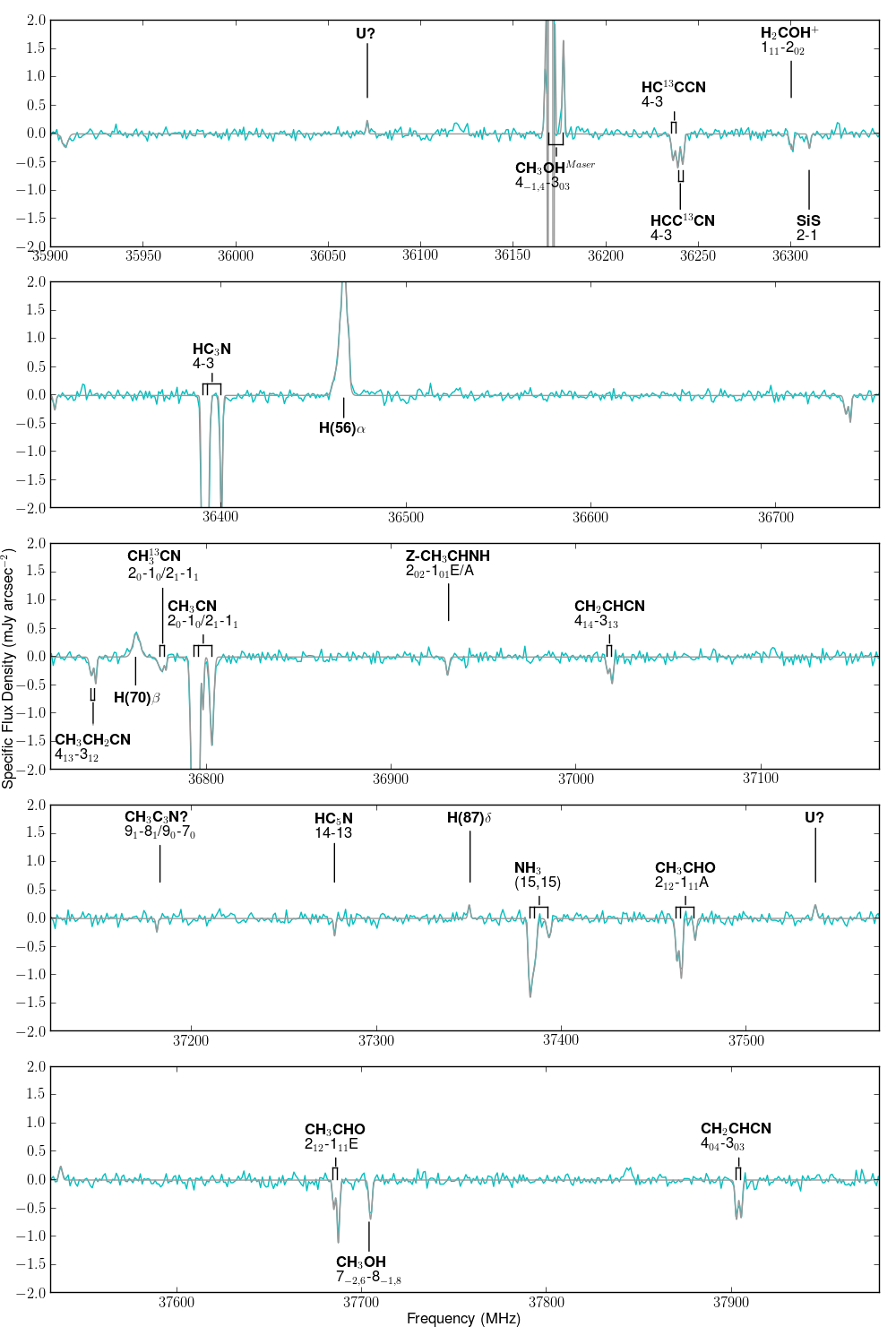}
\caption{Spectrum Towards K4.}
\end{figure*}

\begin{figure*}
\includegraphics[width=6.0in]{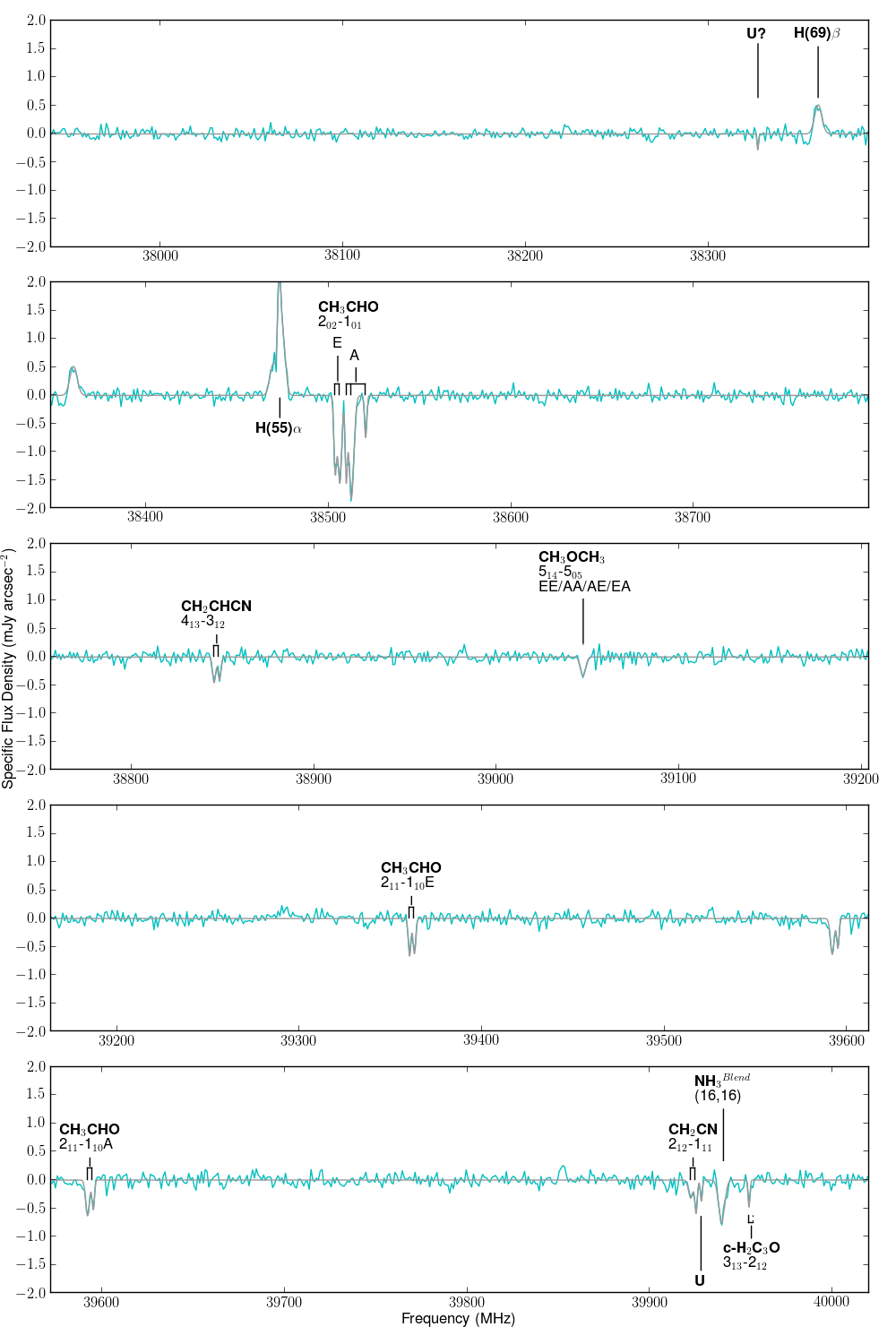}
\caption{Spectrum Towards K4.}
\end{figure*}

\begin{figure*}
\includegraphics[width=6.0in]{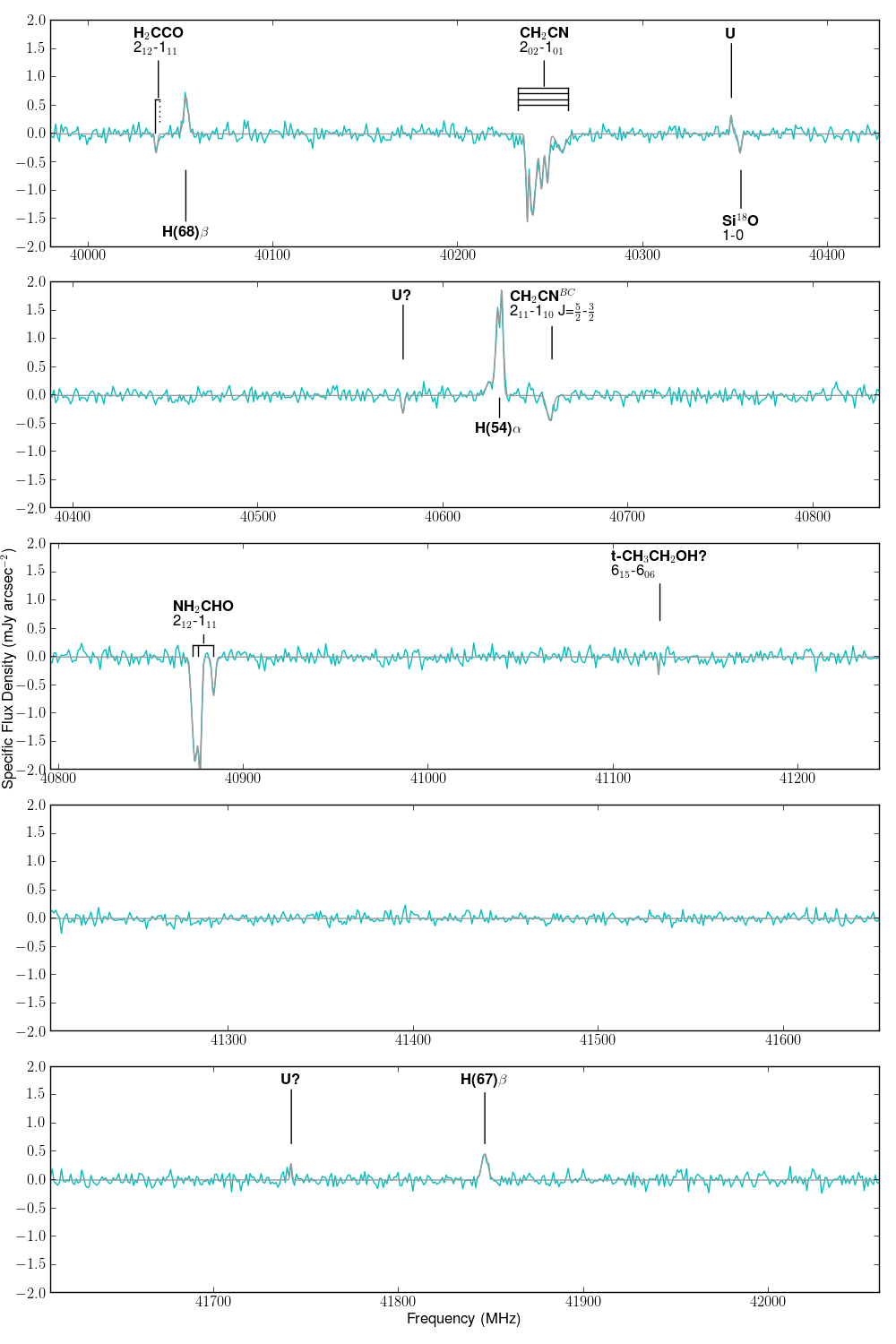}
\caption{Spectrum Towards K4.}
\end{figure*}

\begin{figure*}
\includegraphics[width=6.0in]{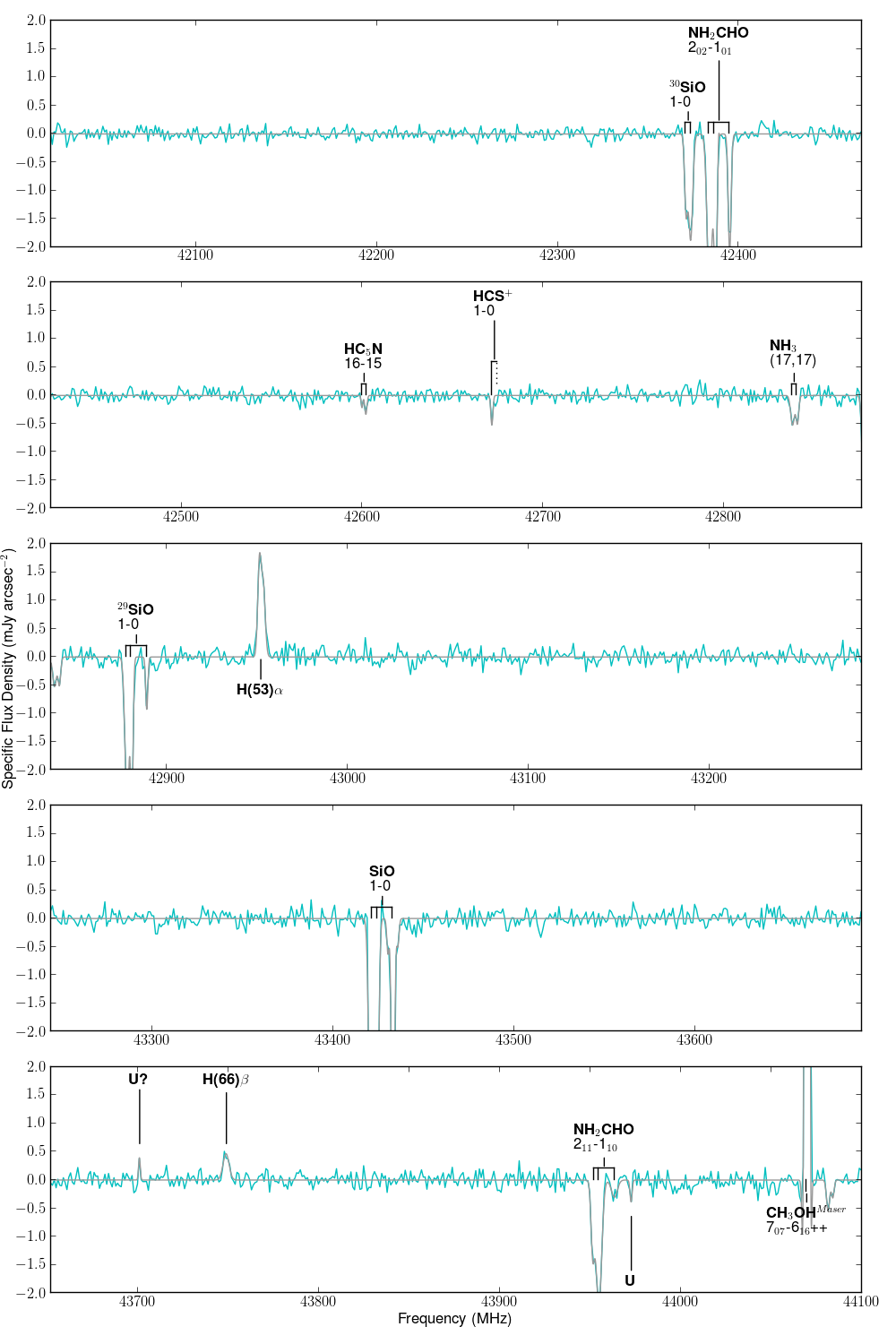}
\caption{Spectrum Towards K4.}
\end{figure*}

\begin{figure*}
\includegraphics[width=6.0in]{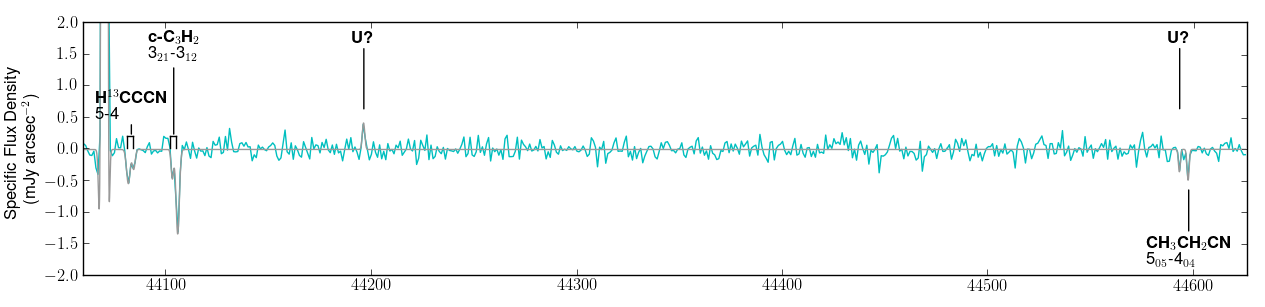}
\caption{Spectrum Towards K4.}
\end{figure*}

\clearpage

\section*{Appendix C\\Line Identification and Gaussian Fit Tables} \label{Ap:AppendixC}

Line identifications and Gaussian fit parameters, including velocity center, height, width, and integrated flux towards K6, L,
and K4 are provided. The following conventions are used in all tables.  
Tentative identifications are noted with a `?', as are unidentified lines (U-lines) suspected to be artifacts.  
Strong masing transitions are marked with `$Maser$', blended lines are marked with `$B$', 
and line fits that have been adjusted from the output of the line fitter are marked with `$Adj$'.
In cases where the line-fitter determined a 1-component fit that we suspect to be a 2-component transition
with insufficient signal-to-noise to be identified as such by the fitter, the species is marked with `$BC$'.
Footnotes, linked from the velocity column, provide notes on any other notable aspects of lines.

We have applied the following rules to assigning the rest frequencies and velocities to U-lines.  U-lines detected towards 
K6 are treated as follows:
\begin{enumerate}
 \item 1-component transitions detected towards K6 are assumed to be at a velocity of 64.0~\kmss.
 \item For 2-component transitions, we assume that the lower velocity component is at 64.0~\kmss.
\end{enumerate}
 U-lines detected towards L and K4 are treated as follows:
\begin{enumerate}
 \item If the U-line is detected towards K6, we use the rest frequency determined above. 
 \item If the U-line is not detected towards K6, we assume the velocity of the low velocity gas component, namely 56 and 62~\kms for L and K4 respectively.\
\end{enumerate}

For cases in which a transition has blended spectral 
structure (such as blended transitions of A- and E-states), the listed transition indicates what components are expected to be blended, 
with the first listed component corresponding to the transition rest frequency in the first column.

\onecolumn

\newpage
\begin{center}
\footnotesize{ 
\begin{longtable}{lllrrrr}
%\fontsize{10}{12}
%\centering
\caption{Line identifications and fit parameters towards K6.}
\label{bigTable_K6}\\  
 \hline \hline
Line Rest & Species & Transition & Fit Velocity  & Height & Width & $\int I_v dv$\\  
  Freq \footnotesize{(MHz)} &  &  & \footnotesize{(\kmss)} & \footnotesize{(mJy~arcsec$^{-2}$)} & \footnotesize{(\kmss)} & \footnotesize{(mJy~arcsec$^{-2}$ \hspace{.7mm} } \\  
  &&&&&& \footnotesize{ $\times$\,\kmss)}\\ 
\hline 
 \endfirsthead
 
\multicolumn{7}{c} 
 {{\bfseries \tablename\ \thetable{} -- Line Identifications and Fit Parameters towards K6}} \\  
 \hline
 
Line Rest & Species & Transition & Fit Velocity  & Height & Width & $\int I_v dv$\\  
  Freq \footnotesize{(MHz)} &  &  & \footnotesize{(\kmss)} & \footnotesize{(mJy~arcsec$^{-2}$)} & \footnotesize{(\kmss)} & \footnotesize{(mJy~arcsec$^{-2}$ \hspace{.7mm} } \\  
 &&&&&& \footnotesize{ $\times$\,\kmss)}\\ 
 \hline 
 \endhead
 
\hline \multicolumn{7}{|r|}{{Continued on next page}} \\  \hline 
 \endfoot
 
\hline \hline
 \endlastfoot 
 
29853.31 & U & \hspace{4mm}------   & 64.0 & 0.20 & 19.7 & 4.3 \\   
29901.40 & CH$_3$OCH$_3$ & 1$_{10}$-1$_{01}$EE/AA/AE/EA & 64.2 & -0.64 & 20.2 & -13.9 \\   
29914.49 & NH$_3$ & (11,11) & 83.4 & -2.51 & 12.5 & -29.5 \\   
 &  &  & 64.6 & -4.96 & 15.9 & -83.5 \\   
 &  &  & 11.3 & -0.52 & 18.6 & -11.4 \\   
29956.11 & U &  \hspace{4mm}------   & 64.0 & -0.23 & 20.1 & -4.9 \\   
29971.48 & Hydrogen & H(75)$\beta$ & 84.6 & 1.20 & 26.2 & 33.6 \\   
 &  &  & 61.8 & 1.82 & 32.3 & 62.7 \\
\hline
\end{longtable}
}

%%%%%%%%%%%%%%%%%%%%%%%%%%%%%%%%%%%%%%%%%%%%%%%%%%%%%%%%%%%%%%%%%%%%%%%%%%%%%%%%%%%%%%%%%%%%%%%%%%%%%%%%%%%%%%%%%%%%%%%%
%%%%%%%%%%%%%%%%%%%%   BEGINNING L 
\vspace{1.4in}
\footnotesize{ 
\begin{longtable}{lllrrrr}
%\fontsize{10}{12}
%\centering
\caption{Line identifications and fit parameters towards L.}
\label{bigTable_L}\\  
 \hline \hline
Line Rest & Species & Transition & Fit Velocity  & Height & Width & $\int I_v dv$\\  
  Freq \footnotesize{(MHz)} &  &  & \footnotesize{(\kmss)} & \footnotesize{(mJy~arcsec$^{-2}$)} & \footnotesize{(\kmss)} & \footnotesize{(mJy~arcsec$^{-2}$ \hspace{.7mm} } \\  
  &&&&&& \footnotesize{ $\times$\,\kmss)}\\ 
\hline 
 \endfirsthead
 
\multicolumn{7}{c} 
 {{\bfseries \tablename\ \thetable{} -- Line Identifications and Fit Parameters towards L}} \\  
 \hline
 
Line Rest & Species & Transition & Fit Velocity  & Height & Width & $\int I_v dv$\\  
  Freq \footnotesize{(MHz)} &  &  & \footnotesize{(\kmss)} & \footnotesize{(mJy~arcsec$^{-2}$)} & \footnotesize{(\kmss)} & \footnotesize{(mJy~arcsec$^{-2}$ \hspace{.7mm} } \\  
 &&&&&& \footnotesize{ $\times$\,\kmss)}\\ 
 \hline 
 \endhead
 
\hline \multicolumn{7}{|r|}{{Continued on next page}} \\  \hline 
 \endfoot
 
\hline \hline
 \endlastfoot 
 
29901.40 & CH$_3$OCH$_3$ & 1$_{10}$-1$_{01}$EE/AA/AE/EA & $^{[1]}$61.2 & -0.28 & 28.7 & -8.5 \\   
29914.49 & NH$_3$ & (11,11) & 74.0 & -1.21 & 14.9 & -19.2 \\   
 &  &  & 54.9 & -1.16 & 15.3 & -19.0 \\   
 &  &  & 20.2 & -0.31 & 14.5 & -4.8 \\   
29971.48 & Hydrogen & H(75)$\beta$ & 76.6 & 1.42 & 26.2 & 39.5 \\   
29979.38 & {\it t-}CH$_3$CH$_2$OH & 3$_{12}$-3$_{03}$ & 59.1 & -0.49 & 15.8 & -8.2 \\   
30001.55 & SO & 1$_{0}$-0$_{1}$ & 73.7 & -4.10 & 17.1 & -70.1 \\   
 &  &  & 56.1 & -8.86 & 14.2 & -133.1 \\
\hline
\end{longtable}
}
\end{center}

%%%%%%%%%%%%%%%%%%%%%%%%%%%%%%%%%%%%%%%%%%%%%%%%%%%%%%%%%%%%%%%%%%%%%%%%%%%%%%%%%%%%%%%%%%%%%%%%%%%%%%%%%%%%%%%%%%%%%%%%
%%%%%%%%%%%%%%%%%%%%   BEGINNING K4 
\vspace{1.4in}
\footnotesize{ 
\begin{longtable}{lllrrrr}
%\fontsize{10}{12}
%\centering
\caption{Line identifications and fit parameters towards K4.}
\label{bigTable_K4}\\  
 \hline \hline
Line Rest & Species & Transition & Fit Velocity  & Height & Width & $\int I_v dv$\\  
  Freq \footnotesize{(MHz)} &  &  & \footnotesize{(\kmss)} & \footnotesize{(mJy~arcsec$^{-2}$)} & \footnotesize{(\kmss)} & \footnotesize{(mJy~arcsec$^{-2}$ \hspace{.7mm} } \\  
  &&&&&& \footnotesize{ $\times$\,\kmss)}\\ 
\hline 
 \endfirsthead
 
\multicolumn{7}{c} 
 {{\bfseries \tablename\ \thetable{} -- Line Identifications and Fit Parameters towards K4}} \\  
 \hline
 
Line Rest & Species & Transition & Fit Velocity  & Height & Width & $\int I_v dv$\\  
  Freq \footnotesize{(MHz)} &  &  & \footnotesize{(\kmss)} & \footnotesize{(mJy~arcsec$^{-2}$)} & \footnotesize{(\kmss)} & \footnotesize{(mJy~arcsec$^{-2}$ \hspace{.7mm} } \\  
 &&&&&& \footnotesize{ $\times$\,\kmss)}\\ 
 \hline 
 \endhead
 
\hline \multicolumn{7}{|r|}{{Continued on next page}} \\  \hline 
 \endfoot
 
\hline \hline
 \endlastfoot 
%\begin{longtable}{lllrrrr}
%\caption{Line identifications and fit parameters towards K4.}
%\label{bigTable_K4}\\  
% \hline \hline
% Line Rest & Species & Transition & Fit Velocity  & Height & Width & $\int I_v dv$ \\  
%  Freq \footnotesize{(MHz)} &  &  & \footnotesize{(\kmss)} & \footnotesize{(mJy\,arcsec$^{-2}$)} & \footnotesize{(\kmss)} & \footnotesize{(mJy\,arcsec$^{-2}$  \hspace{.7mm}} \\  
%  &&&&&& \footnotesize{ $\times$\,\kmss)}\\ 
% \hline 
% \endfirsthead
 
%\multicolumn{7}{c} 
% {{\bfseries \tablename\ \thetable{} -- Line Identifications and Fit Parameters towards K4.}} \\  
% \hline
 
% Line Rest & Species & Transition & Fit Velocity  & Height & Width & $\int I_v dv$ \\  
%  Freq \footnotesize{(MHz)} &  &  & \footnotesize{(\kmss)} & \footnotesize{(mJy\,arcsec$^{-2}$)} & \footnotesize{(\kmss)} & \footnotesize{(mJy\,arcsec$^{-2}$ \hspace{.7mm}} \\  
%  &&&&&& \footnotesize{ $\times$\,\kmss)}\\ 
% \hline 
% \endhead
 
%\hline \multicolumn{7}{|r|}{{Continued on next page}} \\  \hline 
% \endfoot
 
%\hline \hline
% \endlastfoot 
29901.40 & CH$_3$OCH$_3$ & 1$_{10}$-1$_{01}$EE/AA/EA/AE & 62.4 & -0.24 & 22.7 & -5.8 \\   
29914.49 & NH$_3$ & (11,11) & 82.9 & -1.64 & 15.9 & -27.8 \\   
 &  &  & 63.3 & -2.49 & 20.2 & -53.4 \\   
 &  &  & 0.1 & -0.54 & 17.5 & -10.1 \\   
29971.48 & Hydrogen & H(75)$\beta$ & 67.9 & 0.87 & 30.2 & 27.8 \\   
29979.38 & {\it t-}CH$_3$CH$_2$OH & 3$_{12}$-3$_{03}$ & $^{[1]}$83.0 & 0.24 & 11.9 & 3.0 \\   
 &  &  & 4.7 & -0.25 & 12.2 & -3.2 \\   
30001.55 & SO & 1$_{0}$-0$_{1}$ & 83.3 & -7.41 & 15.8 & -115.6 \\
\hline
\end{longtable}
}

\end{document}